\documentclass[prx,superscriptaddress,aps,amsmath,amssymb,floatfix,reprint,raggedbottom]{revtex4-2}
    \usepackage[margin=1in]{geometry}
    \usepackage{graphicx}
    \usepackage[subpreambles=true]{standalone}
    \usepackage{balance}
    \usepackage{hyperref}
    
    \hypersetup{
        colorlinks=true,
        linkcolor=blue,
        citecolor=blue,
        urlcolor=black
    }
    
    \usepackage{amsmath}
    \usepackage{times}
    
    \usepackage{dcolumn}
    \usepackage{xcolor}
    \usepackage{lipsum}
    \usepackage{soul}
    \usepackage{braket}
    \usepackage{amsfonts}
    \usepackage{multirow}
    \usepackage[utf8]{inputenc}
    
    \usepackage[ruled]{algorithm2e}

    \DeclareUnicodeCharacter{2009}{\,}

    \makeatletter
    \renewcommand{\fnum@figure}{\textbf{Fig.~\thefigure}}
    \makeatother
    
    \AtBeginDocument{%
    \makeatletter
    \long\def\@makecaption#1#2{%
      \vskip\abovecaptionskip
      {\footnotesize #1.
    #2\par}%
      \vskip\belowcaptionskip}%
    \makeatother
    }
    
    \makeatletter
    \def\bbordermatrix#1{\begingroup \m@th
      \@tempdima 4.75\p@
      \setbox\z@\vbox{%
        \def\cr{\crcr\noalign{\kern2\p@\global\let\cr\endline}}%
        \ialign{$##$\hfil\kern2\p@\kern\@tempdima&\thinspace\hfil$##$\hfil
          &&\quad\hfil$##$\hfil\crcr
          \omit\strut\hfil\crcr\noalign{\kern-\baselineskip}%
          #1\crcr\omit\strut\cr}}%
      \setbox\tw@\vbox{\unvcopy\z@\global\setbox\@ne\lastbox}%
      \setbox\tw@\hbox{\unhbox\@ne\unskip\global\setbox\@ne\lastbox}%
      \setbox\tw@\hbox{$\kern\wd\@ne\kern-\@tempdima\left[\kern-\wd\@ne
        \global\setbox\@ne\vbox{\box\@ne\kern2\p@}%
        \vcenter{\kern-\ht\@ne\unvbox\z@\kern-\baselineskip}\,\right]$}%
      \null\;\vbox{\kern\ht\@ne\box\tw@}\endgroup}
    \makeatother
    
    \emergencystretch=\maxdimen
    \usepackage[compact]{titlesec}
    \titlespacing{\section}{0pt}{*3}{*2}
    \titlespacing{\subsection}{0pt}{*2}{*2}
    \titlespacing{\subsubsection}{0pt}{*2}{*2}
    
    \titleformat{\section}{\normalfont\small\bfseries}{\thesection.}{1em}{\MakeUppercase}

    \usepackage{booktabs}

\usepackage{placeins}

\newcommand{\beginsupplement}{%
        \setcounter{subsection}{0}%
        \setcounter{subsubsection}{0}%
        \renewcommand{\thesubsection}{S\arabic{subsection}}%
        \renewcommand{\thesubsubsection}{\thesubsection.\arabic{subsubsection}}%
        \setcounter{table}{0}%
        \renewcommand{\thetable}{S\arabic{table}}%
        \setcounter{figure}{0}%
        \renewcommand{\thefigure}{S\arabic{figure}}%
        \renewcommand{\theHfigure}{S\arabic{figure}}%
        \renewcommand{\theHtable}{S\arabic{table}}%
        \renewcommand{\theHequation}{S.\arabic{equation}}%
        \setcounter{equation}{0}%
        \renewcommand{\theequation}{S.\arabic{equation}}%
        \setcounter{section}{0}%
        \renewcommand{\thesection}{}%
     }
    
\begin{document}
    
    \title{Programmable Probabilistic Computer with 1,000,000 p-bits}
    
\author{Navid Anjum Aadit$^{\dagger}$}
\email{navidanj@stanford.edu}
\altaffiliation[Present address: ]{Department of Electrical Engineering,
Stanford University, Stanford, CA 94305, USA}
\affiliation{Department of Electrical \& Computer Engineering, University of California, Santa Barbara, Santa Barbara, CA 93106, USA}

\author{Xiuqi Zhang}
\thanks{These authors contributed equally to this work.}
\affiliation{Department of Electrical \& Computer Engineering, University of California, Santa Barbara, Santa Barbara, CA 93106, USA}

 \author{Shuvro Chowdhury}
    \affiliation{Department of Electrical \& Computer Engineering, University of California, Santa Barbara, Santa Barbara, CA 93106, USA}
    
    \author{Kevin Callahan-Coray}
    \affiliation{Department of Electrical \& Computer Engineering, University of California, Santa Barbara, Santa Barbara, CA 93106, USA}
    
    \author{Kyle Lee}
    \affiliation{Department of Electrical \& Computer Engineering, University of California, Santa Barbara, Santa Barbara, CA 93106, USA}
    
    \author{Saleh Bunaiyan}
    \affiliation{Department of Electrical \& Computer Engineering, University of California, Santa Barbara, Santa Barbara, CA 93106, USA}
    
    \affiliation{Electrical Engineering Department, King Fahd University of Petroleum \& Minerals,
    Dhahran 31261, Saudi Arabia}
    
    \author{Sanjay Seshan}
    \affiliation{Department of Electrical and Computer Engineering, Carnegie Mellon University, Pittsburgh, PA 15213, USA}
    
    \author{Clayton Thomas}
    \affiliation{Siemens Digital Industries Software, USA}
    
    \author{Jason Twigg}
    \affiliation{Siemens Digital Industries Software, USA}

    \author{Andrew Seawright}
    \affiliation{Siemens Digital Industries Software, USA}

    \author{Forrest Brewer}
    \affiliation{Department of Electrical \& Computer Engineering, University of California, Santa Barbara, Santa Barbara, CA 93106, USA}
    
    \author{Tathagata Srimani}
    \affiliation{Department of Electrical and Computer Engineering, Carnegie Mellon University, Pittsburgh, PA 15213, USA}

    \author{Kerem Y. \c{C}amsar{\i}}
    \email{Corresponding author: camsari@ucsb.edu}
    \affiliation{Department of Electrical \& Computer Engineering, University of California, Santa Barbara, Santa Barbara, CA 93106, USA}
    
      \begin{abstract}
    Many problems in physics and machine learning require sampling the collective states of large numbers of interacting variables. On modern graphics processors, repeated movement of interaction data between memory and compute units can limit how efficiently these calculations can scale. Here we demonstrate a programmable probabilistic computer with one million interacting probabilistic bits (p-bits) across 18 field-programmable gate arrays, sustaining \(10^{12}\) probabilistic updates per second. Coupling weights remain beside the update circuits in on-chip memory, and only single-bit boundary states pass between devices. Experiments on three-dimensional and quasi-two-dimensional spin glasses show that the ratio of boundary-exchange frequency to local update frequency controls the effect of communication delays on optimization. A bound derived from the graph partition and interconnect identifies operating conditions that recover the residual-energy decay of an unpartitioned GPU reference.
    Infrequent exchange reduces the decay exponent, although faster local updates can shorten the time to reach a target energy. A cluster mean-field model reproduces this dependence, while ferromagnetic sampling experiments reveal a corresponding dependence of magnetization relaxation on the same frequency ratio. These results establish a million-p-bit computing architecture in which processing and local memory scale together, and provide a quantitative design principle for exploiting tolerance to communication delays in larger probabilistic computers.
    \end{abstract}
    \maketitle

    \section{Introduction}
    \label{sec:introduction}
   Large computing systems combine many chips to increase processing
capacity and memory~\cite{radway2021illusion}. In conventional
hardware, including GPUs, moving data between memory and computing
units can constrain both performance and energy
efficiency~\cite{sze2017efficient,boroumand2018google,horowitz20141,dayo2025future,gholami2024ai}.
Quantum processors~\cite{King2023quantum,king2025beyond,mohseni2024build} also face limits on the number of interacting units that can be
accommodated on a single device. Distributing these units across
chips increases capacity, but keeping their interactions current
requires communication to keep pace with local computation.
Probabilistic workloads and computing offer great scope to relax this requirement.
Stochastic units can continue updating while information travels
between chips, provided the resulting changes in their collective
behaviour remain acceptable for the task. Establishing how delayed
information affects sampling and optimization would therefore allow
communication to be provisioned according to the accuracy required,
while permitting faster local updates.
    
Combinatorial optimization problems can be expressed as searches
for low-energy states of an Ising
Hamiltonian~\cite{kirkpatrick1983optimization,lucas2014ising}.
Ising machines implement these searches in
hardware~\cite{mohseni2022ising,finocchio2024roadmap}.
Probabilistic implementations, including p-bit-based machines,
can also sample Boltzmann distributions and therefore support
probabilistic inference~\cite{singh2024cmos}.
We examine the effect of delayed communication using
three-dimensional Edwards-Anderson (EA)
spin glasses~\cite{edwards1975theory} and spin glasses on
the quasi-two-dimensional Pegasus graph~\cite{jauma2024exploring}.
Ground-state search in the three-dimensional model is
NP-hard~\cite{barahona1982computational}.
For both graph families, we measure relaxation towards low-energy
states during annealing. We also study a ferromagnet at fixed
temperature, following its magnetization as it approaches
equilibrium to examine the effect of delayed communication
on a sampling observable.   
    
    \begin{figure*}[t]
      \centering
      \includegraphics[width=\textwidth]{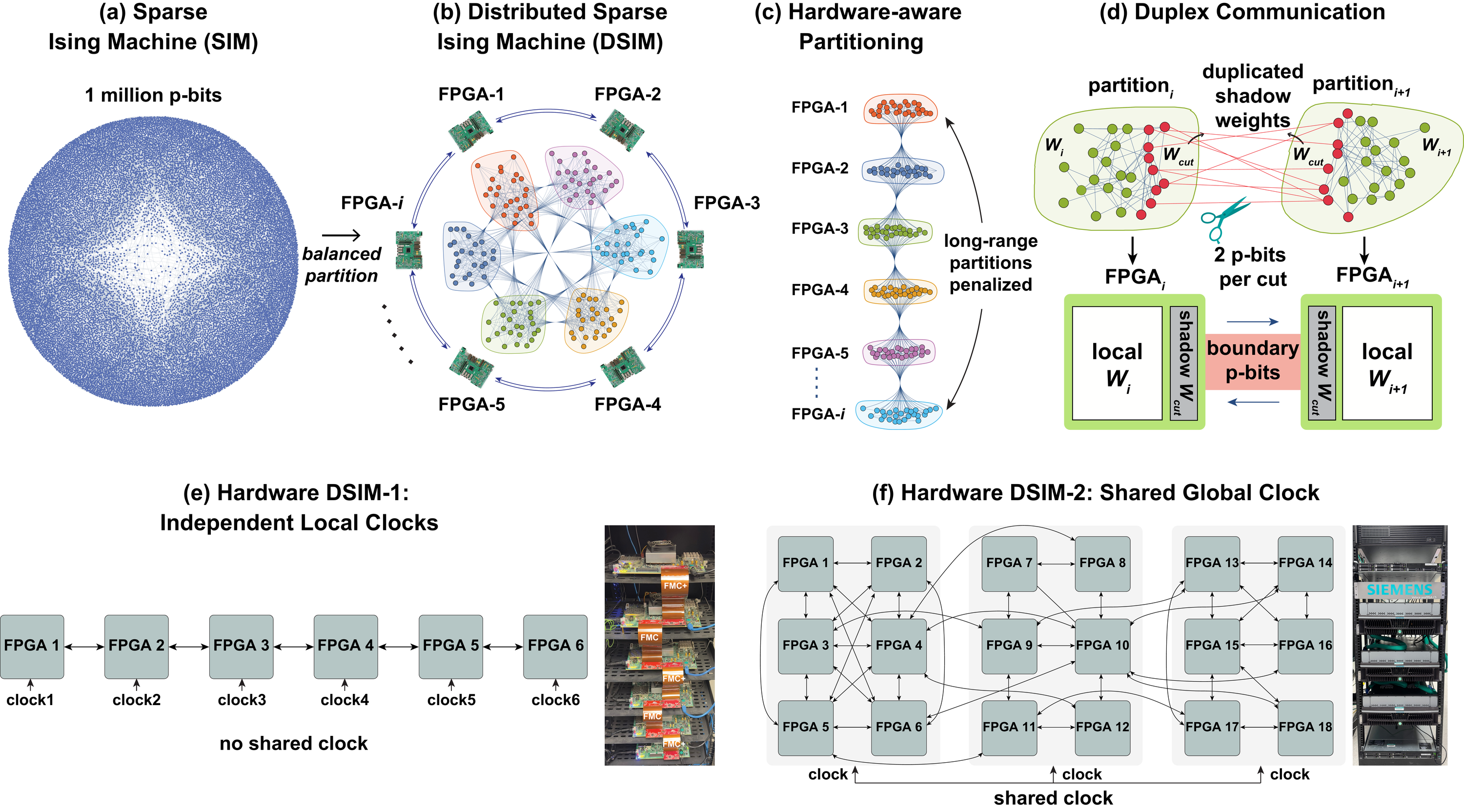}
      \caption{
      \textbf{Distributed sparse Ising machines (DSIMs).}
      \textbf{(a)} A sparse Ising graph at the million-p-bit scale (fewer
      nodes rendered for clarity).
      \textbf{(b)} A balanced min-cut partition maps the graph onto multiple
      FPGAs, each updating its local subgraph from on-chip weight memory.
      \textbf{(c)} Hardware-aware partitioning: a Potts cost function
      penalizes placing strongly connected partitions on distant devices,
      concentrating cut edges at short hop distances (Supplementary
      Sec.~\ref{sec:distance_distribution_app}).
      \textbf{(d)} Duplex boundary exchange: cut-edge weights are duplicated
      as shadow weights on both sides of each cut, so only 1-bit boundary
      states ever cross device boundaries.
      \textbf{(e)} DSIM-1: six FPGAs in a nearest-neighbor chain over FPGA Mezzanine Card (FMC)
      links, each running an independent local clock.
      \textbf{(f)} DSIM-2: eighteen FPGAs on a Siemens Veloce proFPGA CS platform
      sharing a single global clock, with denser interconnect.}
      \label{fig:extreme_fig1}
    \end{figure*}
    
   On the three-dimensional EA benchmark, quantum annealers have
demonstrated coherent critical dynamics with approximately
$2{,}700$ logical spins~\cite{King2023quantum}.
Replica-based Monte Carlo algorithms on probabilistic computers
have matched or improved upon the residual-energy scaling of
quantum annealing on the same
instances~\cite{chowdhury2025pushing}.
For studies of spin-glass statistical mechanics, the Janus
field-programmable gate array (FPGA)
computers~\cite{belletti2008janus,baity2014janus} and optimized
GPU codes~\cite{lulli2015highly,bernaschi2024qisg} have reached
millions of spins on regular lattices.
Coherent Ising machines have reached $10^5$
spins~\cite{honjo2021100}, and complementary metal-oxide-semiconductor
(CMOS) annealing chips tens of thousands~\cite{takemoto20214}.
Simulated bifurcation has also been implemented in
hardware~\cite{goto2019combinatorial,goto2021high} and scaled
across multiple FPGAs~\cite{tatsumura2021scaling}.
For distributed stochastic hardware, we ask how far boundary
exchange can be reduced while local updates continue, and
what cost this imposes on sampling and optimization.
    
  Here we quantify how delayed boundary information affects
optimization and sampling using two distributed sparse Ising
machines (DSIMs; Fig.~\ref{fig:extreme_fig1}).
DSIM-2 combines 18 FPGAs under a shared clock to realize one
million p-bits at $10^{12}$ updates per second.
Both machines keep coupling weights in local on-chip memory
and exchange only single-bit boundary states, with graph
coloring allowing non-adjacent p-bits within each FPGA to
update in parallel~\cite{aadit2022massively,nikhar2024all}.
DSIM-1 uses six FPGAs with independent clocks to vary boundary
exchange and local updates separately.
These clocks permit occasional concurrent updates across chip
boundaries, as in asynchronous p-bit
networks~\cite{selcuk2025dac}.
Experiments on three-dimensional and quasi-two-dimensional spin
glasses show that the ratio of boundary-exchange frequency to
local update frequency controls the measured optimization
dynamics.
A bound derived from the graph partition and interconnect
identifies operating conditions that recover the residual-energy
decay of an unpartitioned GPU reference.
Sufficiently infrequent exchange reduces the fitted power-law
decay exponent over the measured range, although faster local
updates can shorten the time to reach a target energy.
A cluster mean-field
model~\cite{oguchi1951statistics,bethe1935statistical,pelizzola2005cluster,yamamoto2009ccmf,xing2012gmf}
reproduces this dependence on exchange frequency, supporting an
interpretation in terms of partitioned stochastic dynamics
(Supplementary Sec.~\ref{sec:CMFT}).
Fixed-temperature sampling of a ferromagnet shows a corresponding
dependence of magnetization relaxation on the same timing ratio.
These results connect communication requirements to the
collective dynamics, providing a quantitative basis for scaling
probabilistic computers with local weight storage.
    
    
    \begin{figure*}[t]
      \centering
      \includegraphics[width=\textwidth]{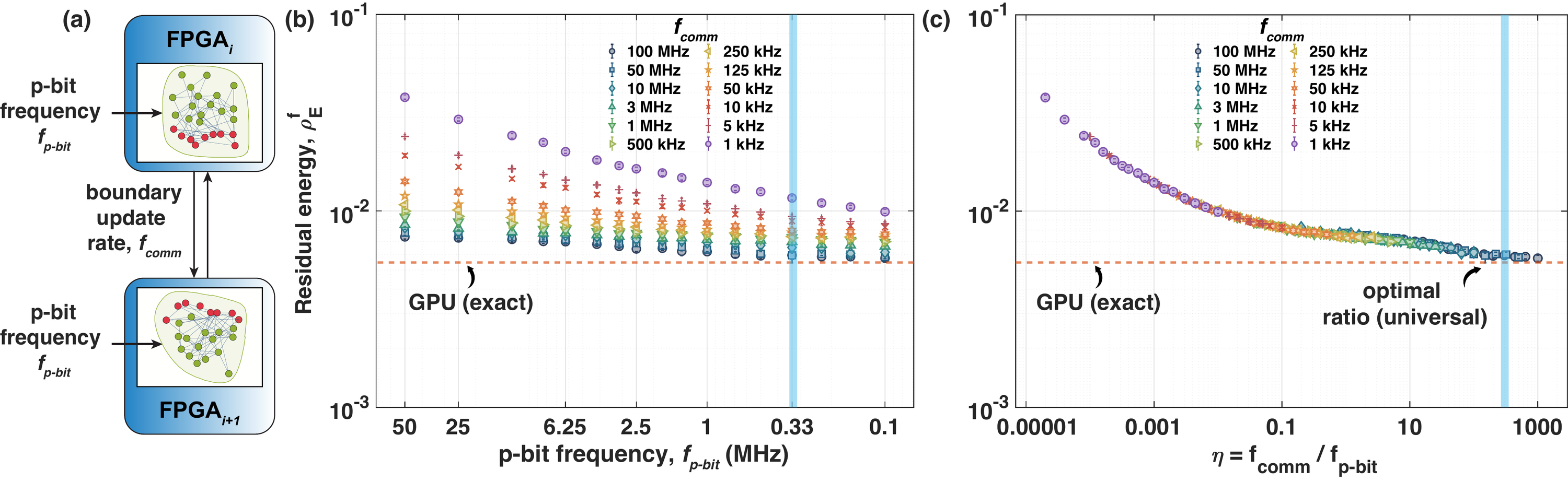}
      \caption{
    \textbf{A single timing ratio controls optimization quality
    ($L^3=37^3$, DSIM-1).}
    \textbf{(a)} Neighboring partitions exchange 1-bit boundary states at
    $f_{\mathrm{comm}}$ while local p-bits update at
    $f_{\mathrm{p\text{-}bit}}$.
    \textbf{(b)} Final residual energy per spin $\rho_E^f$ versus
    $f_{\mathrm{p\text{-}bit}}$ for $f_{\mathrm{comm}}$ from 1~kHz to
    100~MHz, at a fixed budget of $10^6$ sweeps per run (one sweep updates every p-bit once); the dashed line is
    the monolithic GPU baseline (exact, unpartitioned, where ``exact'' denotes boundary information current at every update). The p-bit frequency decreases to the right, and the vertical band marks $f_{\mathrm{p\text{-}bit}} \approx 0.33$~MHz, which corresponds to the band of (c) at $f_{\mathrm{comm}} = 100$~MHz.
    \textbf{(c)} The same data replotted against
    $\eta=f_{\mathrm{comm}}/f_{\mathrm{p\text{-}bit}}$ collapse onto a single
    curve. The vertical band near $\eta \approx 300$ marks the onset of
    saturation, consistent with Eq.~\eqref{eq:extreme_fpbit_bound_main}.
    Error bars: 95\% bootstrap confidence intervals over 10 instances
    $\times$ 10 runs.}
      \label{fig:extreme_fig2}\vspace{-10pt}
    \end{figure*}

    \section{Distributed sparse Ising machines}
    \label{sec:dsim}

   We form a distributed sparse Ising machine by partitioning a
sparse, undirected Ising graph across devices, each of which
updates its local p-bits (Fig.~\ref{fig:extreme_fig1}).
For coupling weights $\{J_{ij}\}$ and biases $\{h_i\}$, the
local field at p-bit $i$ is
$I_i = \beta\bigl(h_i + \sum_j J_{ij}\,m_j\bigr)$,
where $\beta$ is the inverse temperature.
The p-bit updates according to
$m_i = \mathrm{sgn}[\tanh(I_i) + r]$,
where $r$ is a uniform random number in $(-1,+1)$.
    
   A balanced min-cut partition, obtained with tools such as
METIS~\cite{karypis1998software} or
KaHIP~\cite{Sanders2013KaHIP}, reduces the number of edges
crossing between devices and the storage required to duplicate
their weights. Each cut-edge weight is stored on both sides
as a \emph{shadow weight}, so every device can compute its local
fields using on-chip weights, its local p-bit states and the
latest boundary states it has received. This keeps weight
storage local as more devices are added. Only boundary p-bit
states cross between devices during execution: 1 bit per
boundary p-bit, exchanged in both directions because each side
needs the other's states (Fig.~\ref{fig:extreme_fig1}d).

   Boundary states may pass through intermediate devices before
reaching their destination, adding delay and loading shared
links. Minimizing the number of cut edges alone does not account
for these routes. We therefore introduce a Potts cost function
that penalizes placing heavily connected partitions many hops
apart, favoring shorter communication paths
(Fig.~\ref{fig:extreme_fig1}c and Supplementary
Sec.~\ref{sec:potts_partitioning_app}).
To quantify the communication load, we define a worst-case
congestion metric $C_{\max}$ from the boundary sizes, hop
distances and available pins. We use this metric, together
with the communication rate and coloring schedule, to derive
a conservative bound on the local update rate for completing
boundary exchange within one color-update interval
(Supplementary Sec.~\ref{sec:comm_cost_app}).
    
  Each p-bit is updated at a rate $f_{\mathrm{p\text{-}bit}}$,
while boundary states are transmitted using a communication
clock $f_{\mathrm{comm}}$. We define the dimensionless timing
ratio
\begin{equation}
\eta = \frac{f_{\mathrm{comm}}}{f_{\mathrm{p\text{-}bit}}},
\label{eq:eta_def_main}
\end{equation}
which counts the communication clock cycles available per p-bit
update. For a fixed partition and interconnect, larger $\eta$
allows more frequent boundary refresh relative to local updates,
while smaller $\eta$ leaves boundary states unchanged for more
updates.
    
    We implement the architecture on two FPGA platforms
(Fig.~\ref{fig:extreme_fig1}e,f).
DSIM-1 connects six FPGAs in a nearest-neighbor chain with
independent local clocks and source-synchronous duplex links,
allowing the local update and communication rates to be adjusted
independently (Supplementary Sec.~\ref{sec:bus_chain_app}
and Fig.~\ref{fig:supp_dsim1_photo}).
DSIM-2 uses 18 AMD VP1902 FPGAs~\cite{amd_vp1902} on a Siemens
Veloce proFPGA CS prototyping platform, with a shared global
clock and denser interconnect
(Supplementary Fig.~\ref{fig:supp_dsim2_photo}).
On this platform, $f_{\mathrm{p\text{-}bit}}$ and
$f_{\mathrm{comm}}$ cannot be adjusted independently however
overclocking can be studied by raising the global clock beyond the frequency
at which the design meets its timing constraints.
    
    \begin{figure*}[t]
      \centering
      \includegraphics[width=\textwidth]{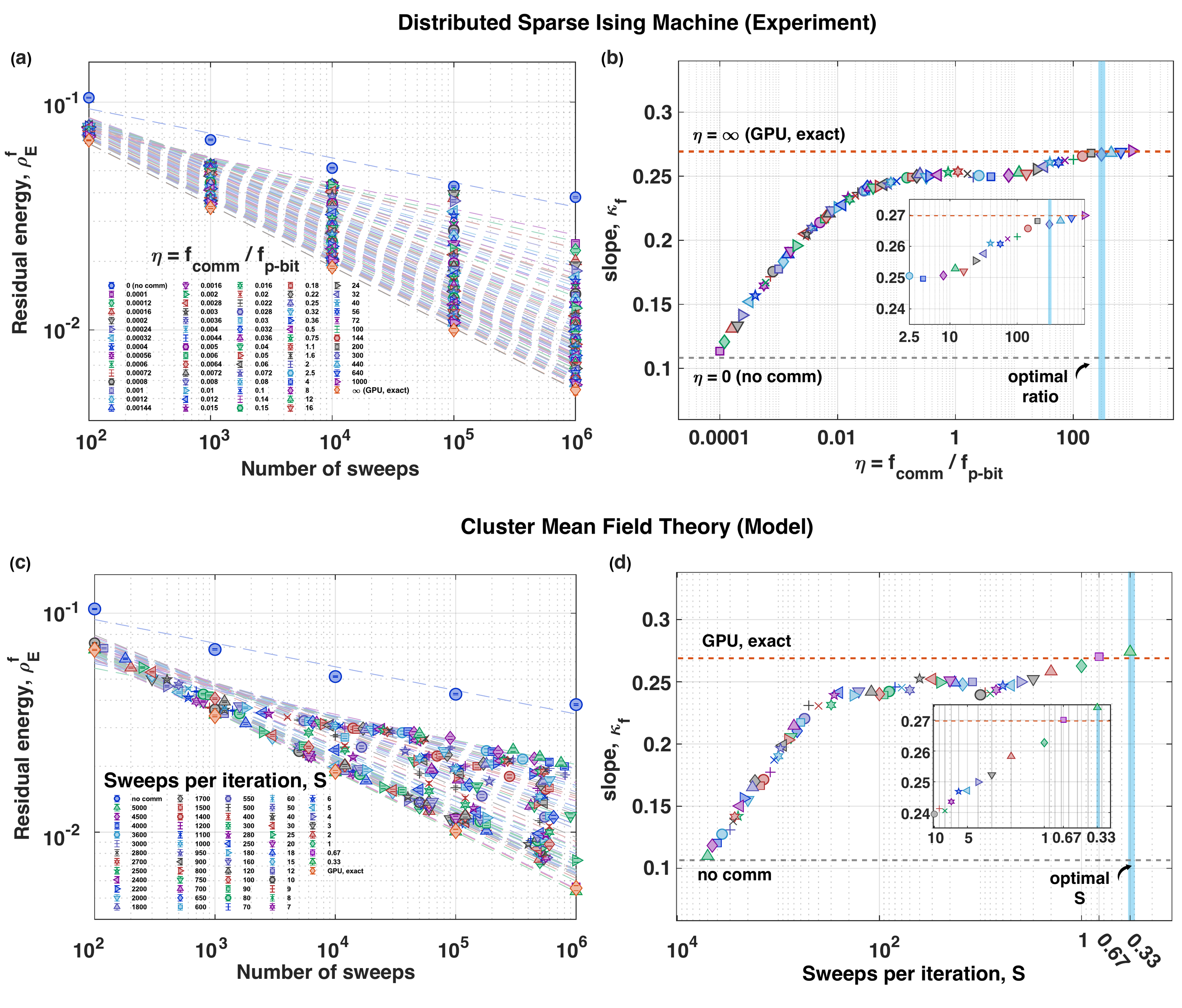}
    \caption{
    \textbf{Stale boundaries reduce the power-law exponent in the same way in
    hardware and in theory ($L^3=37^3$).}
    \textbf{(a)} DSIM-1 residual energy $\rho_E^f$ versus sweeps; each curve
    is one $(f_{\mathrm{p\text{-}bit}},f_{\mathrm{comm}})$ combination,
    spanning $\eta$ from $0$ (no communication) to the exact limit
    (unpartitioned GPU as monolithic-reference).
    \textbf{(b)} Fitted exponent $\kappa_f$ of the power-law decay of $\rho_E^f$ with sweeps, versus $\eta$, saturating near
    $\kappa_f \approx 0.27$. Inset: large-$\eta$ region; the band marks the optimal ratio of Eq.~\eqref{eq:extreme_fpbit_bound_main}, at the onset of saturation.
    \textbf{(c,d)} The same observables from the CMFT model on a GPU, with
    identical partitioning, instances, and $\beta$ schedule. The control
    parameter is $S$, the number of internal sweeps between boundary
    exchanges; large $S$ corresponds to small $\eta$. The number of sweeps per iteration extends below one, to $S = 2/3$ and $S = 1/3$,
    where boundary states are exchanged after every two color-group updates and after every color-group update (Supplementary Sec.~\ref{sec:CMFT}). The CMFT exponent
    saturates near $\kappa_f \approx 0.27$ for $S \le 1$, the unpartitioned value.
    Inset in \textbf{(d)}: small-$S$ region with the optimal-$S$ marker.
    Dashed lines in \textbf{(b)} and \textbf{(d)}: exact and no-communication
    limits. Error bars in \textbf{(a)} and \textbf{(c)}: 95\% bootstrap
    confidence intervals over 10 instances $\times$ 10 runs.}
      \label{fig:extreme_fig3}
    \end{figure*}
    
    \section{Results}
    \label{sec:results}
    
   At a fixed annealing budget, the final residual energy on
DSIM-1 collapses onto a single curve when plotted against
the timing ratio $\eta$ (Fig.~\ref{fig:extreme_fig2}b,c).
We anneal $L^3=37^3$ EA spin glasses ($N=50{,}653$) for
$10^6$ Monte Carlo sweeps (MCS) per run, varying
$f_{\mathrm{p\text{-}bit}}$ at several $f_{\mathrm{comm}}$
values (Fig.~\ref{fig:extreme_fig2}a).
One sweep updates every p-bit once.
The final residual energy per spin is
$\rho_E^f=(E^f-E_{\mathrm{ground}})/N$, where $E^f$ is
the final energy and $E_{\mathrm{ground}}$ is a putative
ground energy (Methods).

The residual energy saturates near $\eta\approx300$,
consistent with the ratio $\eta\approx305$ we calculate
from the partition and physical mapping using the
conservative bound
\begin{equation}
  f_{\mathrm{p\text{-}bit}} \le
  \frac{f_{\mathrm{comm}}}{2\,N_{\mathrm{color}}\,C_{\max}},
  \label{eq:extreme_fpbit_bound_main}
\end{equation}
where $N_{\mathrm{color}}$ is the number of color groups.
We call this sufficient ratio for recovering the
unpartitioned reference the optimal ratio
(Supplementary Sec.~\ref{sec:comm_cost_app} and
Table~\ref{tab:eta_threshold_app}).

Below this ratio, the measured trajectories determine the
onset and extent of degradation, with fitted exponents and
prefactors setting the time-to-target tradeoff
(Fig.~\ref{fig:extreme_fig3}b).
We also test the bound on quasi-two-dimensional (quasi-2D)
Pegasus spin glasses~\cite{King2023quantum,jauma2024exploring}
(Supplementary Sec.~\ref{sec:quasi2d_app}).
    
    \begin{figure*}[t]
      \centering
      \includegraphics[width=\textwidth]{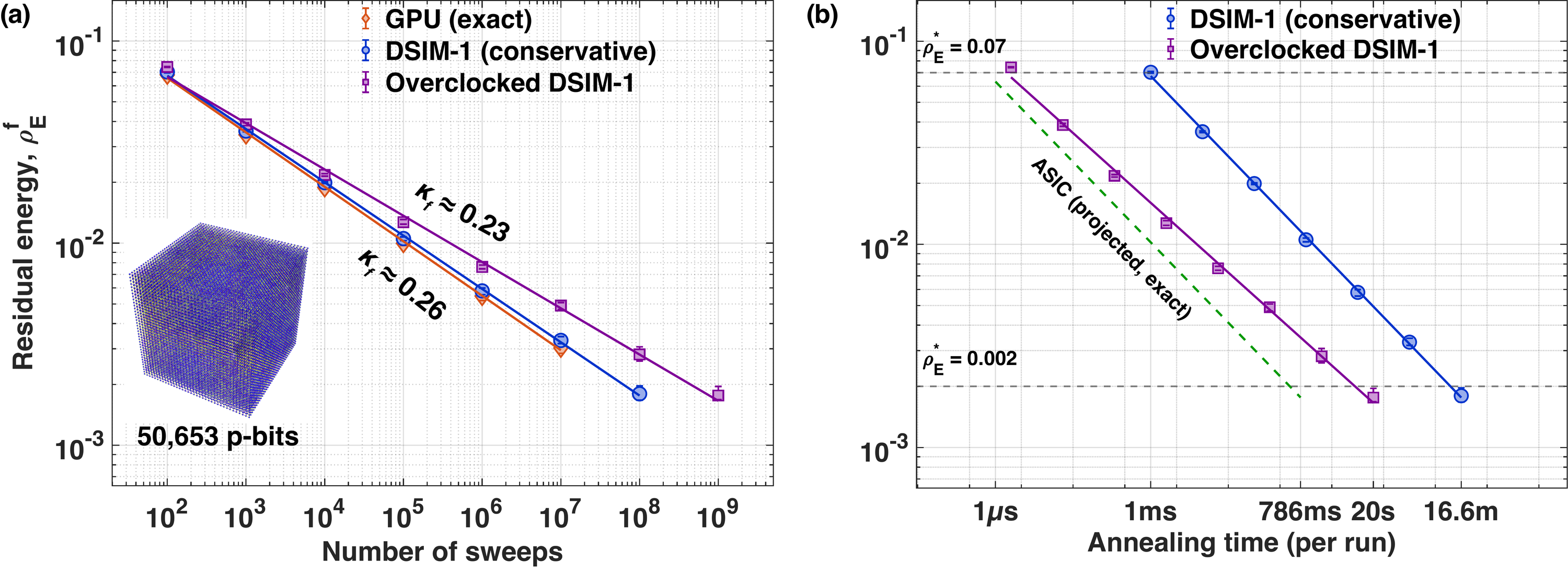}
    \caption{
    \textbf{Time-to-target at $50{,}653$ p-bits ($L^3=37^3$, DSIM-1).}
    \textbf{(a)} Residual energy $\rho_E^f$ versus sweeps: GPU (exact,
    $\kappa_f \approx 0.2693$), conservative DSIM-1 at $0.10$~MHz ($\kappa_f \approx 0.2637$), and overclocked DSIM-1 at $50$~MHz ($\kappa_f \approx 0.2289$); ``exact'' means boundary information is
    current (fresh) at every update, so that every p-bit is sampled from
    its conditional distribution given the present states of all its
    neighbors, as in exact Gibbs sampling of the full
    graph~\cite{aadit2022massively}, and ``inexact'' that boundary states are
    stale, reaching the update with a delay, much like Hogwild-style approximate Gibbs sampling~\cite{johnson2013analyzing}.  On DSIM-1, independent clocks allow occasional concurrent updates of neighboring p-bits on different FPGAs, so its agreement with the GPU is empirical, and its two modes are labeled conservative and overclocked. The small GPU--DSIM-1 gap reflects platform differences in random-number
    generation (on-chip linear-feedback shift registers, LFSRs~\cite{singh2024cmos}, versus
    Philox~\cite{salmon2011parallel}) and arithmetic precision (fixed point versus floating point, see Methods), not staleness. Data extend to $10^9$
    sweeps; putative ground energies come from runs up to $10^{10}$ sweeps. These longer searches do not
    exclude reference-energy errors that could affect the late-time
    residuals.
    \textbf{(b)} The same data versus on-chip annealing time (one-time
    weight-load of $0.5$~s excluded). Measured flip rates: $5.1\times
    10^{9}$/s at $0.10$~MHz and $2.53\times 10^{12}$/s at $50$~MHz. The
    overclocked mode reaches the target residual energy $\rho_E^\star=0.07$ (dashed) $410.5\times$
    faster and $\rho_E^\star=0.002$ $62.07\times$ faster than the
    conservative mode. These speedups compare the two DSIM-1 modes.
    Green dashed: projected multi-chip system built from 7~nm partitions
    with die-to-die links sized by Eq.~\eqref{eq:extreme_fpbit_bound_main},
    at $100$~MHz (exact, Supplementary
    Sec.~\ref{sec:ucie_feasibility_app}).
    Error bars: 95\% bootstrap confidence intervals over 10 instances
    $\times$ 10 runs.}
      \label{fig:extreme_fig4}
    \end{figure*}
    
   On DSIM-1, boundary staleness reduces the exponent of the
approximate power-law decay $\rho_E^f\propto t_a^{-\kappa_f}$
(Fig.~\ref{fig:extreme_fig3}a), where $t_a$ counts sweeps and
$\kappa_f$ is obtained from log-log fits. The exponent rises
with $\eta$ and saturates in the few-hundred range
(Fig.~\ref{fig:extreme_fig3}b), consistent with
Eq.~\eqref{eq:extreme_fpbit_bound_main}.
These exponents describe finite fitting windows and may
change at longer annealing budgets. At $\eta=0$, disconnected
partitions solve a different optimization problem and the full-system residual energy
approaches a nonzero floor.

Throughout this work, an NVIDIA RTX 6000 Ada GPU provides
the unpartitioned reference on the same instances and
annealing schedule. Here it yields $\kappa_f\approx0.2693$,
compared with $0.2637$ for conservative DSIM-1 operation at
$0.10$~MHz. We attribute the small gap to the hardware's
linear-feedback shift register (LFSR)~\cite{singh2024cmos} and s$\{4\}\{1\}$ fixed-point
arithmetic compared with the GPU's Philox
generator~\cite{salmon2011parallel} and floating-point
arithmetic (Methods).
Overclocking DSIM-1 to $50$~MHz reduces $\kappa_f$ to $0.2289$.
To test the local update circuits, we disconnect the
inter-FPGA links. Each subgraph then anneals independently,
with energy curves unchanged across all tested clock rates
(Supplementary Fig.~\ref{fig:supp_subgraph_energy} and
Sec.~\ref{sec:subgraph_energy_app}).
We therefore attribute the reduced exponent in coupled
DSIM-1 runs to stale boundary information.

    \begin{figure*}[t]
      \centering
      \includegraphics[width=\textwidth]{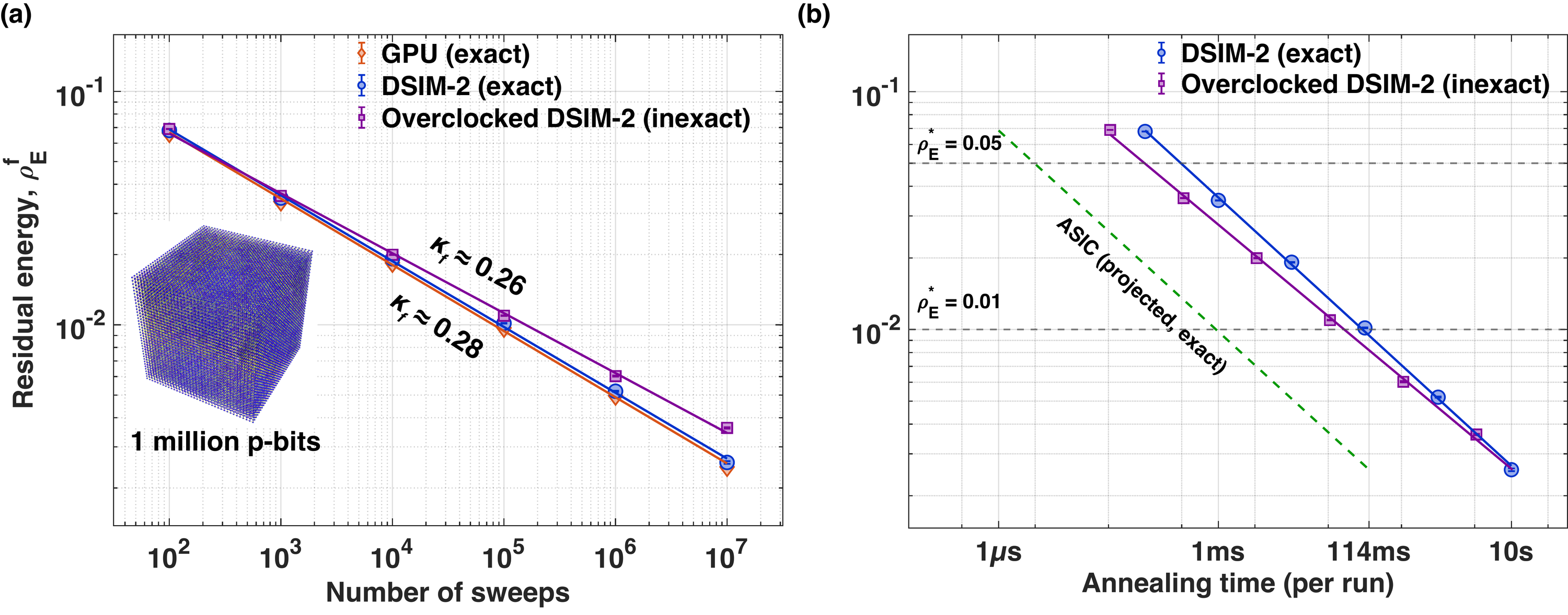}
    \caption{
    \textbf{Time-to-target at one million p-bits ($L^3=100^3$, DSIM-2).}
    \textbf{(a)} Residual energy $\rho_E^f$ versus sweeps: GPU (exact,
    $\kappa_f \approx 0.2836$), conservative DSIM-2 at $1$~MHz (exact,
    $\kappa_f \approx 0.2820$), and overclocked DSIM-2 at $3$~MHz (inexact,
    $\kappa_f \approx 0.2565$). ``Exact'' and ``inexact'' are defined in
    Fig.~\ref{fig:extreme_fig4}. The residual GPU gap reflects platform differences in random-number
    generation and arithmetic precision (Methods), not staleness. Timing closes at $1$~MHz; the
    $3$~MHz point is the same synthesis globally overclocked, and beyond
    $3$~MHz read/write synchronization fails. The reduced exponent at $3$~MHz is likely due to stale boundary information, because timing violations strike the
    longest paths first (main text). Data shown extend to $10^7$ sweeps; putative ground energies were obtained from FPGA runs up to $10^9$ sweeps (Methods).
    \textbf{(b)} The same data versus on-chip annealing time (one-time
    weight-load of $2$~s excluded). Measured flip rates: $10^{12}$/s at
    $1$~MHz and $3\times 10^{12}$/s at $3$~MHz. The overclocked mode
    reaches $\rho_E^\star=0.05$ (dashed) $2.82\times$ faster and
    $\rho_E^\star=0.01$ $2.23\times$ faster than the conservative mode,
    with an estimated crossover near $\rho_E^\star \approx 0.004$. These speedups
    compare the two DSIM-2 modes. Green dashed: projected multi-chip system
    built from 7~nm partitions with die-to-die links sized by
    Eq.~\eqref{eq:extreme_fpbit_bound_main}, at $100$~MHz (exact, Supplementary
    Sec.~\ref{sec:ucie_feasibility_app}).
    Error bars: 95\% bootstrap confidence intervals over 10 instances
    $\times$ 10 runs.}
      \label{fig:extreme_fig5}
    \end{figure*}
    
    A parallel cluster mean-field theory (CMFT) reproduces the
reduction and saturation of the hardware decay exponent
(Fig.~\ref{fig:extreme_fig3}c,d). We implement the model on
the same GPU with identical partitions, instances and
annealing schedule (Supplementary Sec.~\ref{sec:CMFT}).
Each cluster performs exact local Markov chain Monte Carlo
(MCMC) updates using boundary mean fields held fixed for $S$
internal sweeps. Large $S$ corresponds to small $\eta$.
Over the range studied, residual energy decays approximately
as a power law whose exponent decreases smoothly as exchange
becomes less frequent. Exchanging after every two color groups
($S=2/3$) or every color group ($S=1/3$) recovers the saturated
exponent near $0.27$, matching the hardware and unpartitioned
reference.

The CMFT trajectories are calculated without hardware input.
Comparing them with hardware requires a fitted relation between
$S$ and $\eta$ (Supplementary Fig.~\ref{fig:eta_sample_alignment}),
whose transfer to another design remains untested.
The observed gradual degradation does not establish convergence
or unbiased sampling. The correspondence supports a cluster
mean-field description of partitions evolving between
exchanges, even though the hardware exchanges individual
boundary states rather than averages.
    
   Overclocked DSIM-1 reaches $\rho_E^\star=0.07$ $410.5\times$
faster and $\rho_E^\star=0.002$ $62.07\times$ faster than its
conservative mode (Fig.~\ref{fig:extreme_fig4}b).
For $N=50{,}653$, the measured update rate
$Nf_{\mathrm{p\text{-}bit}}$ rises from $5.1\times10^9$
updates/s at $0.10$~MHz to $2.53\times10^{12}$ updates/s
at $50$~MHz, with the entire six-FPGA platform drawing
150 to 180~W. These rates count update attempts, whether
or not the state changes (Methods). Faster updates consume
staler boundary information, so the throughput gain comes
with a smaller decay exponent. The conservative and
overclocked modes follow single power laws over six and
seven decades of sweeps, respectively, from $10^2$ to
$10^8$ and $10^2$ to $10^9$ (Fig.~\ref{fig:extreme_fig4}a).
The exponent difference reduces the speedup at lower
target energies. Extrapolating the fitted power laws
predicts a crossover beyond the measured range,
conditional on those fits continuing to hold.

Fig.~\ref{fig:extreme_fig4}b also shows a projection for
a 7~nm application-specific integrated circuit (ASIC)
implementing the same partition. The design closes timing
at 100~MHz with an area of 0.66~mm$^2$ and power of 248~mW
per partition (Supplementary Sec.~\ref{sec:ucie_feasibility_app}
and Table~\ref{tab:supp_asic_summary}).
The local graph structure is size-independent, and the
projection assumes the conservative-mode exponent at
the higher clock.
    
  DSIM-2 operates one million p-bits ($L^3=100^3$) across
18 FPGAs at $10^{12}$ updates/s.
At $f_{\mathrm{p\text{-}bit}}=1$~MHz, the design meets its
timing constraints and its residual-energy decay agrees
with the unpartitioned GPU reference within the reported
confidence interval (Fig.~\ref{fig:extreme_fig5}a).
Globally overclocking the same implementation to
$f_{\mathrm{p\text{-}bit}}=3$~MHz raises throughput to
$3\times10^{12}$ updates/s with a smaller decay exponent.
It reaches $\rho_E^\star=0.05$ and $0.01$ faster than the
1~MHz mode by factors of $2.82$ and $2.23$, respectively
(Fig.~\ref{fig:extreme_fig5}b).
The timing bound places the 1~MHz point near the optimal
ratio and the 3~MHz point well below it
(Eq.~\eqref{eq:extreme_fpbit_bound_main} and Supplementary
Sec.~\ref{sec:comm_cost_app}).
Implementation-specific GPU time and energy comparisons
are reported in Supplementary Sec.~\ref{sec:gpu_time_app}.

The same qualitative dependence on $\eta$ appears in
quasi-2D $\pm1$ Pegasus spin glasses, with up to 15 neighbors
per spin compared with six on the cubic lattice.
On DSIM-1, $25{,}088$-spin instances show an exponent that
rises from its disconnected value and saturates near the
unpartitioned reference under two annealing schedules
(Supplementary Sec.~\ref{sec:quasi2d_app}).
On DSIM-2, $251{,}328$-spin instances recover the GPU's
residual-energy decay and retain their exponent at higher
throughput when overclocked to half the optimal ratio
(Supplementary Sec.~\ref{sec:p103_app}).

On the larger Pegasus design, we test disconnected partitions as a control. We find that local-subgraph energy curves are unchanged across the
tested clock rates. Higher clock rates cause read/write
synchronization to completely fail 
(Supplementary Sec.~\ref{sec:p103_app}). Even though there is no direct way to separate local and communication clocks in DSIM-2, the local-subgraph controls suggest the local clocks do not fail and that communication and synchronization
are the main timing limitation. 

\begin{figure*}[t]
      \centering
      \includegraphics[width=\textwidth]{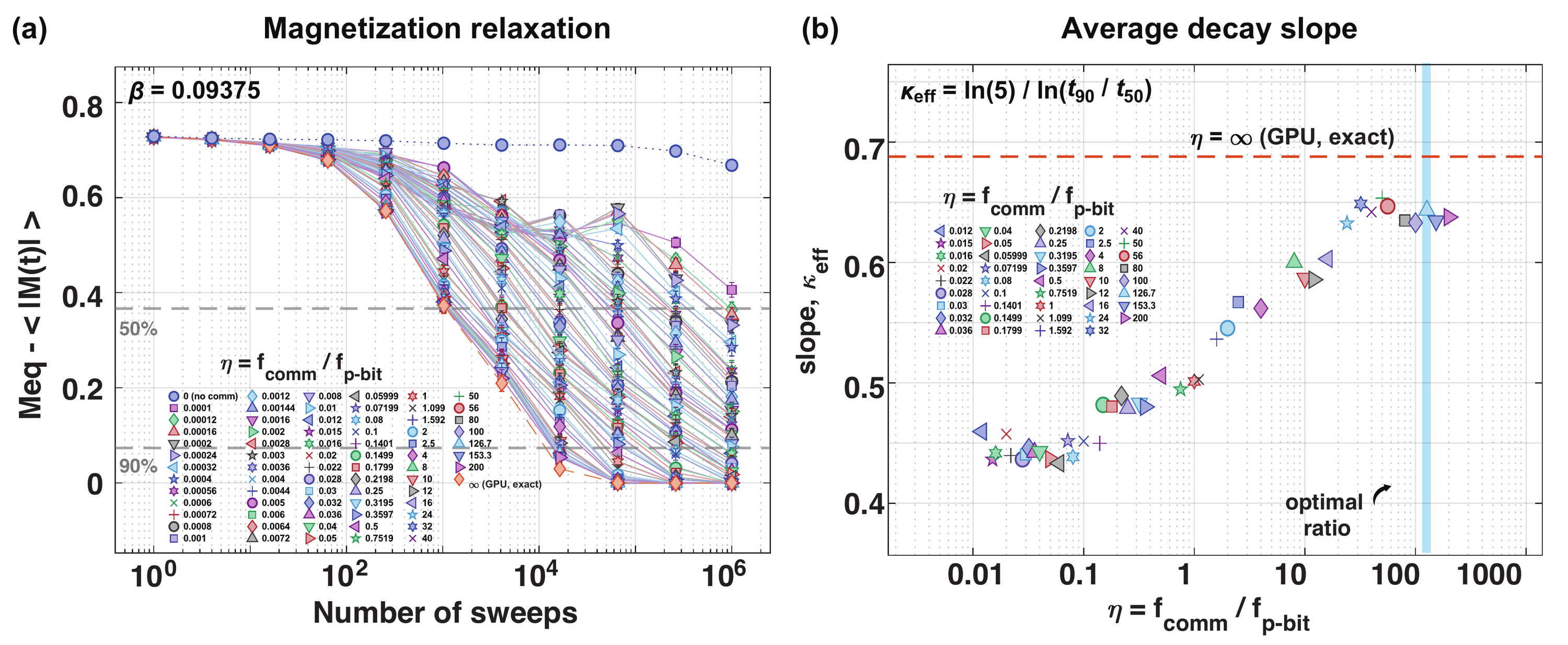}
    \caption{
    \textbf{Sampling at fixed temperature on DSIM-1.}
    A ferromagnet on the Pegasus graph P25
    ($14{,}208$ p-bits, $J_{ij} = +1$, zero field) is partitioned over the six FPGAs of DSIM-1
    (Fig.~\ref{fig:extreme_fig1}e), with 1-bit boundary states exchanged at
    $f_{\mathrm{comm}}$ and local updates at $f_{\mathrm{p\text{-}bit}}$.
    Each run starts with three partitions at $+1$ and three at $-1$, so
    $M(0) = 0$, at fixed $\beta = 0.09375$, the heat-capacity peak.
    \textbf{(a)} Gap $M_{\rm eq} - \langle |M(t)| \rangle$ versus sweeps for
    $71$ values of $\eta$ from $10^{-4}$ to $200$, averaged over $1{,}000$
    runs, with the exact GPU reference ($\eta = \infty$, $M_{\rm eq} = 0.7334$) and
    the disconnected system ($\eta = 0$), whose magnetization stays near zero because each partition keeps its initial sign. Horizontal lines mark the gaps at which
    $\langle |M| \rangle$ reaches $50\%$ and $90\%$ of $M_{\rm eq}$. Error bars are $95\%$ confidence intervals over
    runs.
    \textbf{(b)} Average decay slope from $50\%$ to $90\%$,
    $\kappa_{\rm eff} = \ln 5/\ln(t_{90}/t_{50})$, versus $\eta$, for the
    $44$ values of $\eta \geq 0.012$ at which both levels are
    reached within $10^6$ sweeps. The disconnected system never reaches the
    $50\%$ level and therefore has no $\kappa_{\rm eff}$. The dashed line is the exact GPU
    value and the vertical band the optimal ratio of
    Eq.~\eqref{eq:extreme_fpbit_bound_main} for this partition, $\eta \approx 124$
    (Supplementary Sec.~\ref{sec:sampling_app}).}
      \label{fig:extreme_fig6_sampling}
    \end{figure*}
    
We test relaxation at fixed temperature using a ferromagnet
on the Pegasus P25 graph~\cite{King2023quantum}, with $14{,}208$
p-bits and $103{,}664$ couplings distributed over the six FPGAs
of DSIM-1 (Fig.~\ref{fig:extreme_fig6_sampling}).
All couplings have $J_{ij}=+1$, and zero external field makes
the all-$+1$ and all-$-1$ states equally likely.
At $\beta=0.09375$, the heat-capacity peak where fluctuations
are largest, the unpartitioned GPU gives an ordered equilibrium
with $M_{\rm eq}=0.7334$.
We initialize three partitions at $+1$ and three at $-1$,
so each is ordered but the total magnetization is zero.
Reaching the reference requires the partitions to align
through their exchanged 1-bit boundary states.
The ordering time therefore measures how well they communicate.

We follow the average absolute magnetization
$\langle |M(t)| \rangle$, since either sign can order at zero
field, over $1{,}000$ independent runs of $10^6$ sweeps.
We compare the unpartitioned reference ($\eta=\infty$),
the disconnected control ($\eta=0$) and $71$ values of $\eta$
from $10^{-4}$ to $200$.
Without communication, partitions retain their initial signs
and $\langle |M| \rangle<0.07$ throughout the run, so the
system does not order.
For every measured $\eta\ge0.012$, magnetization reaches
$90\%$ of $M_{\rm eq}$ within $10^6$ sweeps, requiring more
sweeps as $\eta$ decreases
(Fig.~\ref{fig:extreme_fig6_sampling}a).

For each setting, we define $t_{50}$ and $t_{90}$ as the
sweep counts at which magnetization reaches $50\%$ and $90\%$
of $M_{\rm eq}$.
The remaining gap falls fivefold, giving an average logarithmic
decay slope $\kappa_{\rm eff}=\ln 5/\ln(t_{90}/t_{50})$.
This is an average between fixed levels, not a fitted
power-law exponent.
The disconnected control has no $\kappa_{\rm eff}$ because
it never reaches $50\%$.
The slope rises with $\eta$ toward the unpartitioned GPU value
(Fig.~\ref{fig:extreme_fig6_sampling}b).
A uniform stretching of time would leave $t_{90}/t_{50}$
unchanged, so the changing slope shows that boundary delays
alter the relaxation profile.
Slow exchange reduces both this slope and the optimization
exponent $\kappa_f$, while frequent exchange recovers their
unpartitioned values (Fig.~\ref{fig:extreme_fig3}b and
Supplementary Table~\ref{tab:opt_vs_sampling}).
Because exact Boltzmann-distribution comparisons are limited
to toy problems, magnetization serves as a proxy for the
full distribution.
Its sensitivity to boundary exchange extends the observed
communication dependence from spin-glass optimization to
a fixed-temperature sampling observable
(Methods and Supplementary Sec.~\ref{sec:sampling_app}). 
The same qualitative dependence on $\eta$ in optimization
and fixed-temperature sampling suggests that the communication
tradeoff is a broader property of partitioned stochastic dynamics.

The same architecture applies to other sparse graphs that
can be colored and partitioned. We demonstrate this
programmability on three further optimization problems,
keeping all weights on-chip and exchanging only 1-bit
boundary states.
On the $20{,}000$-node G81 Max-Cut instance~\cite{Gset},
DSIM-1 independently reaches the certified optimum of
$14{,}060$, first found by
Cosm~\cite{zick2025performance,zick2026cosm}
(Supplementary Sec.~\ref{sec:g81_app}). The replicas of adaptive parallel tempering are processed
sequentially on the same hardware by time-division multiplexing
(Supplementary Sec.~\ref{sec:g81_app}), so the result is obtained
with a single copy of the machine.
DSIM-2 also solves planted Pegasus and Zephyr instances
with $39{,}040$ and $80{,}800$ p-bits, respectively, whose
ground states are known by
construction~\cite{navid2022nano,hen2019equation}
(Supplementary Sec.~\ref{sec:pegasus_zephyr_app}).
For a random three-literal satisfiability (3SAT) instance
encoded with $250{,}011$ p-bits, the best DSIM-2 run
satisfies $55{,}416$ of $55{,}558$ clauses ($99.74\%$)
after $10^9$ sweeps, closely tracking the optimized GPU
baseline (Supplementary Sec.~\ref{sec:3sat_app}).

   \section{Conclusion}
\label{sec:conclusion}

We have built a programmable million-p-bit computer from
networked FPGAs, sustaining $10^{12}$ updates/s with local
coupling weights and 1-bit boundary exchange.
Across 3D and quasi-2D spin glasses,
$\eta=f_{\mathrm{comm}}/f_{\mathrm{p\text{-}bit}}$
governs the measured tradeoff between boundary freshness
and update throughput.
Infrequent exchange gradually degrades residual-energy
decay over the measured range, while sufficient exchange
recovers the unpartitioned decay.
Faster local updates can nevertheless shorten time to target.
A cluster mean-field model reproduces this behavior of
partitions evolving between exchanges.

A quasi-2D ferromagnet at the heat-capacity peak shows the
same qualitative $\eta$ dependence: slower refresh gradually
lengthens relaxation, while frequent exchange recovers the
unpartitioned equilibrium magnetization
(Fig.~\ref{fig:extreme_fig6_sampling}).
Related tradeoffs occur in Gibbs sampling with stale
reads~\cite{johnson2013analyzing}, distributed machine
learning~\cite{ho2013more} and stochastic-gradient Monte
Carlo~\cite{chen2016stochastic}.
Synchronous implementations that preserve an unpartitioned
machine's iteration dependencies include communication
in each iteration
(Supplementary Sec.~\ref{sec:arch_comparisons_app}).
DSIMs quantify the cost of letting local updates continue
with delayed boundaries.

The bound in Eq.~\eqref{eq:extreme_fpbit_bound_main} uses
the partition, link budget and coloring schedule to specify
a conservative communication requirement.
Independent clocks on DSIM-1 vary $\eta$ over seven orders
of magnitude.
At fixed p-bits per device, update rate, link budget and
color count, short communication routes permit linear
throughput growth while the sufficient ratio remains bounded
(Supplementary Sec.~\ref{sec:scaling_app}).
This provides a route to physical probabilistic computers
built from vast networks of locally updating devices,
adding processing and memory together beyond the capacity
of any single chip.
Dedicated-silicon projections illustrate this direction
(Supplementary Sec.~\ref{sec:ucie_feasibility_app}).
The shared $\eta$ dependence across optimization and sampling
suggests that this freedom to relax communication could
extend to other probabilistic workloads.
    
    \section{Methods}
    \paragraph{Benchmark instances.}
    We study 3D Edwards--Anderson spin glasses on $L\times L\times L$
    lattices with $J_{ij}\in\{\pm 1\}$ drawn independently and uniformly at
    random on nearest-neighbor edges~\cite{King2023quantum,chowdhury2025pushing},
    with periodic boundaries in $z$ and open boundaries in $x$ and $y$.
    We report $L^3=37^3$ and $L^3=100^3$, with 10 disorder instances at each
    size. DSIM-1 experiments use the Potts partitioning introduced here to
    align with the chain topology. For DSIM-2, graph-level and SLR-level
    partitioning was generated with METIS and then mapped onto the platform
    through the Veloce proFPGA CS implementation flow. The quasi-2D spin
    glasses are $\pm1$ instances drawn in the same way on the Pegasus graph,
    $25{,}088$ p-bits (P33) on DSIM-1 (10 instances, partitioned with KaHIP~\cite{Sanders2013KaHIP} into 6 parts) and $251{,}328$ p-bits (P103) on DSIM-2 (10 instances, partitioned with METIS into 72 parts), with the
    best energy found in long anneals as the reference energy
    (Supplementary Secs.~\ref{sec:quasi2d_app} and~\ref{sec:p103_app}).
    \paragraph{Update schedule and sweep definition.}
    P-bits are updated by graph coloring with $N_{\mathrm{color}}$ color
    groups; one sweep (one MCS) updates all groups once.
    $N_{\mathrm{color}}=3$ for $L^3=37^3$, $2$ for $L^3=100^3$, $4$ for
    Pegasus P41, $6$ for Zephyr Z50, $4$ for 3SAT, $2$ for G81, and $4$ for the
    quasi-2D Pegasus spin glasses and the P25 ferromagnet.
    $f_{\mathrm{p\text{-}bit}}$ is the rate at which each p-bit is updated:
    one period $1/f_{\mathrm{p\text{-}bit}}$ contains one complete sweep,
    during which the $N_{\mathrm{color}}$ color groups are processed one
    after the other, so that all $N$ p-bits are updated once per period and
    the flip rate is $N f_{\mathrm{p\text{-}bit}}$. On DSIM-1 the whole graph is colored once and every FPGA uses that coloring, but each FPGA steps through the color groups on its own local clock, so the color schedules are not synchronized across FPGAs. On DSIM-2 all partitions share the global clock, every FPGA updates the same color group at the same time, and the color groups advance one per global-clock cycle, so
    $f_{\mathrm{p\text{-}bit}} = f_{\mathrm{global}}/N_{\mathrm{color}}$.
    With $N_{\mathrm{color}}=2$ at $L^3=100^3$, the $1$~MHz and $3$~MHz
    operating points correspond to global clocks of $2$ and $6$~MHz.
    \paragraph{Annealing schedule and numeric format.}
    EA results use simulated annealing with $\beta=0.5,1.0,\ldots,5.0$;
    Pegasus, Zephyr, and 3SAT use $\beta=0.5,0.625,\ldots,10$. The quasi-2D Pegasus spin glass on DSIM-1 uses 10 equally spaced values from $0$ to $15.5$, rounded to the weight precision, and a geometric schedule in the robustness test (Supplementary Sec.~\ref{sec:quasi2d_app}), and the quasi-2D spin glass on DSIM-2 uses $\beta=0.5,1.0,\ldots,15.5$ (Supplementary Sec.~\ref{sec:p103_app}). Each $\beta$ value is held for an equal share of the sweep budget of
    the run.
    Every point on a residual-energy curve is a separate run with its own
    sweep budget $t_a$. These schedules were chosen empirically and applied identically
    across the FPGA and GPU runs. The reported power laws are empirical fits over the ranges shown in the figures. GPU references use floating-point arithmetic.
    Hardware uses fixed point: s$\{4\}\{1\}$ (signed, 4 integer bits, 1 fractional bit) for EA and the quasi-2D Pegasus spin glasses, s$\{4\}\{3\}$ for the planted Pegasus and Zephyr instances and 3SAT, s$\{4\}\{6\}$ for G81 to accommodate the distinct inverse temperatures of adaptive parallel tempering, and s$\{3\}\{6\}$ for the sampling experiment to represent its inverse temperature at the heat-capacity peak, $\beta = 6/64$. Pseudorandom numbers are generated on-chip with linear-feedback shift registers (LFSRs)~\cite{singh2024cmos} on
    both DSIM-1 and DSIM-2, except in the sampling experiment, which uses a xoroshiro generator~\cite{blackman2021scrambled} because correlations between LFSR outputs can bias the sampled distribution~\cite{singh2024cmos}, while the GPU baselines use the Philox generator~\cite{salmon2011parallel}. For the quasi-2D spin glasses and the sampling experiment, the GPU references reproduce the hardware arithmetic instead: P33 on DSIM-1 and P103 on DSIM-2 run the same p-bit as the hardware, with its LFSR and tanh lookup table, and the ferromagnet runs the fixed-point format of the hardware with the Philox generator.
    \paragraph{GPU baseline.}
    All GPU references run on an NVIDIA RTX 6000 Ada with the same
    instances and annealing schedules, on a single device with no
    partitioning. The GPU code, like the machine, runs any sparse graph with multi-bit weights. The GPU serves as the unpartitioned reference for the decay
    exponent $\kappa_f$. Its time per sweep on the same instances and
    schedule, and the resulting time and energy to target, are given in
    Supplementary Sec.~\ref{sec:gpu_time_app}. We
    note that the three million unique weights at $L^3=100^3$ occupy about
    3~MB at 8-bit precision, within a single modern GPU's L2 cache (e.g.,
    NVIDIA Blackwell B200~\cite{tirumala2024nvidia}).
    \paragraph{DSIM-2 operating points.}
    The DSIM-2 design closes timing at $f_{\mathrm{p\text{-}bit}}=1$~MHz.
    Globally overclocking the same synthesis succeeds up to $3$~MHz; beyond
    that, read/write synchronization across the 18 FPGAs fails from
    accumulated timing violations, which is why Fig.~\ref{fig:extreme_fig5}
    contains two operating points. The quasi-2D design on DSIM-2 closes
    timing at $0.47$~MHz and was overclocked to $0.75$ and $1.25$~MHz
    (Supplementary Sec.~\ref{sec:p103_app}). Access to the platform was limited,
    which is why the $L^3=100^3$ analysis uses shorter sweep budgets than
    $L^3=37^3$ (see below).
    \paragraph{Sampling experiment.}
    The ferromagnet of Fig.~\ref{fig:extreme_fig6_sampling} is the Pegasus
    graph P25 ($14{,}208$ p-bits, $103{,}664$ couplings, degree up to $15$)
    with $J_{ij} = +1$ on every edge and $h_i = 0$. It is split with
    KaHIP~\cite{Sanders2013KaHIP} into six partitions of $2{,}368$ p-bits,
    one per FPGA of DSIM-1. All runs are at the fixed inverse temperature
    $\beta = 6/64 = 0.09375$, with no annealing, and $\eta$ is set mainly
    through $f_{\mathrm{comm}}$, with $f_{\mathrm{p\text{-}bit}}$ between
    $0.18$ and $0.5$~MHz. Each run starts with every p-bit of three
    partitions at $+1$ and every p-bit of the other three at $-1$. Across runs, we cycle through all $20$ assignments of three up and three down partitions. The GPU reference uses the same initial assignments. Before
    each run the boundary states held on every FPGA are refreshed to this initial assignment, so the first updates already see the correct
    neighbor states. For run $r$ the magnetization is
    $M_r(t) = N^{-1}\sum_i m_i^{(r)}(t)$, and the plotted quantity is
    $\langle |M(t)| \rangle = R^{-1}\sum_{r=1}^{R} |M_r(t)|$ with
    $R = 1{,}000$ independent runs, recorded at $t = 0, 1, 2, 4, \ldots,
    2^{19}$ and $10^6$ sweeps along one continuous chain per run. The absolute value is taken run by run,
    before the average, because at zero field the two ordered states are
    equivalent. The equilibrium value $M_{\rm eq} = 0.7334$ is that of the
unpartitioned GPU sampler on the same graph at the same $\beta$, and the
same $M_{\rm eq}$ is used for every $\eta$, so a bias in the mean
absolute magnetization would appear as a gap that does not close. $t_{50}$ and $t_{90}$
    are read from the averaged curve as the sweeps at which
    $\langle |M| \rangle$ reaches $0.5\,M_{\rm eq}$ and $0.9\,M_{\rm eq}$,
    and the same definition applied to the GPU curve gives the reference
    value of $\kappa_{\rm eff}$.
    \paragraph{Putative ground energies.}
    Exact ground energies are not known at these sizes. Following standard
    practice~\cite{belletti2008janus,King2023quantum}, for each instance we
    define a putative ground energy as the minimum energy observed across
    all platforms and timing settings. GPU baselines were generated up to
    $10^7$ sweeps at both sizes, while FPGA runs were extended to $10^{10}$
    sweeps for $L^3=37^3$ and $10^9$ for $L^3=100^3$. This gives the
    analysis window (up to $10^9$ and $10^7$ sweeps, respectively) at least
    one order of magnitude of buffer for the reference energies $E_{\mathrm{ground}}$. 
    \paragraph{Confidence intervals.}
    Error bars on residual energies are $95\%$ bootstrap confidence intervals
    of the mean over the 10 instances, each instance entering as the mean of
    its 10 runs, with $1{,}000$ bootstrap resamples. The sampling experiment
    uses $95\%$ confidence intervals over its $1{,}000$ runs.
    
    \paragraph{Flip-rate measurement.}
    A flip is an update attempt: each p-bit evaluates its local field,
    draws a random number, and sets its state once per sweep, whether or
    not the state changes. Flip rates are measured
    on-chip~\cite{aadit2022massively}. Each color
    group carries a counter with a fixed preset (e.g., $50{,}000$ cycles at
    a $125$~MHz reference clock), defining a known time window during which
    p-bit flips accumulate. At the preset, a broadcast disable shared
    across all FPGAs stops all counters simultaneously; the per-color
    counts are aggregated and divided by the elapsed time.
    \paragraph{Sweep counting and reported times.}
    Runs are controlled by a common counter with a programmable preset at a
    $125$~MHz reference clock, providing both the completed MCS count and
    the wall-clock annealing time. On DSIM-2 the broadcast disable arrives
    synchronously and readout states are exactly aligned; on DSIM-1,
    independent local clocks introduce in principle a small stop-time
    uncertainty across partitions, with no measurable effect on
    optimization quality since updates are slow at high $\beta$. Reported
    annealing times reflect on-chip computation only: the one-time
    weight-load over Peripheral Component Interconnect Express (PCIe) (about $0.5$~s for $L^3=37^3$, $2$~s for
    $L^3=100^3$) and host read/write do not scale with sweep count and are
    excluded.
    
    \section*{Acknowledgments}
    
    We acknowledge Siemens Digital Industries Software for access to the Veloce proFPGA CS platform, and for the tool access, platform support, and engineering time provided through N.A.A.'s internship.
    We thank Subhasish Mitra and Robert Radway for helpful discussions that improved the manuscript.
    We acknowledge support from the Office of Naval Research (ONR) Young Investigator Program grant, the National Science Foundation (NSF) CAREER Award under grant number CCF 2106260, the Army Research Laboratory under grant number W911NF-24-1-0228, the Semiconductor Research Corporation (SRC) grant, and the ONR-MURI grant N000142312708.
    Use was made of computational facilities purchased with funds from the National Science Foundation (CNS-1725797) and administered by the Center for Scientific Computing (CSC).
    The CSC is supported by the California NanoSystems Institute and the Materials Research Science and Engineering Center (MRSEC; NSF DMR 2308708) at UC Santa Barbara.
    
    \section*{Author contributions}
    
    N.A.A.\ and K.Y.\c{C}.\ conceived the project.
    N.A.A.\ designed and built the $50{,}653$-node distributed p-computer at UCSB and the million-node p-computer at Siemens, led all hardware experiments and data analysis, and designed the DSIM-2 experiments. N.A.A. and K.Y.\c{C}. wrote the manuscript. F.B. provided feedback on problems, architecture choices and FPGA implementation.  X.Z.\ designed and carried out the parallel cluster mean-field theory study and the quasi-2D spin-glass and sampling experiments on DSIM-1.
    S.B.\ developed and ran all GPU benchmarks.
    K.C.-C.\ and K.L.\ developed the Potts model-based topology-aware partitioning.
    S.C.\ advised on 3D spin-glass physics, instance generation, and scaling analysis.
    S.S.\ performed the architecture-level study and 7\,nm ASIC projection under the supervision of T.S.
    C.T., J.T., and A.S.\ provided the proFPGA CS platform infrastructure, interconnect, and compile and mapping methodology at Siemens, and C.T.\ carried out the DSIM-2 runs, with A.S.\ managing the Siemens team.
    K.Y.\c{C}.\ supervised the project, guided the experimental design and interpretation, contributed to the writing, and shaped the discussion and overall direction of the work.
    All authors discussed the results and edited the manuscript.
    
    \section*{Competing interests}
    
    The authors declare no competing interests.
    
    \section*{Data and code availability}
    
    Data and code necessary to reproduce the main plots will be made available in a public repository upon publication.
    Additional artifacts required to regenerate FPGA bitstreams are available upon reasonable request subject to platform licensing constraints.

\clearpage

\onecolumngrid
\beginsupplement

\begin{center}
{\large\bfseries Supplementary Information}\\[10pt]
{\large Programmable Probabilistic Computer with 1,000,000 p-bits}
\end{center}
\vspace{1.5em}

\setcounter{subsection}{0}

\subsection{Overview}

This Supplementary Information provides the full details behind the distributed sparse Ising machine (DSIM) framework presented in the main text.
The main text identifies two limits that force distribution: the problem can outgrow a single chip's capacity, and the coupling weights may not fit in on-chip memory, so that off-chip memory bandwidth caps the update rate.
The DSIM addresses both by partitioning the graph across devices and using shadow weights to keep all weights on-chip, so that only the 1-bit states of boundary probabilistic bits (p-bits) need to cross device boundaries.
The timing ratio \(\eta = f_{\mathrm{comm}}/f_{\mathrm{p\text{-}bit}}\) then governs the behavior: saturation toward a monolithic baseline when \(\eta\) is large, a useful overclocked mode when \(\eta\) is smaller, and a close correspondence with a theoretical cluster mean-field model.
The supplementary sections are arranged as follows.
Section~\ref{sec:background_app} provides additional device-level and algorithmic background.
Section~\ref{sec:CMFT} gives the full parallel cluster mean-field theory formulation and explains why it reproduces the hardware behavior in Fig.~\ref{fig:extreme_fig3}c,d.
Section~\ref{sec:comm_cost_app} develops the communication-cost metric and the conservative clocking bound (Eq.~\eqref{eq:extreme_fpbit_bound_main}), and Section~\ref{sec:scaling_app} gives its scaling with the number of devices.
Section~\ref{sec:distance_distribution_app} shows how physical topology affects congestion and how topology-aware Potts partitioning mitigates it.
Section~\ref{sec:bus_chain_app} describes source-synchronous boundary transport on DSIM-1.
Section~\ref{sec:subgraph_energy_app} presents the disconnected-links control experiment.
Section~\ref{sec:quasi2d_app} tests the $\eta$ dependence on a quasi-two-dimensional (quasi-2D) spin glass, Section~\ref{sec:p103_app} repeats the test at $251{,}328$ p-bits on DSIM-2, and Section~\ref{sec:sampling_app} complements the sampling experiment of the main text.
Section~\ref{sec:gpu_time_app} reports the runtime of the graphics processing unit (GPU) reference and the time and energy to target.
Section~\ref{sec:ucie_feasibility_app} translates the timing rule into a Universal Chiplet Interconnect Express (UCIe) feasibility check and reports 7~nm application-specific integrated circuit (ASIC) metrics.
Section~\ref{sec:g81_app} reports the G81 Max-Cut result.
Section~\ref{sec:slr_partitioning_app} explains Super Logic Region (SLR) driven partitioning on DSIM-2.
Sections~\ref{sec:pegasus_zephyr_app}--\ref{sec:3sat_app} present Pegasus, Zephyr, and 3-literal satisfiability (3SAT) results on DSIM-2.
Section~\ref{sec:arch_comparisons_app} compares the DSIM with other multi-chip architectures.
Throughout, we use the same residual-energy definition as the main text: for a run returning final energy \(E^f\), the final residual energy per spin is
\begin{equation}
\rho_E^f = \frac{E^f - E_{\mathrm{ground}}}{N},
\label{eq:residual_energy_def_supp}
\end{equation}
where \(N\) is the total number of p-bits and \(E_{\mathrm{ground}}\) is a putative ground energy established by aggregating the best energies found across extensive runs.

\subsection{Background}
\label{sec:background_app}

The main text introduces p-bits, the sparse Ising machine architecture, and the broader Ising machine context.
This section adds device-level and algorithmic details that support the supplementary derivations.

Since the initial demonstration of stochastic magnetic tunnel junction (sMTJ) based p-bits~\cite{borders2019integer}, the device space has grown rapidly.
Heterogeneous complementary metal-oxide-semiconductor (CMOS) plus sMTJ prototypes~\cite{singh2024cmos,selcuk2025dac,duffee2025integrated} and on-chip p-bit cores using stochastic MTJs paired with 2D transistors~\cite{daniel2024experimental} have brought p-bit hardware closer to monolithic integration.
On the architecture side, the sparse Ising machine (SIM) has been extended to support all-to-all reconfigurability through multiplexed sparse topologies~\cite{nikhar2024all} and satisfiability solvers~\cite{grimaldi2022spintronics}, and recent work showed that probabilistic computers (p-computers) can match and exceed quantum-annealer scaling exponents on hard 3D spin-glass benchmarks~\cite{chowdhury2025pushing}.

When a problem exceeds a single device, related multi-chip Ising efforts have partitioned the computation across networked field-programmable gate arrays (FPGAs) with autonomous synchronization~\cite{tatsumura2021scaling,kashimata2024efficient}, scaled analog Ising machines to multi-chip configurations~\cite{sharma2022increasing}, and extended CMOS annealing chips beyond single-die limits~\cite{takemoto20214}.
More generally, studies of networked computing dies  have identified inter-chip bandwidth as the primary multi-chip scaling constraint~\cite{radway2021illusion,srimani2024next}.
The DSIM follows the same broad motivation but differs in how the cost of distribution is characterized: the main text shows that performance depends on how fresh the boundary information is, summarized by the timing ratio $\eta$.

Cluster mean-field theory (CMFT) provides a direct algorithmic counterpart to that hardware picture.
Yamamoto's correlated cluster mean-field theory~\cite{yamamoto2009ccmf} improved on earlier formulations~\cite{weiss1907,oguchi1951statistics,bethe1935statistical,pelizzola2005cluster} by making the effective boundary fields depend on the cluster's own configuration.
Our parallel-CMFT implementation adapts this framework to a distributed optimization setting, as detailed in the next section.

\pagebreak
\subsection{Parallel cluster mean-field theory}
\label{sec:CMFT}

Parallel cluster mean-field theory (parallel-CMFT) is the algorithmic model compared with the hardware in Fig.~\ref{fig:extreme_fig3}c,d of the main text.
We give the full formulation below and then explain why a graphics processing unit (GPU) based CMFT model reproduces the optimization dynamics of the DSIM hardware.

\subsubsection{Formulation}

The exact Ising graph is partitioned into disjoint clusters (also called partitions in the hardware context), where spins are updated using stochastic Monte Carlo dynamics within each cluster.
A boundary p-bit is one that has at least one neighbor in a different cluster.
The coupling weights $J_{ij}$ are unchanged, and the only approximation is that boundary p-bits in neighboring clusters are represented by their mean-field averages rather than their instantaneous states~\cite{weiss1907,bethe1935statistical,oguchi1951statistics,pelizzola2005cluster,yamamoto2009ccmf,zimmer2016quantum}.

\begin{figure}[h]
  \centering
  \includegraphics[width=1\textwidth]{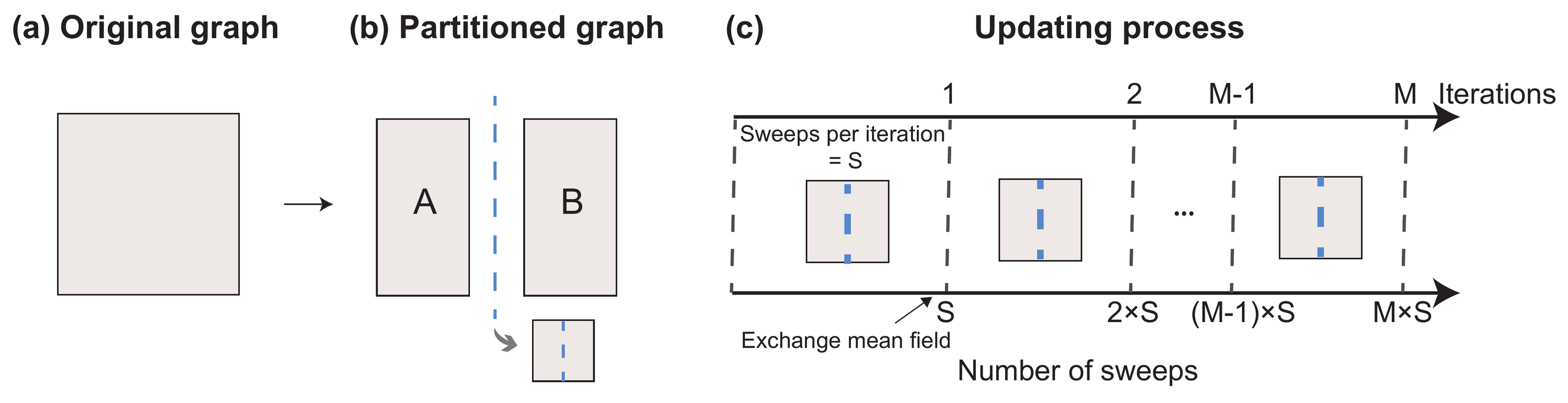}
\caption{
\textbf{Parallel cluster mean-field framework.}
\textbf{(a)} Original interaction graph.
\textbf{(b)} Partition of the graph into two clusters, $A$ and $B$.
\textbf{(c)} Iterative update with boundary exchange: each cluster performs $S$ internal sweeps before exchanging boundary mean-field values with its neighbor. This procedure is repeated for a total of $M$ iterations.
}
  \label{fig:supp_CMFT}
\end{figure}

To clarify the mechanism, we describe the method using a simplified two-cluster example, $A$ and $B$, as shown in Fig.~\ref{fig:supp_CMFT}.
Clusters $A$ and $B$ are updated in parallel.
During one cluster iteration, each cluster performs $S$ sweeps, where one sweep is a full Markov chain Monte Carlo (MCMC) update over all p-bits in both clusters.
Within these $S$ sweeps, each cluster updates its own p-bits exactly, but when computing local fields for boundary p-bits, the states of their neighbors in the other cluster are not available in real time. Instead, those contributions are approximated using mean-field values computed at the end of the previous iteration.
Each p-bit takes a bipolar state $m_i \in \{-1,+1\}$. 
For a boundary p-bit, the exchanged mean-field is computed as the average over the $S$ sweeps within one iteration.
If $m_i^{(t)}$ is the state of p-bit $i$ at the $t$-th Monte Carlo sweep, then the mean-field value after $S$ sweeps is given by
\[
\langle m_i \rangle = \frac{1}{S}\sum_{t=1}^{S} m_i^{(t)} .
\]

After completing the $S$ sweeps, the last configuration is retained as the updated state, and the newly computed boundary mean-fields are exchanged between clusters.
This completes one cluster iteration. These exchanged mean-field values remain fixed during the next iteration and are updated only after the subsequent set of $S$ sweeps.
Accordingly, the new local field for p-bit $i$ in cluster $A$ is
\[
I_i^{\mathrm{new}}
=
\beta \biggl(
h_i
+
\sum_{j \in A} J_{ij} m_j
+
\sum_{k \in B} J_{ik} \langle m_k \rangle
\biggr),
\]
where $h_i$ is the original bias of p-bit $i$, $J_{ij}$, $J_{ik}$ are the weight strengths between the respective p-bit pairs, and $\beta$ is the inverse temperature.

\subsubsection{Why CMFT reproduces the hardware behavior}

In CMFT, the number of sweeps per iteration, $S$, plays the same role as $\eta = f_{\mathrm{comm}} / f_{\mathrm{p\text{-}bit}}$ in hardware: both control how often boundary information is refreshed relative to local updates (main text Fig.~\ref{fig:extreme_fig3}a,b). Large $S$ (infrequent exchange) is analogous to small $\eta$, and $S=1$ (exchange after every sweep) is analogous to large $\eta$.

Both systems evolve local subgraphs with boundary information
held between exchanges. In hardware, finite communication time
delays the arrival of new 1-bit boundary states. CMFT instead
holds mean-field averages fixed across $S$ sweeps and performs
exact local Gibbs updates conditional on these fields. In both
systems, the residual-energy decay exponent $\kappa_f$ decreases
with infrequent exchange and saturates near the unpartitioned
reference when exchange is frequent
(Fig.~\ref{fig:extreme_fig3}b,d).

\begin{figure}[t]
  \centering
  \includegraphics[width=1\textwidth]{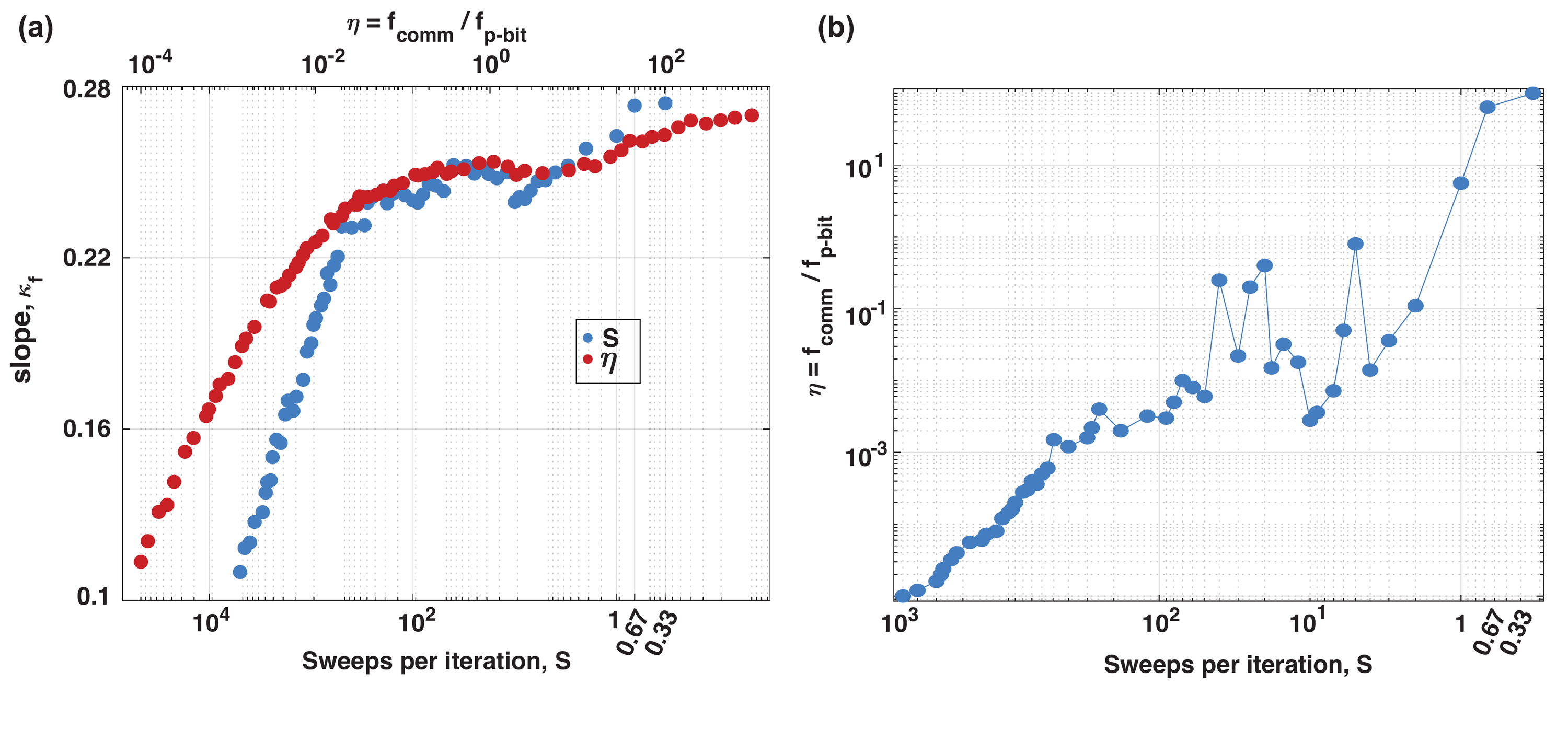}
\caption{
\textbf{Alignment of decay exponents from hardware and algorithmic control parameters.}
\textbf{(a)} Slope $\kappa_f$ extracted from the residual-energy decay as a function of the timing ratio $\eta = f_{\mathrm{comm}}/f_{\mathrm{p\text{-}bit}}$ (red, from DSIM-1 experiments) and the number of sweeps per iteration $S$ (blue, from parallel-CMFT).
The horizontal axes are related through a least-squares fit performed in log space, which rescales the sweep axis such that the two datasets collapse onto the same scaling curve.
The collapse indicates that both systems are shaped by the same mechanism: delayed boundary information in partitioned Gibbs dynamics.
The CMFT points include $S = 2/3$ and $S = 1/3$ sweeps per iteration (Sec.~\ref{sec:CMFT}), at which the CMFT exponent has saturated at the unpartitioned value.
\textbf{(b)} Mapping between $S$ and $\eta$ derived from \textbf{(a)}, where each point pairs the two control parameters that yield the same $\kappa_f$ value.
The mapping is monotonic at large $S$ (left) where mean-field averages are well converged, and becomes noisier at small $S$ (right) where both exponent curves are nearly flat, so that pairing by equal $\kappa_f$ is poorly conditioned.}
\label{fig:eta_sample_alignment}
\end{figure}

This is confirmed by the collapse of the fitted exponent curves from the two systems onto the same functional form once the $\eta$ axis is aligned with the $S$ axis (Fig.~\ref{fig:eta_sample_alignment}a). The explicit $\eta$--$S$ mapping is shown in Fig.~\ref{fig:eta_sample_alignment}b. Its scatter at small $S$ arises where both exponent curves are nearly flat, so that pairing values of $S$ and $\eta$ by equal $\kappa_f$ is poorly conditioned there.

The CMFT model takes no input from the hardware. From the graph, the partition and the annealing schedule, it computes the exponent as a function of the number of sweeps per iteration $S$, including where the exponent saturates, its disconnected limit and how both move when the partition changes. At $S = 1/N_{\mathrm{color}}$ it reduces to Gibbs sampling of the unpartitioned graph. 

The hardware enters only when we compare the model with a particular machine. The fit of Fig.~\ref{fig:eta_sample_alignment}a relates $S$ to the hardware ratio $\eta$ and places the model curve on the hardware axis, without changing the computed exponents. A fit is needed because for $S > 1$ the model exchanges mean-field averages over $S$ sweeps, while the hardware exchanges 1-bit states held between refreshes. We have not established that this mapping transfers unchanged to a new design.

\subsubsection{Fewer than one sweep per iteration ($S<1$)}

\begin{figure}[t]
  \centering
  \includegraphics[width=0.55\textwidth]{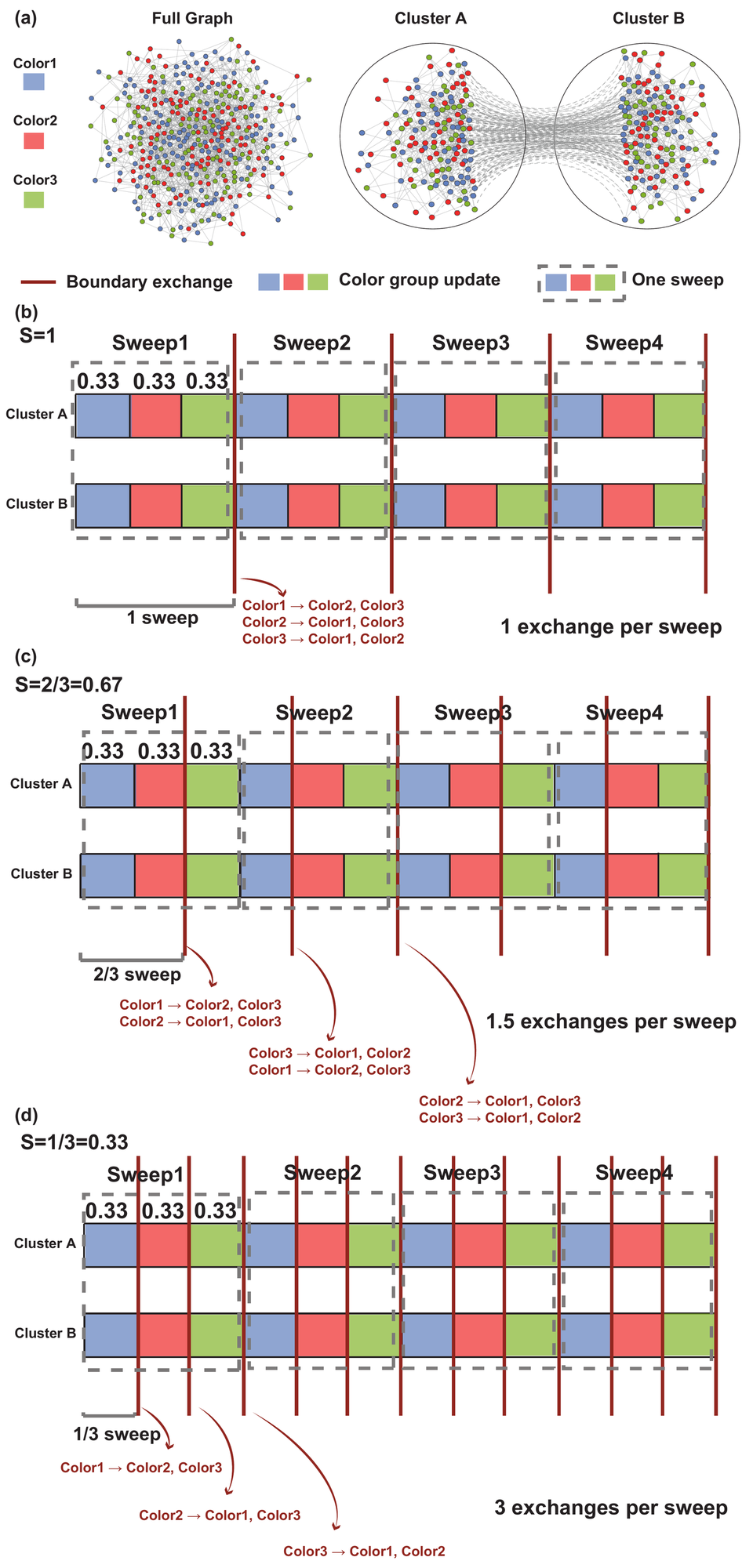}
\caption{
\textbf{Parallel-CMFT boundary exchange for $S \leq 1$ sweeps per iteration.} \textbf{(a)} A graph colored with three color groups and partitioned into clusters $A$ and $B$, with the couplings between clusters dashed. Each Monte Carlo sweep consists of three sequential color-group updates. \textbf{(b)} For $S=1$, boundary states are exchanged after every full sweep (one exchange per sweep). \textbf{(c)} For $S=2/3$, an exchange is performed after every two color-group updates (1.5 exchanges per sweep). \textbf{(d)} For $S=1/3$, an exchange is performed after every color-group update (three exchanges per sweep), the finest granularity used here. At each exchange event the current boundary states are communicated between clusters.}
  \label{fig:supp_subsweep}
\end{figure}

Above the optimal ratio the hardware delivers every boundary state within one color phase, so it refreshes boundary information more than once per sweep. We extended CMFT to do the same. At $L^3 = 37^3$ a sweep is $N_{\mathrm{color}} = 3$ color groups updated one after another (Fig.~\ref{fig:supp_subsweep}). Exchanging after every two color groups gives $S = 2/3$, and exchanging after every color group gives $S = 1/N_{\mathrm{color}} = 1/3$. With $S \le 1$ no boundary p-bit is updated twice between exchanges. What is exchanged is then its current 1-bit state, the same quantity the hardware sends. At $S = 1/3$ every color group sees current boundary states, so the CMFT dynamics are those of the unpartitioned graph, like the hardware above the optimal ratio.

Main text Fig.~\ref{fig:extreme_fig3}c,d includes these points. As $S$ decreases, the CMFT exponent rises monotonically and saturates near $\kappa_f \approx 0.27$ for $S \le 1$. The hardware saturates at $0.2637$, and the GPU gives $0.2693$. CMFT therefore reproduces the saturation the hardware shows near $\eta \approx 300$, and the model follows the hardware over the full range of $\eta$.

\subsection{Communication-cost metric and conservative clocking bound}
\label{sec:comm_cost_app}
\label{sec:freq_bound_details_app}

We now develop the communication-cost metric $C_{\max}$ that enters the conservative clocking bound of the main text, Eq.~\eqref{eq:extreme_fpbit_bound_main}, and then derive the bound itself.

A DSIM is defined by two resources that scale very differently.
Local field updates are bandwidth-limited by on-chip weight access, while the evaluation of cross-partition terms in those fields is rate-limited by how fast boundary \emph{states} can be exchanged.
The DSIM design goal is therefore not to move weights across devices, but to keep every weight local, and to pay only for exchanging 1-bit boundary p-bit states.
Once the graph is partitioned into \(K\) clusters, the system runs the same Ising MCMC dynamics as a monolithic machine, but boundary p-bit states arrive with some delay rather than being available instantaneously.

At the hardware level, boundary traffic is carried on a constrained physical interconnect.
For chain-like wiring, boundary packets that logically connect clusters \(a\) and \(b\) may need to traverse multiple hops, and the sustainable throughput is limited by the narrowest (most pin-limited) link along the route.
This section defines a scalar cost metric that quantifies the worst-case boundary-transport bottleneck.
That number is what ultimately appears inside the conservative clocking bound of the main text.

\subsubsection{Boundary sizes, hop distance, and pin budget}

Consider a sparse, undirected Ising graph with symmetric weights \(J_{ij}=J_{ji}\).
Let the vertex set be partitioned into \(K\) clusters, and label clusters by \(a,b\in\{1,\dots,K\}\).
For each unordered pair \((a,b)\), define
\begin{itemize}
\item \(b_{ab}\): the number of boundary p-bit \emph{states} that must be shipped from cluster \(a\) to cluster \(b\).
Operationally, \(b_{ab}\) is the number of p-bits in cluster \(a\) that participate in cut edges with cluster \(b\) (equivalently, the number of 1-bit state values that cluster \(b\) needs from \(a\) to evaluate cross-partition terms in its local fields). For an unordered pair, \(b_{ab}\) is the larger of the two directional counts.
\item \(d_{ab}\): the hop distance between the devices hosting clusters \(a\) and \(b\), defined as the number of physical links that a boundary packet must traverse.
For a linear chain, \(d_{ab}=|a-b|\) once an ordering has been chosen.
\item \(P_{ab}\): the number of usable data pins on the narrowest link along the route from \(a\) to \(b\).
When a route crosses multiple links, the bottleneck link determines \(P_{ab}\).
\end{itemize}

Note that \(b_{ab}\) is a property of the \emph{partition}, while \(d_{ab}\) and \(P_{ab}\) are properties of the \emph{physical mapping and wiring}.
A partitioning tool can minimize the total number of cut edges, but it does not, by itself, guarantee that those edges are aligned with short physical paths.

\subsubsection{Total communication cost and worst-path cost}

We define two related costs.
The first is a total communication cost:
\begin{equation}
  C_{\text{tot}} = \sum_{a<b} b_{ab}\,\frac{d_{ab}}{P_{ab}}.
\label{eq:comm_cost_sum_app}
\end{equation}
This quantity is useful for comparing different mappings and partitions because it increases when a large number of boundary bits are forced onto long or narrow paths.
For clocking feasibility, however, the critical object is the worst (most demanding) route.
We therefore define a worst-path cost:
\begin{equation}
  C_{\max} = \max_{a<b}\left( b_{ab}\,\frac{d_{ab}}{P_{ab}} \right).
\label{eq:cworst_def_app}
\end{equation}

\(C_{\max}\) identifies the cluster pair that creates the tightest boundary-transport bottleneck once the design is placed on the physical interconnect. To keep the estimate simple and general, we use the most demanding
FPGA pair without adding the bits from other pairs that share a
link. Counting those additional bits slightly changes the estimated
ratios but does not change our conclusions. The values depend on
the chosen partition and the pins assigned to each link. For custom
link designs, more detailed estimates can account for how bits
are packed and forwarded.

\subsubsection{Why permutation matters (even when the partition is fixed)}

\begin{figure}[t]
  \centering
  \includegraphics[width=\textwidth]{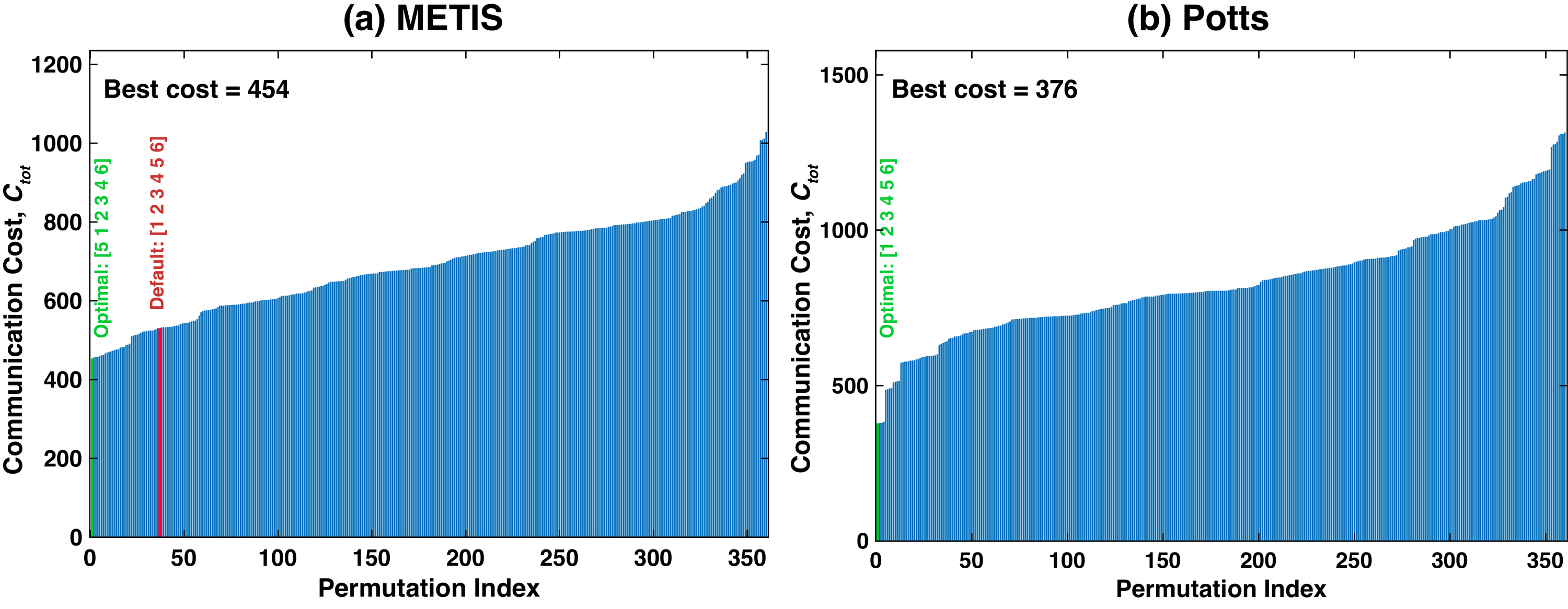}
  \caption{
  \textbf{Permutation sensitivity of communication cost on a 6-FPGA chain.}
  Communication cost \(C_{\text{tot}}\) from Eq.~\eqref{eq:comm_cost_sum_app} versus permutation index for six clusters of the \(L^3 = 37^3\) Edwards--Anderson (EA) spin glass.
  \textbf{(a)} METIS partitioning: the default ordering is suboptimal and an explicit mapping search is needed.
  \textbf{(b)} Potts partitioning: the topology-aware partition is naturally aligned with the chain, so the canonical ordering (or its reverse) already achieves the minimum.}
  \label{fig:supp_comm_cost}
\end{figure}

For standard partitioners such as METIS~\cite{karypis1998software} and KaHIP~\cite{Sanders2013KaHIP}, the cluster labels are arbitrary.
A partition defines membership, but it does not define an ordering.
On a chain, the physical slot order \([1,2,\dots,K]\) is a separate choice, and changing that order changes \(d_{ab}\), and therefore changes both \(C_{\text{tot}}\) and \(C_{\max}\).
For \(K=6\), there are \(6!/2=360\) distinct orderings up to reversal.
Fig.~\ref{fig:supp_comm_cost}a shows that the choice of ordering alone can change \(C_{\text{tot}}\) by more than a factor of two for a representative \(L^3=37^3\) instance on a 6-node chain, meaning that standard partitioners would require an explicit search over orderings to find the best placement.
The topology-aware Potts partitioning introduced in Section~\ref{sec:distance_distribution_app} avoids this search entirely: because the partition is constructed to align with the chain, the canonical ordering (or its reverse) already achieves the minimum cost (Fig.~\ref{fig:supp_comm_cost}b).

\subsubsection{From cost metric to clocking bound}

Equation~\eqref{eq:extreme_fpbit_bound_main} in the main text bounds the local update clock given $C_{\max}$.
We derive it here.
Consider boundary traffic from cluster \(a\) to cluster \(b\).
At communication clock \(f_{\mathrm{comm}}\), a link with \(P_{ab}\) data pins can deliver \(P_{ab}\) boundary bits per clock cycle.
Shipping \(b_{ab}\) boundary states through one link therefore takes \(b_{ab}/(P_{ab} f_{\mathrm{comm}})\) seconds.
Because \(b_{ab}\) typically exceeds \(P_{ab}\), the boundary bits must be sent in multiple frames that time-division multiplex the available pins, so the end-to-end transfer time scales as \(b_{ab} d_{ab}/(P_{ab} f_{\mathrm{comm}})\).
We include a factor of two to account for the packing and unpacking overhead at the sender and receiver, giving the transport time for one cluster pair:
\begin{equation}
  \tau_{ab} = \frac{2\,b_{ab}\,d_{ab}}{P_{ab}\,f_{\mathrm{comm}}}.
\label{eq:tab_delay}
\end{equation}

\subsubsection{Including coloring and the clocking bound}

Updates use graph coloring~\cite{aadit2022massively} with \(N_{\mathrm{color}}\) independent sets, so that one sweep (one Monte Carlo sweep, MCS) updates all colors once.
For boundary states to remain useful, they must be refreshed before the local p-bits that depend on them are updated again.
To be safe, the communication network needs to handle boundary updates roughly as fast as the system cycles through its color updates.
The worst-case communication time is therefore
\begin{equation}
  \tau_{\text{comm}}
  =
  N_{\mathrm{color}} \max_{a<b}\tau_{ab}
  =
  \frac{2\,N_{\mathrm{color}}\,C_{\max}}{f_{\mathrm{comm}}},
  \label{eq:tcomm_appendix}
\end{equation}
where \(C_{\max}\) is defined in Eq.~\eqref{eq:cworst_def_app}.
The local p-bit update period is \(\tau_{\text{p-bit}} = 1/f_{\mathrm{p\text{-}bit}}\).
A conservative requirement is that the local update period should not be much shorter than the time needed to refresh the most demanding boundary path, \(\tau_{\text{p-bit}} \ge \tau_{\text{comm}}\).
Combining with Eq.~\eqref{eq:tcomm_appendix} gives
\begin{equation}
  f_{\mathrm{p\text{-}bit}}
  \le
  \frac{f_{\mathrm{comm}}}{2\,N_{\mathrm{color}}\,C_{\max}}
  \equiv f_{\mathrm{p\text{-}bit,max}},
  \label{eq:fpbit_bound_appendix}
\end{equation}
which is Eq.~\eqref{eq:extreme_fpbit_bound_main} of the main text.

\subsubsection{Evaluating the bound for DSIM-1}

We now evaluate Eq.~\eqref{eq:fpbit_bound_appendix} for the \(L^3=37^3\) system on the 6-FPGA DSIM-1 chain with topology-aware Potts partitioning (introduced next in Section~\ref{sec:distance_distribution_app}).
The five inter-FPGA links have pin counts \(P = [54,\, 30,\, 54,\, 26,\, 54]\) (from FPGA~1--2 through FPGA~5--6).
For a non-adjacent pair \((a,b)\), the bottleneck pin count is the minimum over all links along the path from \(a\) to \(b\).
From the measured partition, the worst-case cluster pair is \((4,\,6)\), with boundary size \(b_{46} = 660\), hop distance \(d_{46} = 2\), and bottleneck pin count \(P_{46} = \min(26,\,54) = 26\).
This gives
\[
  C_{\max}
  = b_{46}\,\frac{d_{46}}{P_{46}}
  = 660 \times \frac{2}{26}
  \approx 50.8.
\]
With \(N_{\mathrm{color}}=3\), the conservative bound predicts the optimal ratio
\[
  \eta_{\mathrm{threshold}}
  = 2\,N_{\mathrm{color}}\,C_{\max}
  = 2 \times 3 \times 50.8
  \approx 305,
\]
which is consistent with the empirical onset of saturation near \(\eta \approx 300\) observed in Fig.~\ref{fig:extreme_fig2}c of the main text.

\subsubsection{Evaluating the bound for DSIM-2}
\label{sec:dsim2_bound_app}

DSIM-2 runs all its FPGAs from one global clock, so the bound of Eq.~\eqref{eq:fpbit_bound_appendix} fixes one safe clock for the whole machine. We call conservative clock the fastest p-bit clock at which every boundary state reaches its destination within one color phase, even over the longest path between devices. At this clock the boundary information is current at every update, and the machine satisfies Eq.~\eqref{eq:fpbit_bound_appendix}. A faster clock takes the machine below the optimal ratio, and the boundary information arrives late.

We evaluate the bound for the two DSIM-2 designs of this work from their partitions and their placement on the \(18\) FPGAs. In the \(L^3 = 100^3\) design (\(10^6\) p-bits, \(N_{\mathrm{color}} = 2\)), the worst-case pair sends \(b_{ab} = 3{,}607\) boundary states over \(d_{ab} = 3\) hops of \(P_{ab} = 80\) pins, so
\[
  C_{\max} = 3607 \times \frac{3}{80} \approx 135,
  \qquad
  \eta_{\mathrm{threshold}} = 2 \times 2 \times 135 \approx 541.
\]
Each pin runs at \(1.8\)~Gb/s, and every transfer has a fixed overhead of about \(54\)~ns. A pin carries \(45\) boundary states, which sets the effective communication clock at \(f_{\mathrm{comm}} \approx 0.57\)~Gb/s. Eq.~\eqref{eq:fpbit_bound_appendix} then gives \(f_{\mathrm{p\text{-}bit,max}} = 0.57~\mathrm{GHz}/541 \approx 1.05\)~MHz, in line with the \(1\)~MHz at which this design closes timing (main text Fig.~\ref{fig:extreme_fig5}). The \(1\)~MHz point therefore sits at \(\eta \approx 570\), at the optimal ratio, and the \(3\)~MHz point at \(\eta \approx 190\), well below it.

The Pegasus P103 design (\(251{,}328\) p-bits, \(N_{\mathrm{color}} = 4\)) has a smaller worst-case boundary. Its worst-case pair sends \(b_{ab} = 1{,}722\) boundary states over \(d_{ab} = 3\) hops of \(P_{ab} = 80\) pins, so
\[
  C_{\max} = 1722 \times \frac{3}{80} \approx 64.6,
  \qquad
  \eta_{\mathrm{threshold}} = 2 \times 4 \times 64.6 \approx 517.
\]
With \(21.5\) boundary states per pin and the same overhead, \(f_{\mathrm{comm}} \approx 0.33\)~Gb/s, and Eq.~\eqref{eq:fpbit_bound_appendix} gives \(f_{\mathrm{p\text{-}bit,max}} = 0.33~\mathrm{GHz}/517 \approx 0.63\)~MHz. The conservative clock of this design, \(0.47\)~MHz, lies inside this bound, and the overclocked runs at \(0.75\) and \(1.25\)~MHz lie below the optimal ratio (Sec.~\ref{sec:p103_app}). Table~\ref{tab:eta_threshold_app} collects all evaluations of Eq.~\eqref{eq:fpbit_bound_appendix}.

\begin{table*}[t]
\centering
\caption{\textbf{Optimal ratio of Eq.~\eqref{eq:fpbit_bound_appendix}.} For each design, \(b_{ab}\), \(P_{ab}\) and \(d_{ab}\) are the number of boundary states, the number of pins and the number of hops of the worst-case device pair, \(C_{\max} = b_{ab} d_{ab}/P_{ab}\), and \(\eta_{\mathrm{threshold}} = 2 N_{\mathrm{color}} C_{\max}\). The last column gives \(\eta = f_{\mathrm{comm}}/f_{\mathrm{p\text{-}bit}}\) at the operating points reported in this work.}
\label{tab:eta_threshold_app}
{\renewcommand{\arraystretch}{1.25}\setlength{\tabcolsep}{8pt}
\begin{tabular}{@{}lccccccc@{}}
\toprule
System & \(b_{ab}\) & \(P_{ab}\) & \(d_{ab}\) & \(C_{\max}\) & \(N_{\mathrm{color}}\) & \(\eta_{\mathrm{threshold}}\) & \begin{tabular}[b]{@{}c@{}}Operating\\\(\eta\)\end{tabular} \\
\midrule
\multicolumn{8}{@{}l}{\textit{DSIM-1}} \\
\(L^3 = 37^3\) EA spin glass & 660 & 26 & 2 & 50.8 & 3 & 305 & swept, main text Fig.~\ref{fig:extreme_fig2}c \\
Quasi-2D ferromagnet, P25 & 134 & 26 & 3 & 15.5 & 4 & 124 & swept, main text Fig.~\ref{fig:extreme_fig6_sampling} \\
Quasi-2D spin glass, P33 & 258 & 26 & 3 & 29.8 & 4 & 240 & swept, Sec.~\ref{sec:quasi2d_app} \\
\midrule
\multicolumn{8}{@{}l}{\textit{DSIM-2}} \\
\(L^3 = 100^3\) EA spin glass & 3{,}607 & 80 & 3 & 135 & 2 & 541 & \begin{tabular}[t]{@{}c@{}}570 at 1~MHz\\190 at 3~MHz\end{tabular} \\
\addlinespace[2pt]
Quasi-2D spin glass, P103 & 1{,}722 & 80 & 3 & 64.6 & 4 & 517 & \begin{tabular}[t]{@{}c@{}}690 at 0.47~MHz\\435 at 0.75~MHz\\260 at 1.25~MHz\end{tabular} \\
\bottomrule
\end{tabular}}
\end{table*}

\subsection{Scaling with the number of devices}
\label{sec:scaling_app}

Here we derive how the flip rate and the communication each device needs change as devices are added. Let $N$ be the number of p-bits and $K$ the number of devices, so each device holds $N/K$ p-bits.

Every device updates its partition at the p-bit clock $f_{\mathrm{p\text{-}bit}}$, and this clock is set by the bounded neighbor count of the sparse graph, not by $N$ or $K$~\cite{aadit2022massively}. The flip rate is therefore
\begin{equation}
R = N f_{\mathrm{p\text{-}bit}}
\label{eq:rate_scaling_app}
\end{equation}
for every $K$, whether or not the boundary states are current. Partitioning does not cost throughput.

What partitioning changes is where the ratio $\eta = f_{\mathrm{comm}}/f_{\mathrm{p\text{-}bit}}$ sits relative to the optimal ratio of Eq.~\eqref{eq:fpbit_bound_appendix},
\begin{equation}
\eta_{\mathrm{threshold}} = 2 N_{\mathrm{color}} C_{\max},
\qquad
C_{\max} = \max_{a<b} b_{ab}\,\frac{d_{ab}}{P_{ab}},
\label{eq:eta_threshold_app}
\end{equation}
where $a$ and $b$ label devices. Each p-bit of device $a$ adds at most one boundary state to $b_{ab}$, so $b_{ab} \le N/K$ and
\begin{equation}
\eta_{\mathrm{threshold}} \le 2 N_{\mathrm{color}}\,\frac{N}{K}\,\frac{\max_{a<b} d_{ab}}{\min_{a<b} P_{ab}},
\label{eq:eta_bound_app}
\end{equation}
where the maximum and the minimum run over the pairs of devices that exchange boundary states. At fixed $N/K$, the number of devices enters this bound only through $\max_{a<b} d_{ab}$, the longest route a boundary state travels. A partition that ignores physical distance can stretch this route across the whole chain, $K-1$ hops, and the bound then grows linearly with $K$. The bound reaches this order when the two end devices exchange a boundary of order $N/K$, which is what the all-to-all cut pattern of Section~\ref{sec:distance_distribution_app} produces. A topology-aware partition keeps the route short. On DSIM-1 it keeps $\max_{a<b} d_{ab} \le 2$ for the $L^3 = 37^3$ lattice (Fig.~\ref{fig:supp_pyramid}). The bound is then proportional to $N/K$ and does not grow with the number of devices.

Time to target also depends on how the residual energy decays. For a given graph and partition, the exponent follows the measured curve $\kappa_f(\eta)$. It rises from its value at $\eta = 0$, where the partitions are disconnected, to the unpartitioned value $\kappa_f(\eta=\infty)$ at $\eta \ge \eta_{\mathrm{threshold}}$ (main text Fig.~\ref{fig:extreme_fig3}b and Fig.~\ref{fig:eta_sample_alignment}). Over the measured range we write $\rho_E = A(\eta)\, t_a^{-\kappa_f(\eta)}$. Reaching a target $\rho_E^\star$ in that range takes $(A/\rho_E^\star)^{1/\kappa_f}$ sweeps, and each sweep takes one period $1/f_{\mathrm{p\text{-}bit}}$, so the time to target is
\begin{equation}
T(\eta) = \frac{1}{f_{\mathrm{p\text{-}bit}}}\left(\frac{A(\eta)}{\rho_E^\star}\right)^{1/\kappa_f(\eta)}.
\label{eq:tts_scaling_app}
\end{equation}
The first factor is the time per sweep, which does not depend on $N$. The second is the number of sweeps, which grows below the optimal ratio through both the smaller exponent and the prefactor $A(\eta)$ (Section~\ref{sec:quasi2d_app}). A faster p-bit clock cuts the first factor linearly, but below the optimal ratio it raises the second as $\kappa_f$ falls. The operating point that minimizes $T$ therefore lies below $\eta_{\mathrm{threshold}}$ whenever the shorter sweeps outweigh the extra sweeps. This is the overclocked mode of main text Figs.~\ref{fig:extreme_fig4} and~\ref{fig:extreme_fig5}. A disconnected machine, $\eta = 0$, never optimizes the couplings between partitions, whatever its throughput (Section~\ref{sec:subgraph_energy_app}).

If the longest route stays short and $A(\eta)$ and $\kappa_f(\eta)$ do not change, Eq.~\eqref{eq:tts_scaling_app} does not depend on $K$ at fixed $N/K$. The time to target then stays the same as the problem and the number of devices grow together. A monolithic processor behaves differently. Its time per sweep grows in proportion to $N$ once $N$ exceeds its number of parallel units (Section~\ref{sec:gpu_time_app}). At a fixed p-bit clock at or above the optimal ratio, the time-to-target advantage of the distributed machine therefore grows in proportion to $N$, up to the prefactors $A$. Below the optimal ratio, the overclocked machine trades its shorter sweeps against the factor $(\rho_E^\star)^{1/\kappa_f(\eta) - 1/\kappa_f(\eta=\infty)}$, and this trade sets the crossover of main text Figs.~\ref{fig:extreme_fig4}b and~\ref{fig:extreme_fig5}b.

Technology enters only through $f_{\mathrm{p\text{-}bit}}$ and $f_{\mathrm{comm}}$. A faster device lowers $\eta$ at fixed links, and the exponent then follows from $\kappa_f(\eta)$. The machine stays at or above the optimal ratio as long as
\begin{equation}
f_{\mathrm{comm}} \min_{a<b} P_{ab} \ge 2 N_{\mathrm{color}}\,\frac{N}{K}\,\max_{a<b} d_{ab}\, f_{\mathrm{p\text{-}bit}}.
\label{eq:link_scaling_app}
\end{equation}
The link bandwidth per device has to grow with the device clock and with the boundary each link carries, which is at most the number of p-bits per device. Existing die-to-die links meet this condition for the 7~nm partition at $100$~MHz (Section~\ref{sec:ucie_feasibility_app}).

\begin{figure}[t]
  \centering
  \includegraphics[width=0.65\textwidth]{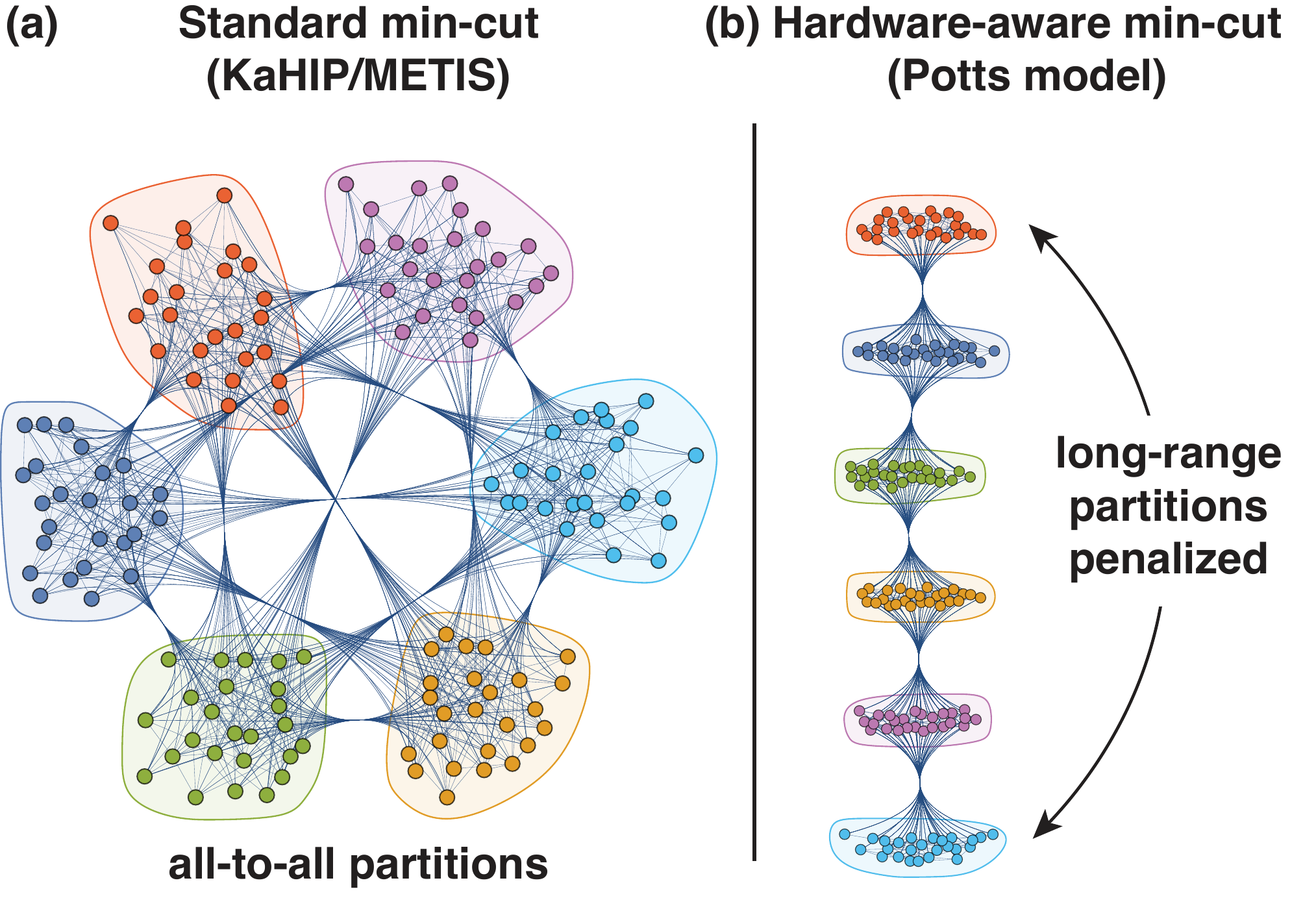}
  \caption{
  \textbf{Standard versus topology-aware partitioning.}
  \textbf{(a)} A distance-blind objective (KaHIP/METIS) minimizes cut size without encoding physical distance, producing all-to-all inter-cluster connections.
  \textbf{(b)} A Potts objective penalizes long-distance interactions in the cluster order, producing partitions that align naturally with a chain and reduce worst-path congestion.}
  \label{fig:supp_potts_cartoon}
\end{figure}

\subsection{Distance distribution and topology-aware Potts partitioning}
\label{sec:distance_distribution_app}
\label{sec:potts_partitioning_app}

The communication-cost metric defined above depends on the hop distance $d_{ab}$ of each cut edge. A small total cut can still be harmful if a noticeable fraction of boundary interactions must traverse two or three hops, because those multi-hop routes concentrate traffic on shared intermediate links.

A direct way to see this is to classify each cut interaction by its distance \(d\).
For a 6-cluster chain, \(d\in\{1,2,3,4,5\}\), and the distribution over \(d\) depends on how partitions align with the chain.
When long-distance traffic is present, it increases both total communication cost and worst-path cost because the same physical links must carry boundary data for many pairs.
This motivates a partitioning strategy that is aware of the physical chain topology.

Standard graph partitioners aim to minimize cut size while balancing cluster sizes, which is the right objective if the hardware topology is all-to-all or if any cluster can communicate with any other at the same cost.
A chain is neither: physical distance matters, and multi-hop routes share links.
To produce partitions that are naturally ordered along a chain, we introduce a Potts objective that penalizes interactions between clusters that are far apart in cluster index.
The objective is used only to compute the cluster assignment and does not change the Ising problem being solved by the p-computer.
Fig.~\ref{fig:supp_potts_cartoon} illustrates the conceptual difference between a distance-blind cut objective and the topology-aware Potts objective.

\begin{figure}[t]
  \centering
  \includegraphics[width=0.9\textwidth]{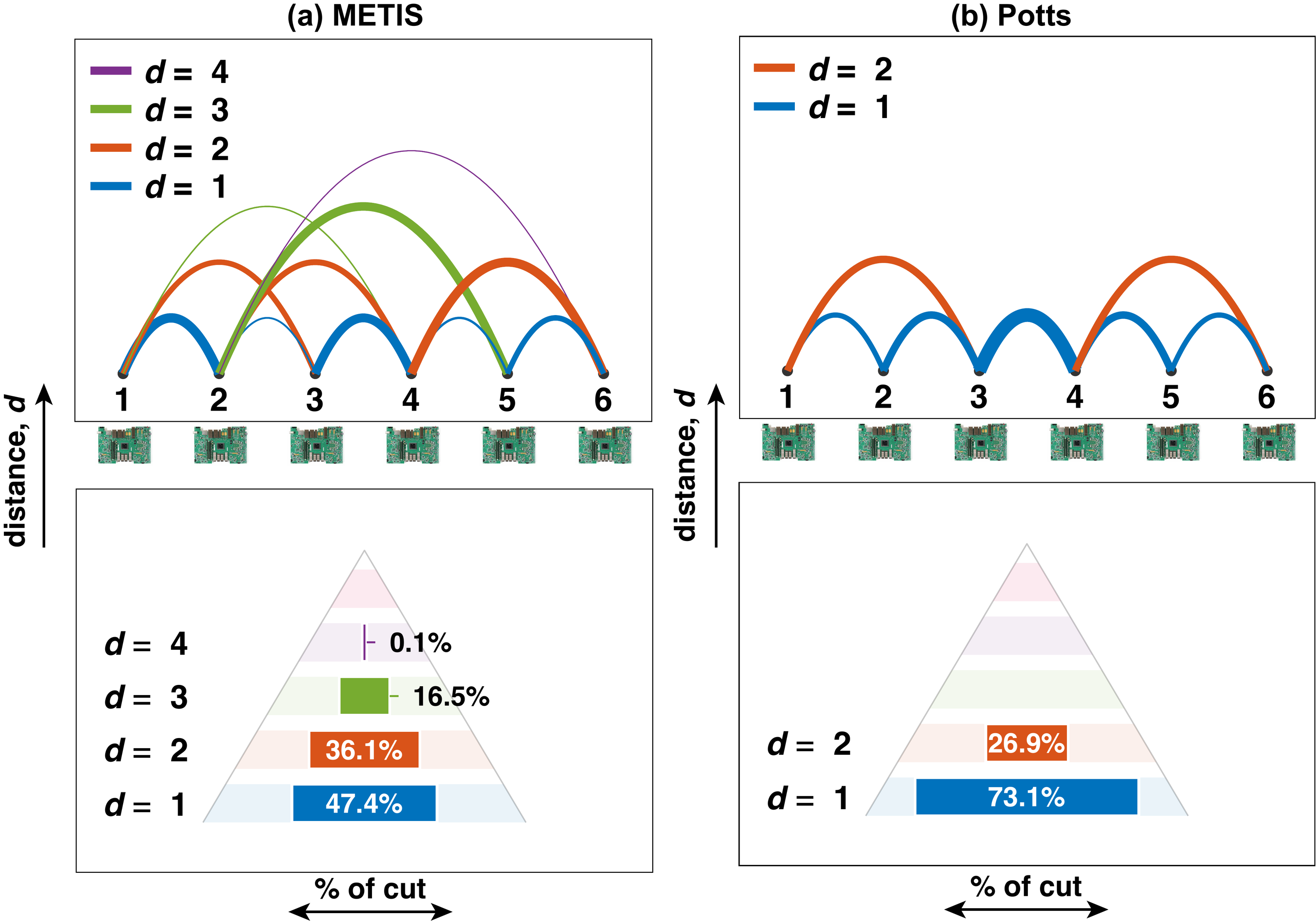}
  \caption{
  \textbf{Distance distribution of cut edges for six clusters on a chain.}
  \textbf{(a)} METIS partitioning produces cuts at distances up to \(d=4\), with only 47.4\% of cut edges at \(d=1\).
  \textbf{(b)} Potts partitioning concentrates 73.1\% of cut edges at \(d=1\) and limits the maximum distance to \(d=2\).
  Concentrating cut interactions at small \(d\) reduces multi-hop pressure on intermediate links and lowers worst-path cost. In both panels, \% of cut denotes the fraction of total cut edges at each distance.}
  \label{fig:supp_pyramid}
\end{figure}

Let \(s_i\in\{1,\dots,K\}\) be the cluster label of node \(i\), where \(K\) is the number of clusters.
We define
\begin{equation}
  H_{\text{Potts}}(\mathbf{s})
  =
  \sum_{(i,j)\in \mathcal{E}} |J_{ij}|\,\kappa(|s_i-s_j|)
  \;+\;
\lambda \sum_{q=1}^{K} \left(n_q - \frac{N}{K}\right)^2,
  \label{eq:potts_energy_appendix}
\end{equation}
where $\mathcal{E}$ is the set of edges in the graph, \(\kappa(d)\) is a distance kernel that penalizes cuts between clusters whose indices differ by \(d = |s_i - s_j|\), \(n_q = \sum_i \mathbf{1}\{s_i=q\}\) is the size of cluster \(q\), \(N\) is the total number of nodes, and the hyperparameter \(\lambda > 0\) controls the strength of the balance penalty.
A simple kernel is
\begin{equation}
  \kappa(d) =
  \begin{cases}
    0, & d=0,\\
    \delta_{\text{near}}, & d=1,\\
    \delta_{\text{far}}, & d\ge 2,
  \end{cases}
  \qquad 0 < \delta_{\text{near}} < \delta_{\text{far}},
  \label{eq:potts_distance_kernel}
\end{equation}
with \(\delta_{\text{far}}/\delta_{\text{near}}\gg 1\).
Minimizing Eq.~\eqref{eq:potts_energy_appendix} suppresses long-range cluster interactions and concentrates the cut at \(d=1\) (and sometimes \(d=2\)) in the cluster index order.
Fig.~\ref{fig:supp_pyramid} confirms this: Potts partitioning concentrates over 73\% of cut edges at \(d=1\), compared with under 48\% for METIS.

\begin{figure}[t]
  \centering
  \includegraphics[width=0.65\textwidth]{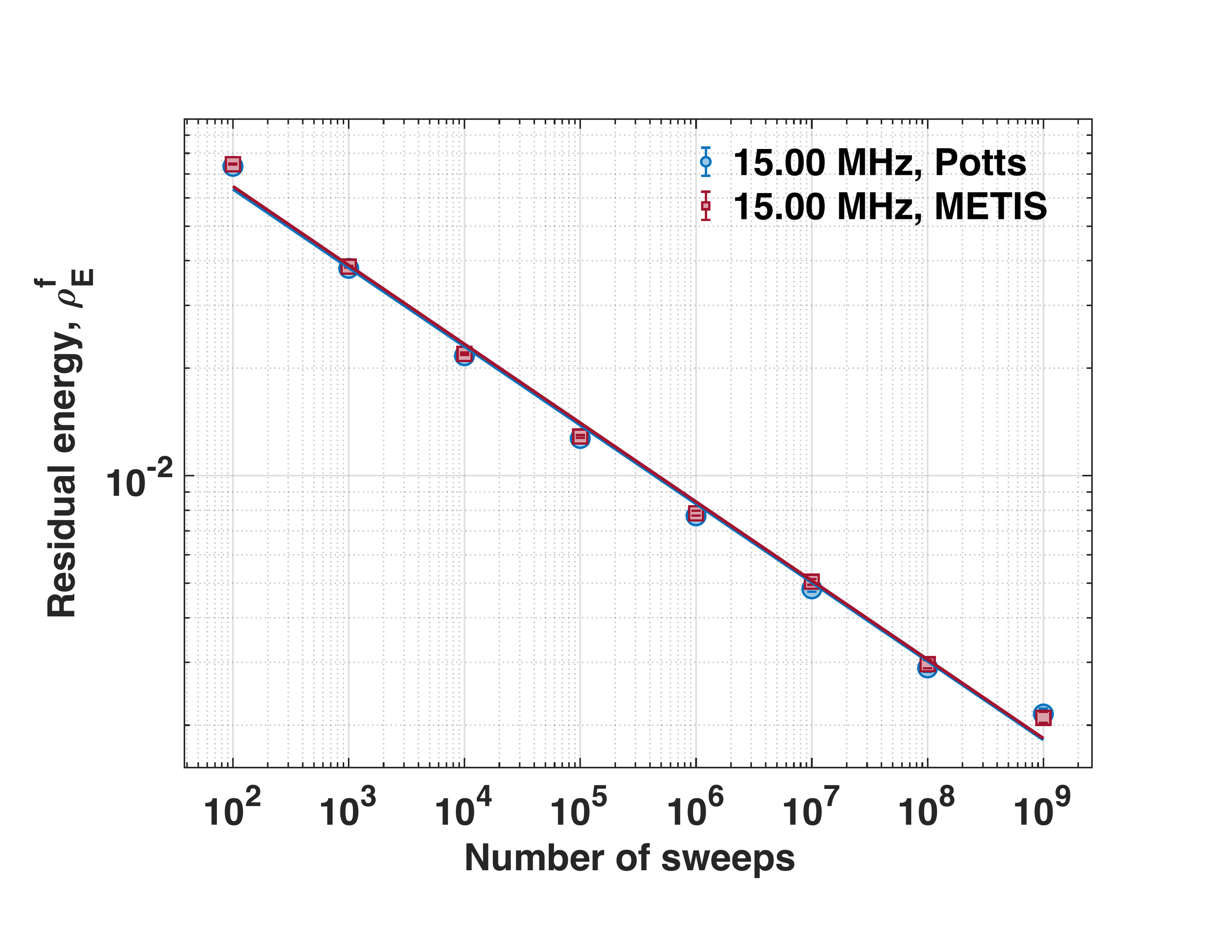}
  \caption{
  \textbf{Solution quality is unchanged under topology-aware partitioning (DSIM-1).}
  Final residual energy \(\rho_E^f\) versus sweeps for the \(L^3 = 37^3\) EA spin glass on DSIM-1 at $f_{\mathrm{p\text{-}bit}} = 15$~MHz using METIS partitions and topology-aware Potts partitions.
The overlap indicates that the Potts objective improves wiring alignment without degrading the optimization result.
All points: 10 instances, 10 runs per instance, error bars are 95\% bootstrap confidence intervals.}
  \label{fig:supp_potts_metis_powerlaw}
\end{figure}

Changing the partition objective changes the physical communication pattern, but it does not change the optimization behavior.
Fig.~\ref{fig:supp_potts_metis_powerlaw} confirms this: the residual-energy decay is statistically indistinguishable between METIS and Potts partitions once the DSIM is operated at the appropriate \(\eta\).
In practice, the Edwards--Anderson (EA) experiments on DSIM-1 use Potts partitioning to align with the linear chain topology, the quasi-2D spin glass and the P25 ferromagnet on DSIM-1 use KaHIP (Section~\ref{sec:quasi2d_app} and main text Methods), and all DSIM-2 experiments use METIS partitioning.

\subsection{Source-synchronous boundary transport on a 6-node chain}
\label{sec:bus_chain_app}

\begin{figure}[htbp]
  \centering
  \includegraphics[width=0.8\textwidth]{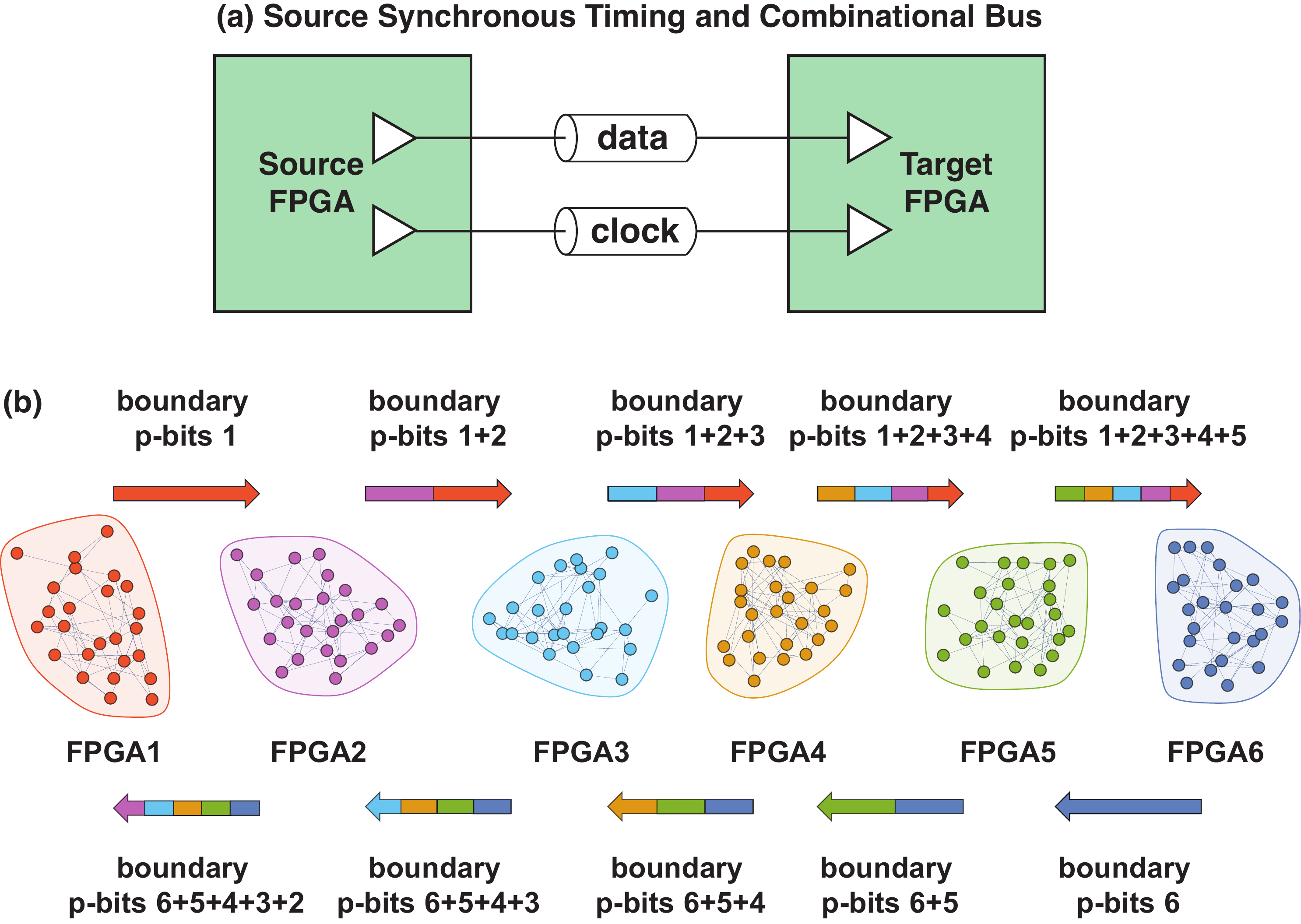}
  \caption{
  \textbf{Source-synchronous boundary transport on a nearest-neighbor chain.}
  \textbf{(a)} A conceptual source-synchronous link: a sender forwards a clock together with packed boundary bits, and the receiver samples relative to the forwarded clock.
  \textbf{(b)} The 6-node chain used in DSIM-1 measurements.
  Non-neighbor boundary interactions are carried as multi-hop traffic that traverses intermediate links, which is the physical origin of hop-dependent congestion captured by \(d_{ab}\) and \(C_{\max}\).}
  \label{fig:supp_bus_chain}
\end{figure}

\begin{figure}[htbp]
  \centering
  \includegraphics[height=0.42\textheight]{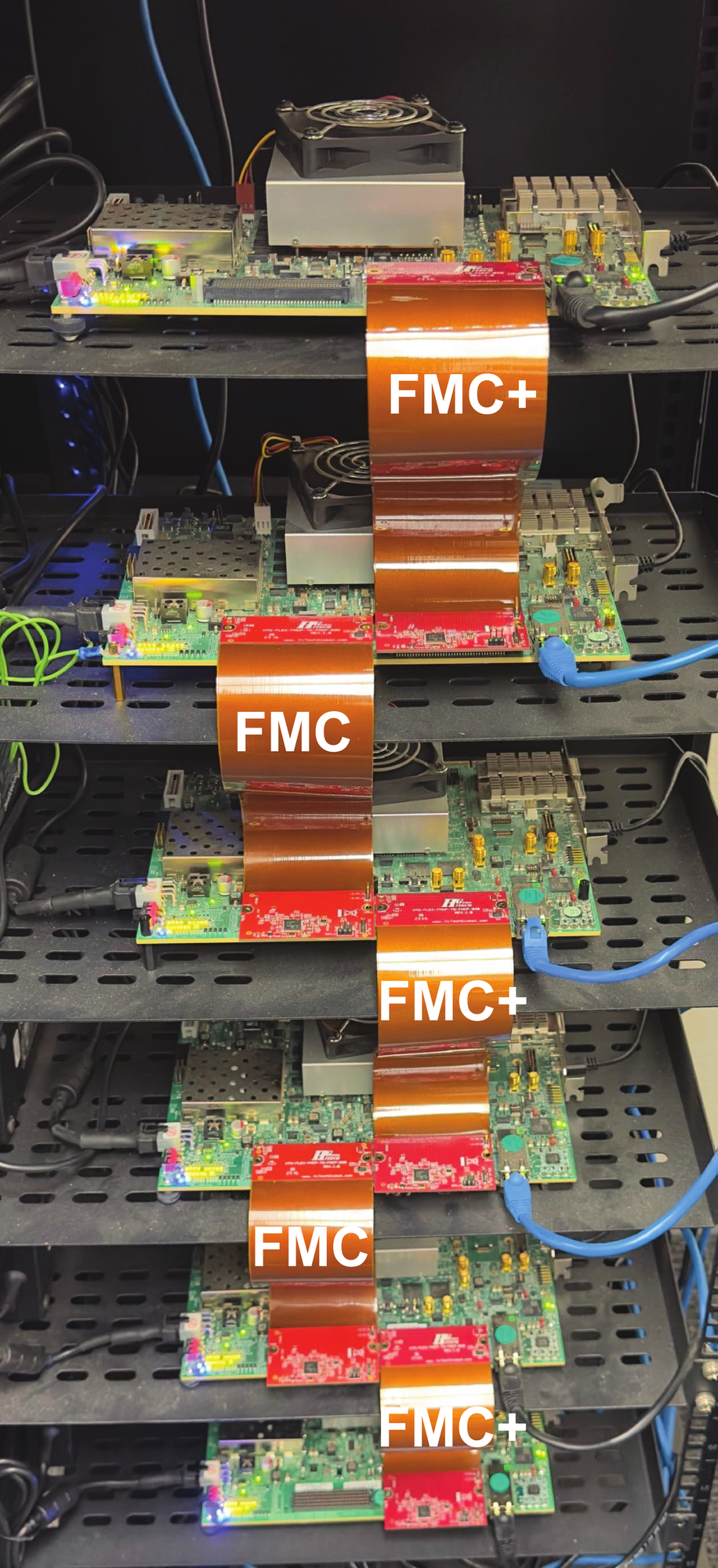}
  \caption{
  \textbf{DSIM-1 hardware.} The six-FPGA nearest-neighbor chain used in the DSIM-1 experiments, with adjacent boards linked over FPGA Mezzanine Card (FMC) and FMC+ connectors.}
  \label{fig:supp_dsim1_photo}
\end{figure}

This section describes the concrete 6-node chain used in the DSIM-1 experiments (Fig.~\ref{fig:extreme_fig1}e of the main text).
Boundary states are transported using a source-synchronous style: data and a forwarded clock travel together, so the receiver samples incoming bits relative to the forwarded clock, without requiring a global phase relationship to the local update clock.
This keeps the boundary channel simple and low-latency at the physical layer, and it matches the DSIM goal of exchanging boundary \emph{states} cheaply.
Fig.~\ref{fig:supp_bus_chain} sketches two complementary views.
Panel~(a) shows the conceptual link, emphasizing that boundary transport is about moving packed 1-bit states with a forwarded clock.
Panel~(b) shows the full chain and emphasizes why hop count matters: boundary information between non-neighboring partitions must traverse multiple intermediate boards, loading the shared links.
Before running any optimization, we verified each link by sending known test patterns and checking that every bit arrived correctly.
Fig.~\ref{fig:supp_dsim1_photo} shows a photograph of the assembled six-FPGA chain.

The whole graph is colored once, before partitioning, and every FPGA uses that coloring. Neighbors therefore never share a color, and within an FPGA two neighbors are never updated at the same time. Across FPGAs the situation is different, because the local clocks are independent: two neighboring p-bits on different FPGAs are occasionally updated at the same time. Only the couplings that cross between FPGAs can be involved, and the min-cut partition keeps these to between $2.5\%$ and $3.6\%$ of all couplings for the three DSIM-1 partitions of this work. The agreement with the unpartitioned reference shows that the sampler tolerates such events, in the same way that asynchronous self-colored p-bits tolerate occasional simultaneous updates~\cite{selcuk2025dac}.

\subsection{Disconnected-links control: isolating the origin of slope loss}
\label{sec:subgraph_energy_app}

\begin{figure}[htbp]
  \centering
  \includegraphics[width=\textwidth]{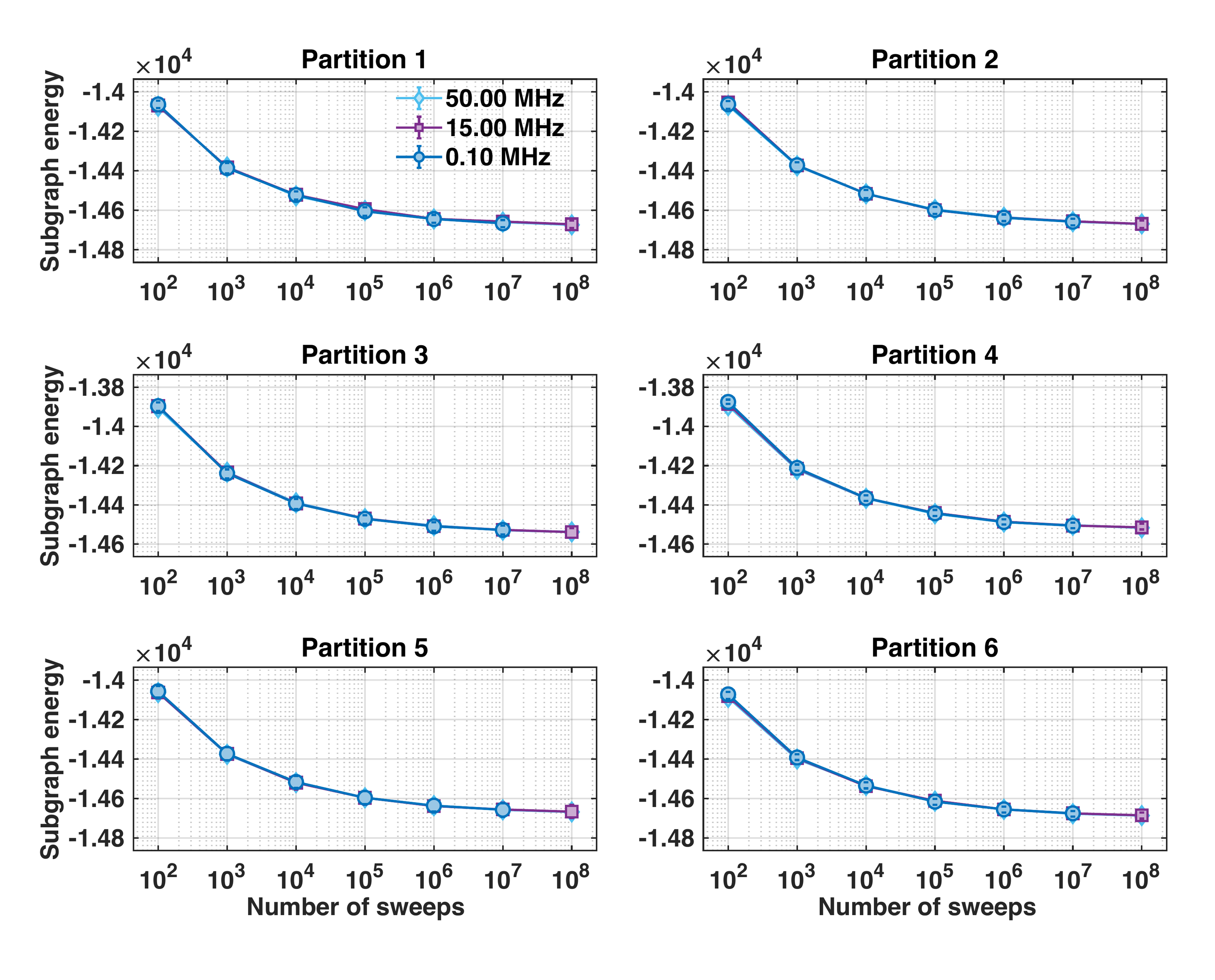}
  \caption{
  \textbf{Local-subgraph energies remain stable when boundary links are disconnected (DSIM-1).}
  Subgraph energy versus sweeps for each of the six partitions of the \(L^3 = 37^3\) EA spin glass on DSIM-1, using the same update logic, annealing schedule (\(\beta = 0.5, 1.0, \ldots, 5.0\)), and fixed-point format s\(\{4\}\{1\}\) as the coupled DSIM-1 runs, but with boundary links disconnected.
  Three local clocks are shown: \(0.10\)~MHz, \(15\)~MHz, and \(50\)~MHz.
  The trajectories overlap for each partition over the full sweep range, supporting the interpretation that slope loss in coupled overclocked operation originates from boundary exchange, not from local update failure.
  All points: 10 instances, 10 runs per instance, error bars are 95\% bootstrap confidence intervals.}
  \label{fig:supp_subgraph_energy}
\end{figure}

A key question raised by the overclocking results (Fig.~\ref{fig:extreme_fig4} of the main text) is whether the modest slope loss originates from local timing violations at high clock rates or specifically from stale boundary information.
To isolate the origin, we run DSIM-1 with boundary links disconnected.
In this mode, each FPGA evolves its own subgraph under the same simulated annealing schedule ($\beta = 0.5, 1.0, \ldots, 5.0$) and fixed-point format s$\{4\}\{1\}$ (signed, 4 integer bits, 1 fractional bit)~\cite{aadit2022massively}, but without receiving any neighbor boundary states.
If local updates were failing at high clock, this would appear as a degradation in local subgraph energy traces even in the disconnected setting.
Fig.~\ref{fig:supp_subgraph_energy} shows that the per-subgraph energy traces at $0.10$~MHz, $15$~MHz, and $50$~MHz remain consistent for each of the six partitions.
The overlap across frequencies indicates that local updates produce correct results at all tested clocks, and therefore the slope loss observed in coupled runs is attributable to boundary exchange dynamics, not to local update errors.

\FloatBarrier
\subsection{Quasi-2D spin glass on DSIM-1}
\label{sec:quasi2d_app}

The main text establishes the $\eta$ dependence on the 3D Edwards--Anderson (EA) lattice. We asked whether it holds on a graph with a different dimension and connectivity. We chose quasi-2D $\pm1$ spin glasses with $25{,}088$ spins on the Pegasus graph P33, the native topology of current quantum annealers~\cite{King2023quantum}. Each spin has up to $15$ neighbors, against six on the cubic lattice. KaHIP splits the graph into six parts of $4{,}180$ to $4{,}182$ p-bits, one per FPGA of DSIM-1, and the graph is colored with \(N_{\mathrm{color}} = 4\). We swept $\eta$ from $10^{-4}$ to $10^{3}$, with 10 instances and 10 runs per point.

The annealing schedule is linear, with $\beta$ taking 10 equally spaced values from $0$ to $15.5$, each held for an equal share of the sweep budget. The reference energy $E_{\mathrm{ref}}$ is the best-known energy of each instance from long ($10^7$-sweep) GPU anneals, not a putative ground energy. We compare against two GPU references on the unpartitioned graph. Both run the same p-bit as the hardware, with its linear-feedback shift register (LFSR) random-number generator and tanh lookup table. The exact reference keeps every coupling ($\eta = \infty$). The no-communication reference drops every coupling between the six partitions during the updates ($\eta = 0$).

\begin{figure}[htbp]
  \centering
  \includegraphics[width=\textwidth]{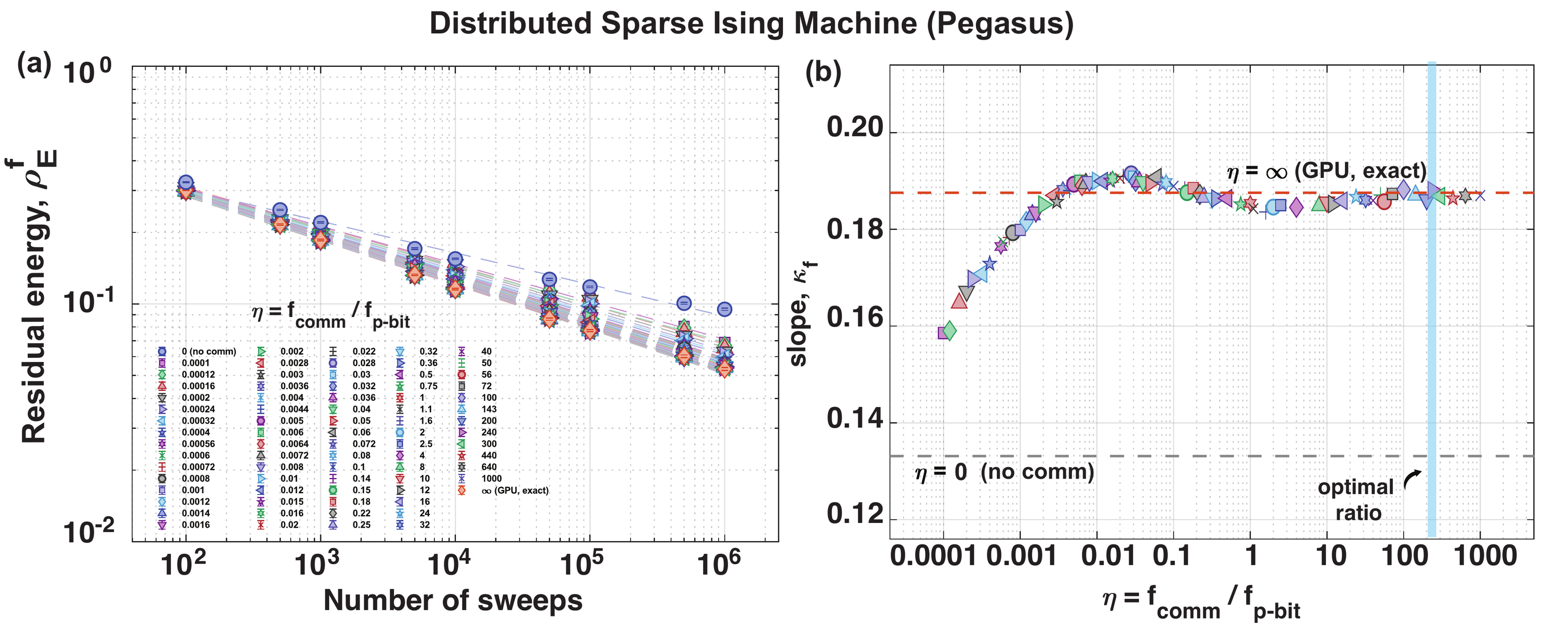}
  \caption{\textbf{Effect of the communication-to-p-bit frequency ratio $\eta$ on annealing a 25{,}088-spin quasi-2D spin glass (Pegasus graph P33) over the six FPGAs of DSIM-1.}
  \textbf{(a)}~Residual energy per spin, $\rho_E^f = (E_f - E_{\mathrm{ref}})/N$, versus the number of sweeps for every measured $\eta = f_{\mathrm{comm}}/f_{\mathrm{p\text{-}bit}}$ (colored symbols). $E_f$ is the energy of the final state of each anneal, and $E_{\mathrm{ref}}$ is the best-known energy of each instance, obtained from long ($10^7$-sweep) GPU anneals. All runs use the same annealing schedule ($\beta$ takes 10 equally spaced values from $0$ to $15.5$, each held for an equal share of the sweep budget). Each point is the mean over 10 instances $\times$ 10 independent runs, and error bars are 95\% confidence intervals of the mean. Dashed lines are power-law fits, $\rho_E^f \propto (\text{sweeps})^{-\kappa_f}$. The two GPU references run the same p-bit (identical LFSR random-number generator and tanh lookup table) on the unpartitioned graph: the exact reference keeps every coupling ($\eta = \infty$), and the no-communication reference drops every coupling between the six partitions during the updates ($\eta = 0$).
  \textbf{(b)}~Fitted exponent $\kappa_f$ versus $\eta$. Dashed lines mark the exact ($\eta = \infty$, red) and no-communication ($\eta = 0$, grey) exponents. At slow communication ($\eta \lesssim 10^{-3}$), $\kappa_f$ falls toward the disconnected limit. It rises with $\eta$ and saturates near the exact value from $\eta \approx 5 \times 10^{-3}$ on: once boundary information is exchanged often enough, the distributed machine anneals with the exponent of the unpartitioned system. The blue band marks the ratio given by Eq.~\eqref{eq:fpbit_bound_appendix} for this partition, $\eta_{\mathrm{threshold}} \approx 240$.}
  \label{fig:supp_quasi2d}
\end{figure}

The quasi-2D spin glass behaves like the EA lattice (Fig.~\ref{fig:supp_quasi2d}). The exponent is reduced at the disconnected limit, rises with $\eta$, and saturates near the unpartitioned GPU value from $\eta \approx 5 \times 10^{-3}$ on. The optimal ratio follows from the partition alone, without the measured curves. On the DSIM-1 chain the links have \(P = [54,\, 30,\, 54,\, 26,\, 54]\) pins. The worst-case cluster pair is \((3,\,6)\), with \(b_{36} = 258\) boundary states over \(d_{36} = 3\) hops, through a bottleneck of \(P_{36} = \min(54,\,26,\,54) = 26\) pins. This gives
\[
  C_{\max} = 258 \times \frac{3}{26} \approx 29.8,
  \qquad
  \eta_{\mathrm{threshold}} = 2 \times 4 \times 29.8 \approx 240
\]
(blue band in Fig.~\ref{fig:supp_quasi2d}b). The exponent saturates well below this ratio, as expected for a sufficient condition.

\begin{figure}[htbp]
  \centering
  \includegraphics[width=\textwidth]{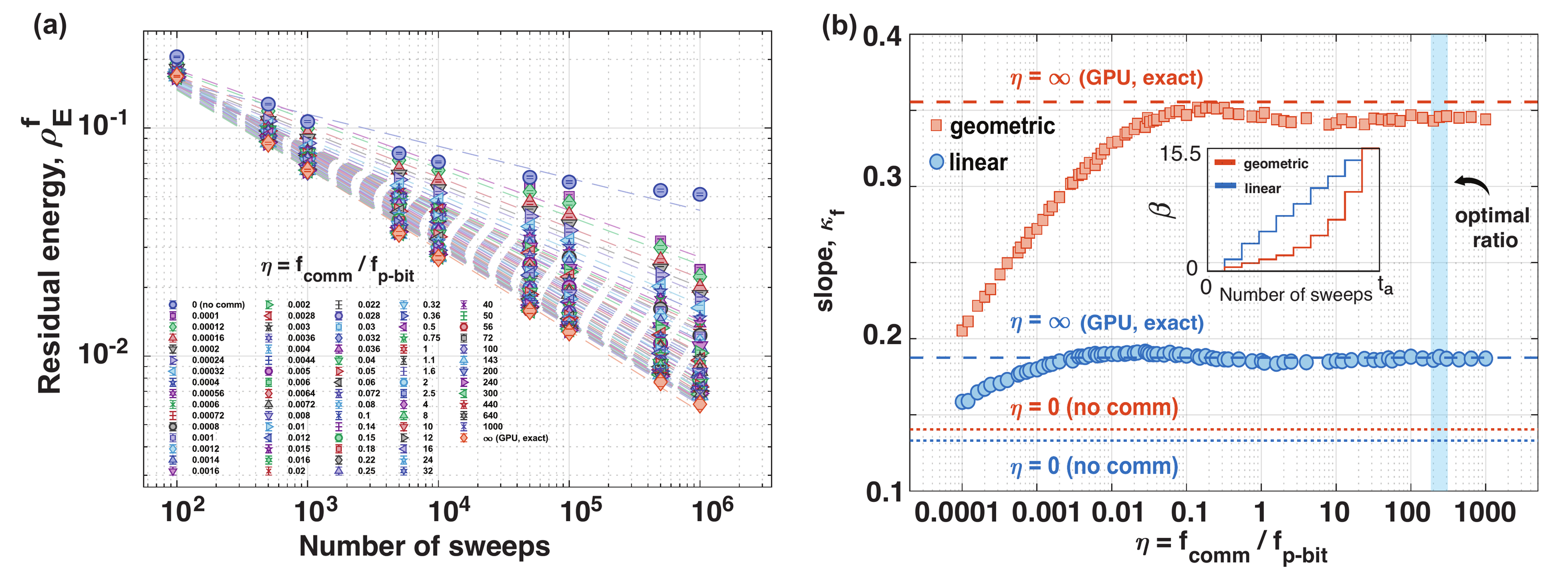}
  \caption{\textbf{Robustness of the $\eta$ dependence to the annealing schedule, 25{,}088-spin quasi-2D spin glass on DSIM-1.}
  \textbf{(a)}~Residual energy per spin versus the number of sweeps for every measured $\eta$ with a geometric annealing schedule, on the same instances, partition and GPU references as Fig.~\ref{fig:supp_quasi2d}, which uses a linear schedule. Each point is the mean over 10 instances $\times$ 10 runs, and error bars are 95\% confidence intervals of the mean. Dashed lines are power-law fits.
  \textbf{(b)}~Fitted exponent $\kappa_f$ versus $\eta$ for the geometric (orange squares) and linear (blue circles, from Fig.~\ref{fig:supp_quasi2d}) schedules, with the GPU references of each schedule in its color, dashed for the unpartitioned limit, $\eta = \infty$, and dotted for no communication, $\eta = 0$. Inset: the two schedules, each holding 10 $\beta$ values for equal shares of the sweep budget $t_a$. At the s$\{4\}\{1\}$ weight precision of the FPGAs they are $\beta = 0, 1.5, 3.5, 5, 7, 8.5, 10.5, 12, 14, 15.5$ (linear) and $\beta = 0, 0.5, 1, 1.5, 2, 3, 4.5, 6.5, 10, 15.5$ (geometric). The absolute exponent depends on the schedule, but the dependence on $\eta$ has the same form: $\kappa_f$ rises from near the no-communication limit and saturates near the unpartitioned value, from $\eta \approx 5 \times 10^{-3}$ (linear) and $\eta \approx 0.05$ to $0.1$ (geometric) on. The blue band marks $\eta_{\mathrm{threshold}} \approx 240$ from Eq.~\eqref{eq:fpbit_bound_appendix}.}
  \label{fig:supp_quasi2d_schedule}
\end{figure}

To check that this behavior does not depend on the schedule, we annealed the same instances over the same range of $\eta$ with a geometric schedule, 10 values of $\beta$ spaced geometrically up to $15.5$, each held for an equal share of the sweep budget (Fig.~\ref{fig:supp_quasi2d_schedule}, both schedules in the inset). The two schedules give different unpartitioned exponents. Their dependence on $\eta$ is the same. With both, $\kappa_f$ falls toward its disconnected value at small $\eta$, rises with $\eta$, and saturates near the unpartitioned value of that schedule, well below the optimal ratio (Fig.~\ref{fig:supp_quasi2d_schedule}b).

Two settings with the same $\kappa_f$ can still give different residual energies. Writing the measured trajectories as $\rho_E^f(t_a,\eta) = A(\eta)\, t_a^{-\kappa_f(\eta)}$, they can differ in the prefactor $A(\eta)$, which depends on how early in each anneal the boundary information becomes current. The residual energy at a fixed budget can therefore keep improving with $\eta$ after the slope has saturated.

\FloatBarrier
\subsection{Quasi-2D spin glass at $251{,}328$ p-bits on DSIM-2}
\label{sec:p103_app}

We repeated the quasi-2D test on DSIM-2 at 10 times the size, on the Pegasus graph P103~\cite{King2023quantum}. The instances have $251{,}328$ spins and up to $15$ neighbors per spin, against six on the $L^3 = 100^3$ lattice run on the same machine. They are generated as for the EA lattices. Every coupling is $+1$ or $-1$ with equal probability, and there is no field. METIS splits each graph into $72$ partitions over the $18$ FPGAs (Sec.~\ref{sec:slr_partitioning_app}), and the graph is colored with $N_{\mathrm{color}} = 4$. The hardware uses the fixed-point format s$\{4\}\{1\}$ (main text Methods). With more neighbors, a boundary p-bit can have several of them on other partitions, so more of its input comes from other FPGAs.

For this partition Eq.~\eqref{eq:fpbit_bound_appendix} gives $f_{\mathrm{p\text{-}bit,max}} \approx 0.63$~MHz (Sec.~\ref{sec:comm_cost_app}). We ran at the conservative clock of $0.47$~MHz, inside the bound, where the boundary information is current at every update. We then overclocked to $0.75$ and $1.25$~MHz, above the bound, where this is no longer guaranteed. The annealing schedule steps through $\beta = 0.5, 1.0, \ldots, 15.5$, with an equal share of the sweep budget at each value. The GPU reference runs the same instances and schedule on the unpartitioned graph with the same p-bit as the hardware, including its LFSR random-number generator and tanh lookup table (main text Methods). For each instance, $E_{\mathrm{ref}}$ is the best energy found in all DSIM-2 and GPU runs, including runs of $10^7$ sweeps that are not in the fits.

\begin{figure}[htbp]
  \centering
  \includegraphics[width=0.6\textwidth]{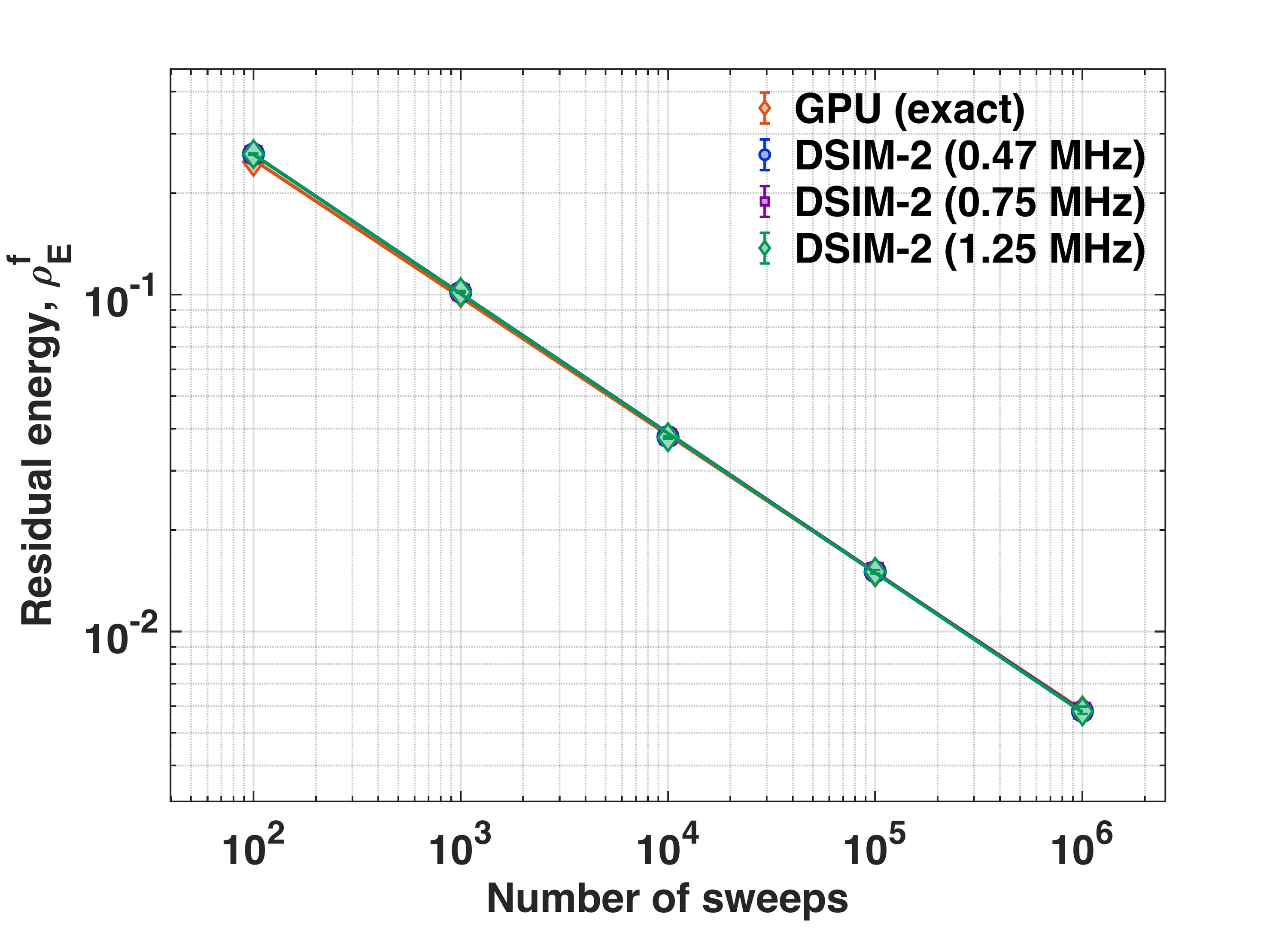}
  \caption{\textbf{Residual energy versus sweeps for a $251{,}328$-spin quasi-2D spin glass on DSIM-2.}
  $\pm1$ spin glass on the Pegasus graph P103, partitioned into $72$ parts over the $18$ FPGAs of DSIM-2 and annealed at the conservative clock $f_{\mathrm{p\text{-}bit}} = 0.47$~MHz ($\kappa_f \approx 0.414$) and overclocked to $0.75$ and $1.25$~MHz ($\kappa_f \approx 0.413$ and $0.414$), against the unpartitioned GPU (exact, $\kappa_f \approx 0.408$) on the same instances and schedule ($\beta = 0.5, 1.0, \ldots, 15.5$).
  Lines are power-law fits, $\rho_E^f \propto (\text{sweeps})^{-\kappa_f}$.
  The GPU reference runs the same p-bit as the hardware (main text Methods).
  ``Exact'' denotes boundary information that is current at every update, as defined in main text Fig.~\ref{fig:extreme_fig4}.
  Each point is the mean over 10 instances $\times$ 10 runs, and error bars are 95\% bootstrap confidence intervals.}
  \label{fig:supp_p103}
\end{figure}

At $0.47$~MHz the distributed machine and the GPU coincide over the four measured decades, with $\kappa_f \approx 0.414$ and $0.408$ (Fig.~\ref{fig:supp_p103}). Here the random numbers and the arithmetic are matched, and the gap seen against the floating-point Philox references of main text Figs.~\ref{fig:extreme_fig4} and~\ref{fig:extreme_fig5} disappears. That gap therefore comes from the platforms and not from partitioning.

At $0.75$ and $1.25$~MHz the residual energy and the exponent stay the same within the error bars, with $\kappa_f \approx 0.413$ and $0.414$. These clocks run at $\eta \approx 435$ and $260$, the second at half the optimal ratio of $517$. DSIM-1 behaves the same way, with an exponent that saturates below the optimal ratio (Sec.~\ref{sec:quasi2d_app}). Across the three clocks the flip rate grows by a factor of $2.7$ at the same exponent, from $1.2 \times 10^{11}$ to $3.1 \times 10^{11}$ flips per second.

\begin{figure}[htbp]
  \centering
  \includegraphics[width=0.95\textwidth]{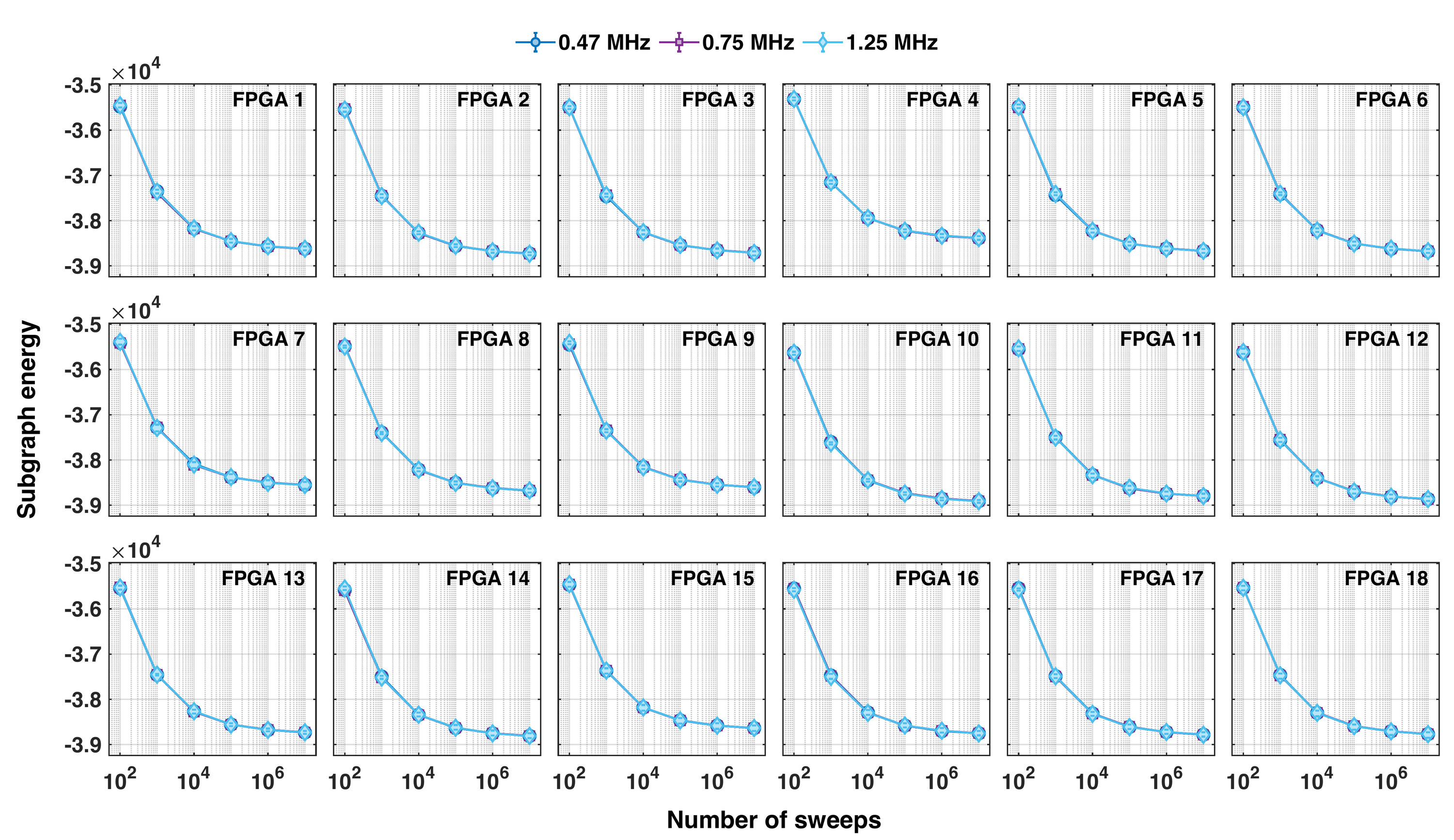}
  \caption{\textbf{Local-subgraph energies with boundary links disconnected (DSIM-2).}
  Subgraph energy of each of the $18$ FPGAs versus sweeps for the P103 design of Fig.~\ref{fig:supp_p103}, with boundary links disconnected. The subgraph energy of an FPGA is the energy of the couplings inside the four partitions it holds, with every coupling to another partition removed. The runs are at the conservative clock $f_{\mathrm{p\text{-}bit}} = 0.47$~MHz and at $0.75$ and $1.25$~MHz.
  Instances, annealing schedule and fixed-point format are those of Fig.~\ref{fig:supp_p103}.
  Each point is the mean over 10 instances $\times$ 10 runs, and error bars are 95\% bootstrap confidence intervals.}
  \label{fig:supp_p103_cut}
\end{figure}

To check the local updates, we ran the disconnected-links control of Sec.~\ref{sec:subgraph_energy_app} at the same three clocks. Each of the $72$ partitions then anneals only its own subgraph. On every one of the $18$ FPGAs the subgraph energy follows the same curve at all three clocks (Fig.~\ref{fig:supp_p103_cut}). The local-subgraph energy curves remain unchanged up to $1.25$~MHz, even though each update sums up to $15$ inputs, compared with six on the lattice. Beyond $1.25$~MHz, read/write synchronization across the $18$ FPGAs fails from accumulated timing violations, as it does beyond $3$~MHz for the $L^3 = 100^3$ design (main text). At the clock set by Eq.~\eqref{eq:fpbit_bound_appendix}, the distributed machine thus anneals a quarter-million-spin problem with the exponent of the unpartitioned system, as the $1$~MHz point does for the $L^3 = 100^3$ lattice (main text Fig.~\ref{fig:extreme_fig5}).

DSIM-2 derives its p-bit clock from the single global clock of the platform, so $\eta$ cannot be tuned on its own as it can on DSIM-1 through $f_{\mathrm{comm}}$. Overclocking the whole machine gives a few operating points, not a sweep. We therefore establish the $\eta$ dependence on DSIM-1, whose independent local clocks set $\eta$ over seven orders of magnitude, and use DSIM-2 to test the large machine at selected points. This is also why the main text concludes that independent local clocks let $\eta$ be set freely on one machine.

\FloatBarrier
\subsection{Sampling at finite $\eta$}
\label{sec:sampling_app}

Main text Fig.~\ref{fig:extreme_fig6_sampling} follows the magnetization of a quasi-2D ferromagnet at fixed temperature as $\eta$ changes (protocol in main text Methods). At the slowest refresh rates, several relaxation curves in main text Fig.~\ref{fig:extreme_fig6_sampling}a level off for a while, with the magnetization near $\langle |M| \rangle \approx 0.2$, before it rises again. This is the plateau expected when each partition has ordered locally but the partitions are not yet aligned with one another.

The vertical band in main text Fig.~\ref{fig:extreme_fig6_sampling}b is the optimal ratio of Eq.~\eqref{eq:fpbit_bound_appendix} for this partition. We evaluate it as for the quasi-2D spin glass (Section~\ref{sec:quasi2d_app}), with \(N_{\mathrm{color}} = 4\) and the same DSIM-1 pin counts. The worst-case cluster pair is again \((3,\,6)\), now with \(b_{36} = 134\) boundary states over \(d_{36} = 3\) hops through \(P_{36} = 26\) pins, so
\[
  C_{\max} = 134 \times \frac{3}{26} \approx 15.5,
  \qquad
  \eta_{\mathrm{threshold}} = 2 \times 4 \times 15.5 \approx 124.
\]
The measurements establish agreement of the mean absolute magnetization with the unpartitioned GPU reference and an average decay slope between two fixed levels. They extend the observed communication dependence from spin-glass optimization to the relaxation of a fixed-temperature sampling observable. Table~\ref{tab:opt_vs_sampling} sets this sampling measurement side by side with the optimization measurement of the main text.

\begin{table}[htbp]
\centering
\caption{\textbf{Optimization and sampling measurements side by side.} The sampling experiment of main text Fig.~\ref{fig:extreme_fig6_sampling} follows the residual-energy analysis of main text Figs.~\ref{fig:extreme_fig3} to~\ref{fig:extreme_fig5} step by step.}
\label{tab:opt_vs_sampling}
\small
{\renewcommand{\arraystretch}{1.2}
\begin{tabular}{@{}lll@{}}
\toprule
\parbox[t]{0.19\textwidth}{\raggedright \strut} & \parbox[t]{0.36\textwidth}{\raggedright \textbf{Optimization (main text Figs.~\ref{fig:extreme_fig3} to~\ref{fig:extreme_fig5})}\strut} & \parbox[t]{0.38\textwidth}{\raggedright \textbf{Sampling (main text Fig.~\ref{fig:extreme_fig6_sampling})}\strut} \\
\midrule
\parbox[t]{0.19\textwidth}{\raggedright System\strut} & \parbox[t]{0.36\textwidth}{\raggedright 3D EA spin glass, $J_{ij} = \pm 1$\strut} & \parbox[t]{0.38\textwidth}{\raggedright Pegasus ferromagnet, $J_{ij} = +1$, zero field\strut} \\
\parbox[t]{0.19\textwidth}{\raggedright Temperature\strut} & \parbox[t]{0.36\textwidth}{\raggedright annealed, $\beta$ from $0.5$ to $5$\strut} & \parbox[t]{0.38\textwidth}{\raggedright fixed, $\beta = 0.09375$ (heat-capacity peak)\strut} \\
\parbox[t]{0.19\textwidth}{\raggedright Reference\strut} & \parbox[t]{0.36\textwidth}{\raggedright putative ground energy\strut} & \parbox[t]{0.38\textwidth}{\raggedright equilibrium magnetization $M_{\rm eq}$ of the unpartitioned sampler\strut} \\
\addlinespace[3pt]
\parbox[t]{0.19\textwidth}{\raggedright Quantity that falls\strut} & \parbox[t]{0.36\textwidth}{\raggedright residual energy per spin, $\rho_E^f(t)$\strut} & \parbox[t]{0.38\textwidth}{\raggedright gap $M_{\rm eq} - \langle |M(t)| \rangle$\strut} \\
\parbox[t]{0.19\textwidth}{\raggedright Plot at each $\eta$\strut} & \parbox[t]{0.36\textwidth}{\raggedright $\rho_E^f$ versus sweeps, both axes logarithmic\strut} & \parbox[t]{0.38\textwidth}{\raggedright gap versus sweeps, logarithmic sweep axis\strut} \\
\parbox[t]{0.19\textwidth}{\raggedright Decay slope\strut} & \parbox[t]{0.36\textwidth}{\raggedright power-law exponent $\kappa_f$, $\rho_E^f \propto t^{-\kappa_f}$\strut} & \parbox[t]{0.38\textwidth}{\raggedright average slope from $50\%$ to $90\%$, $\kappa_{\rm eff} = \ln 5/\ln(t_{90}/t_{50})$\strut} \\
\addlinespace[3pt]
\parbox[t]{0.19\textwidth}{\raggedright Comparison\strut} & \parbox[t]{0.36\textwidth}{\raggedright $\kappa_f$ versus $\eta$ (main text Fig.~\ref{fig:extreme_fig3}b)\strut} & \parbox[t]{0.38\textwidth}{\raggedright $\kappa_{\rm eff}$ versus $\eta$ (main text Fig.~\ref{fig:extreme_fig6_sampling}b)\strut} \\
\parbox[t]{0.19\textwidth}{\raggedright Disconnected limit ($\eta = 0$)\strut} & \parbox[t]{0.36\textwidth}{\raggedright reduced exponent\strut} & \parbox[t]{0.38\textwidth}{\raggedright partitions keep their initial signs, $\langle |M| \rangle < 0.07$\strut} \\
\parbox[t]{0.19\textwidth}{\raggedright Fast communication\strut} & \parbox[t]{0.36\textwidth}{\raggedright $\kappa_f$ reaches the unpartitioned GPU value\strut} & \parbox[t]{0.38\textwidth}{\raggedright $\kappa_{\rm eff}$ approaches the unpartitioned GPU value\strut} \\
\bottomrule
\end{tabular}}
\end{table}

\FloatBarrier
\subsection{GPU time to target}
\label{sec:gpu_time_app}

The GPU curves of main text Figs.~\ref{fig:extreme_fig4}a and~\ref{fig:extreme_fig5}a are the unpartitioned reference for the decay exponent, and here we also report their runtime. The GPU is an NVIDIA RTX 6000 Ada running the same instances and annealing schedule on one device, with no partitioning. A sweep takes $9.39$~$\mu$s at $L^3 = 37^3$ and $20$~$\mu$s at $L^3 = 100^3$, that is, $5.4 \times 10^{9}$ and $5.0 \times 10^{10}$ flips per second. Our GPU code runs a single anneal of one instance at $20$~ps per flip at one million spins. Codes written for the cubic lattice with $\pm 1$ couplings use multi-spin coding and report $3$ to $5$~ps per flip on a GTX Titan~\cite{lulli2015highly}, and less on newer hardware~\cite{bernaschi2026microcanonical}. This technique relies on the binary couplings and the regular lattice of that one model, whereas our GPU code and our machine are both built to run any sparse graph with multi-bit weights, and we therefore regard our GPU code as the apples-to-apples baseline.

Our probabilistic processing units are realized through FPGA logic and programmable routing, while the GPU benefits from a highly optimized silicon implementation. The resulting speed and energy ratios characterize these implementations. They cannot determine the efficiency limits of a dedicated probabilistic computer.

\begin{figure*}[t]
  \centering
  \includegraphics[width=\textwidth]{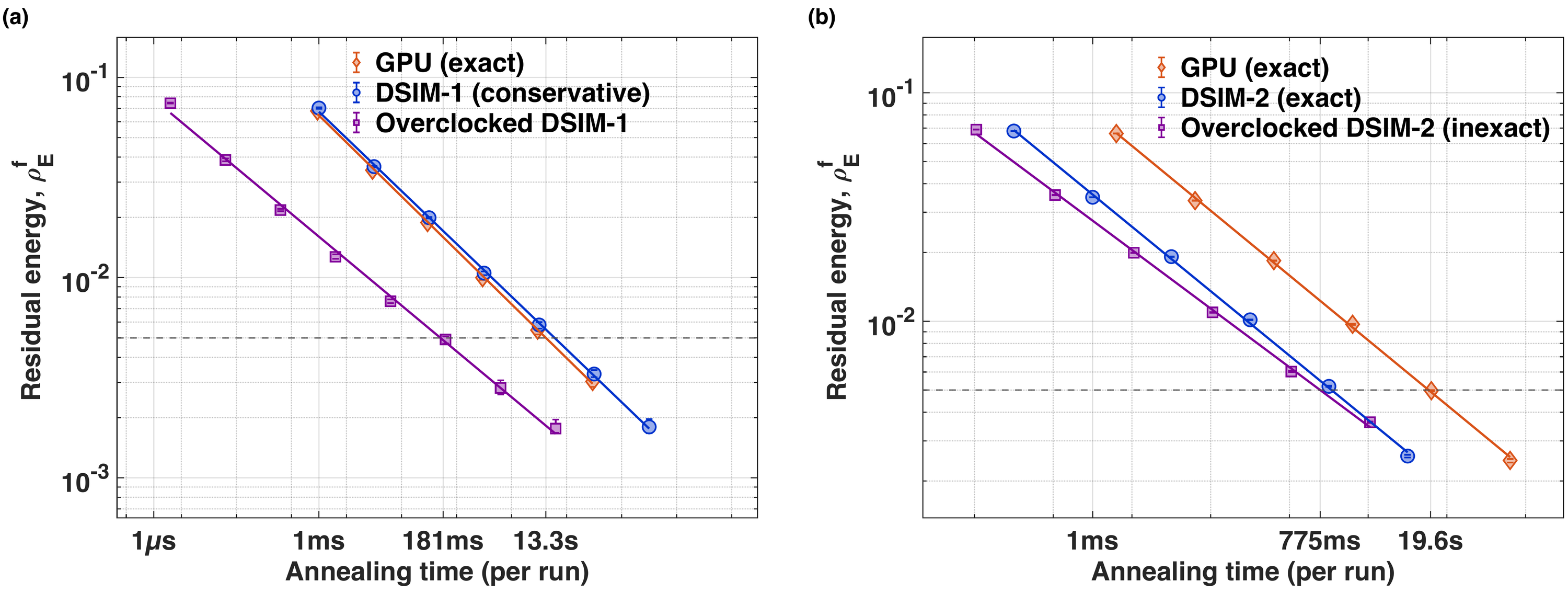}
  \caption{\textbf{Residual energy versus annealing time, each DSIM against the GPU on its own instances.} The two panels are separate experiments at different problem sizes. \textbf{(a)}~$L^3 = 37^3$ on DSIM-1, at $f_{\mathrm{p\text{-}bit}} = 0.10$~MHz (conservative) and $50$~MHz (overclocked), and on the GPU ($9.39$~$\mu$s per sweep), same instances, schedule and putative ground energies as main text Fig.~\ref{fig:extreme_fig4}a. \textbf{(b)}~$L^3 = 100^3$ on DSIM-2, at $1$~MHz (conservative, exact) and $3$~MHz (overclocked, inexact), and on the GPU ($20$~$\mu$s per sweep), same instances, schedule and putative ground energies as main text Fig.~\ref{fig:extreme_fig5}a. The dashed line marks the target $\rho_E^\star = 0.005$ of Table~\ref{tab:gpu_time_supp}, reached where each curve crosses it. The labeled time ticks give the time to target of the overclocked DSIM and of the GPU, and the remaining ticks are reference marks. The time axis is the annealing time per run, sweeps times the measured time per sweep, with fixed per-run costs excluded on all platforms. Error bars: 95\% bootstrap confidence intervals over 10 instances $\times$ 10 runs. ``Exact'' and ``inexact'' denote fresh and stale boundary information, as defined in main text Fig.~\ref{fig:extreme_fig4}.}
  \label{fig:gpu_time_supp}
\end{figure*}

Fig.~\ref{fig:gpu_time_supp} puts the GPU and DSIM curves on the same annealing-time axis. Table~\ref{tab:gpu_time_supp} gives the time and energy to reach $\rho_E^\star = 0.005$ on each platform, read from the measured points, with the projected $7$~nm ASIC of main text Fig.~\ref{fig:extreme_fig4}b. At $L^3 = 37^3$, the conservative DSIM-1 ($0.10$~MHz) runs at about the GPU flip rate with the same exponent, and reaches the target in a comparable time. Its clock is set through Eq.~\eqref{eq:fpbit_bound_appendix} by the bandwidth of the links in the six-board chain, not by the p-bit logic, which runs at $50$~MHz. Overclocked to $50$~MHz, DSIM-1 reaches the target $73\times$ faster than the GPU, despite its reduced exponent. DSIM-2 has faster $80$-pin links, which allow a 10 times higher conservative clock on a problem 20 times larger. At one million p-bits it reaches the target $17\times$ faster than the GPU at $1$~MHz, with the same exponent, and $25\times$ faster at $3$~MHz.

The gap grows with problem size at a fixed p-bit clock, because the DSIM time per sweep is set by $f_{\mathrm{p\text{-}bit}}$ alone while the GPU time per sweep grows with the number of spins. 

These times leave out fixed per-run costs on all platforms. These are the one-time DSIM weight load, $0.5$~s and $2$~s at the two sizes, and the GPU data transfer and setup, $0.23$~s and $1.2$~s per run. With these costs included in full, for a single run on a freshly loaded instance, the ratios at $\rho_E^\star = 0.005$ are $20\times$ for DSIM-1 at $50$~MHz and $6.6\times$ and $7.5\times$ for DSIM-2 at $1$ and $3$~MHz.

Energy to target is power times time to target, with the power of each platform listed in Table~\ref{tab:gpu_time_supp}. Overclocked DSIM-1 reaches the target with under 2\% of the GPU energy, and DSIM-2 with $3.6$ to $5.1$ times less energy than the GPU at one million p-bits. Most of the FPGA power is the static power of these large devices, about $49$~W of the $80$~W of each DSIM-2 FPGA, not the power of the p-bits. The projected ASIC row shows how much further dedicated chips would reduce the energy.

\begin{table*}[t]
\centering
\caption{\textbf{Time and energy to target, each DSIM against the GPU on its own instances and schedule.} The target is $\rho_E^\star = 0.005$. Times are annealing times per run, interpolated in log-log between the measured points of Fig.~\ref{fig:gpu_time_supp}. Speedup and energy saving are the GPU time and energy divided by those of the DSIM at the same size, and energy is power times time to target. GPU power is the mean board power recorded while the GPU anneals. FPGA power is the on-chip power from the implementation reports of each design. The projected ASIC row is read from the projected line of main text Fig.~\ref{fig:extreme_fig4}b, the conservative DSIM-1 power law at $10$~ns per sweep, with the power of six $7$~nm partitions and of die-to-die links at the $0.5$~pJ/bit UCIe target for standard packages (Sec.~\ref{sec:ucie_feasibility_app}).}
\label{tab:gpu_time_supp}
{\small\renewcommand{\arraystretch}{1.15}\setlength{\tabcolsep}{5pt}
\begin{tabular}{@{}lrrrrrrr@{}}
\toprule
Platform & \begin{tabular}[b]{@{}c@{}}Time per\\sweep\end{tabular} & \begin{tabular}[b]{@{}c@{}}Flips per\\second\end{tabular} & \begin{tabular}[b]{@{}c@{}}Time to\\target\end{tabular} & \begin{tabular}[b]{@{}c@{}}Speedup\\vs GPU\end{tabular} & \begin{tabular}[b]{@{}c@{}}Power\\(W)\end{tabular} & \begin{tabular}[b]{@{}c@{}}Energy to\\target\end{tabular} & \begin{tabular}[b]{@{}c@{}}Energy\\saving\end{tabular} \\
\midrule
\multicolumn{8}{l}{$L^3 = 37^3$ ($N = 50{,}653$)} \\
GPU & 9.39~$\mu$s & $5.4 \times 10^{9}$ & 13.3~s & 1$\times$ & 148 & 2.0~kJ & 1$\times$ \\
DSIM-1, 0.10~MHz & 10~$\mu$s & $5.1 \times 10^{9}$ & 18.3~s & 0.7$\times$ & 151 & 2.8~kJ & 0.7$\times$ \\
DSIM-1, 50~MHz & 20~ns & $2.5 \times 10^{12}$ & 0.18~s & 73$\times$ & 170 & 31~J & 64$\times$ \\
Projected 7~nm ASIC, 100~MHz & 10~ns & $5.1 \times 10^{12}$ & 15.1~ms & 880$\times$ & 5.5 & 0.083~J & 24{,}000$\times$ \\
\midrule
\multicolumn{8}{l}{$L^3 = 100^3$ ($N = 10^6$)} \\
GPU & 20~$\mu$s & $5.0 \times 10^{10}$ & 19.6~s & 1$\times$ & 297 & 5.8~kJ & 1$\times$ \\
DSIM-2, 1~MHz & 1~$\mu$s & $1.0 \times 10^{12}$ & 1.13~s & 17$\times$ & 1440 & 1.6~kJ & 3.6$\times$ \\
DSIM-2, 3~MHz & 0.33~$\mu$s & $3.0 \times 10^{12}$ & 0.78~s & 25$\times$ & 1450 & 1.1~kJ & 5.1$\times$ \\
\bottomrule
\end{tabular}}
\end{table*}

\FloatBarrier
\subsection{UCIe feasibility and 7~nm ASIC metrics}
\label{sec:ucie_feasibility_app}

This section translates the conservative bound into a required link-rate envelope using Universal Chiplet Interconnect Express (UCIe) as a concrete short-reach die-to-die example~\cite{ucie_spec}, complementing the multi-chip scaling analysis of Ref.~\cite{srimani2024next}.
Similar analysis applies to other die-to-die interconnects such as Bunch of Wires (BoW)~\cite{ardalan2020bunch}, but the point here is to connect the DSIM timing rule to physical interconnect sizing. The same separation of local computation and boundary exchange
could guide implementations using three-dimensional
integration~\cite{srimani2024next,choi2025foundry}
or nanosecond, sub-femtojoule stochastic magnetic tunnel
junctions~\cite{kaiser2019subnanosecond,singh2024cmos}.
For these devices, the $\eta$ framework relates the local
fluctuation rate to the communication bandwidth required
between partitions.

\subsubsection{Bandwidth requirement expressed as a communication clock}

Starting from Eq.~\eqref{eq:fpbit_bound_appendix}, a conservative requirement for supporting a chosen \(f_{\mathrm{p\text{-}bit}}\) is
\begin{equation}
f_{\mathrm{comm}} \ge 2\,N_{\mathrm{color}}\,C_{\max}\,f_{\mathrm{p\text{-}bit}}.
\end{equation}
To illustrate the scale of the required communication clock, we plug in the values from the worst-case pair of the \(L^3=37^3\) DSIM-1 partition (Section~\ref{sec:comm_cost_app}): \(N_{\mathrm{color}}=3\), boundary size \(b_{ab}=660\), hop distance \(d_{ab}=2\), giving \(C_{\max} = b_{ab}d_{ab}/P_{ab}\). Substituting into the bound yields
\begin{equation}
f_{\mathrm{comm}} = \frac{2\,N_{\mathrm{color}}\,b_{ab}\,d_{ab}}{P_{ab}}\,f_{\mathrm{p\text{-}bit}}.
\end{equation}
For \(f_{\mathrm{p\text{-}bit}}=100\) MHz this becomes
\begin{equation}
f_{\mathrm{comm}} = \frac{792{,}000}{P_{ab}}~\text{MHz}.
\end{equation}
Evaluating at two lane budgets within the UCIe range~\cite{ucie_spec}:
for \(P_{ab}=64\), \(f_{\mathrm{comm}} \approx 12.4\) GHz, and for \(P_{ab}=128\), \(f_{\mathrm{comm}} \approx 6.2\) GHz.
Both rates fall within the standard UCIe operating range, confirming that the conservative bound of Eq.~\eqref{eq:fpbit_bound_appendix} can be satisfied with existing die-to-die interconnect technology. For the one-million-p-bit design of main text Fig.~\ref{fig:extreme_fig5}b, with the partition and mapping of DSIM-2 (18 devices of about \(55{,}600\) p-bits, \(b_{ab} = 3{,}607\), \(d_{ab} = 3\), \(N_{\mathrm{color}} = 2\)), the same bound at \(100\)~MHz requires \(4.33 \times 10^{12}\)~bit/s on the worst-case pair, or \(0.78\) bit per flip of one device.

\subsubsection{Interface power at the required link rate}

We estimate here the interface power of the projected ASIC of the \(L^3=37^3\) system, with \(50{,}653\) p-bits in six \(7\)~nm partitions at \(f_{\mathrm{p\text{-}bit}}=100\)~MHz. Eq.~\eqref{eq:fpbit_bound_appendix} sets the bit rate each link must carry. For the worst-case pair this is \(P_{ab}\,f_{\mathrm{comm}} = 2\,N_{\mathrm{color}}\,b_{ab}\,d_{ab}\,f_{\mathrm{p\text{-}bit}} = 7.92\times 10^{11}\)~bit/s, or \(0.94\) bit per flip of one device. The six partitions form a chain of five links, as in DSIM-1. We size every link for this worst-case rate in both directions and keep it active at all times. The UCIe specification targets \(0.25\)~pJ/bit for advanced packages and \(0.5\)~pJ/bit for standard packages~\cite{ucie_spec}. The interface power is then \(10 \times 7.92\times 10^{11}\)~bit/s times the energy per bit, \(2.0\)~W at \(0.25\)~pJ/bit and \(4.0\)~W at \(0.5\)~pJ/bit. The local logic of the six partitions takes \(1.49\)~W (\(6 \times 248\)~mW, Table~\ref{tab:supp_asic_summary}). With \(50{,}653\) p-bits updated at \(100\)~MHz, or \(5.07\times 10^{12}\) flips per second, the links add \(0.39\) to \(0.78\)~pJ per flip to the \(0.29\)~pJ of the local logic. The estimate is conservative, because Eq.~\eqref{eq:fpbit_bound_appendix} includes a factor of two for packing and unpacking and every link is sized for the worst-case pair.

\subsubsection{A local 100 MHz digital core in 7 nm}

\begin{figure}[t]
  \centering
  \includegraphics[width=0.7\textwidth]{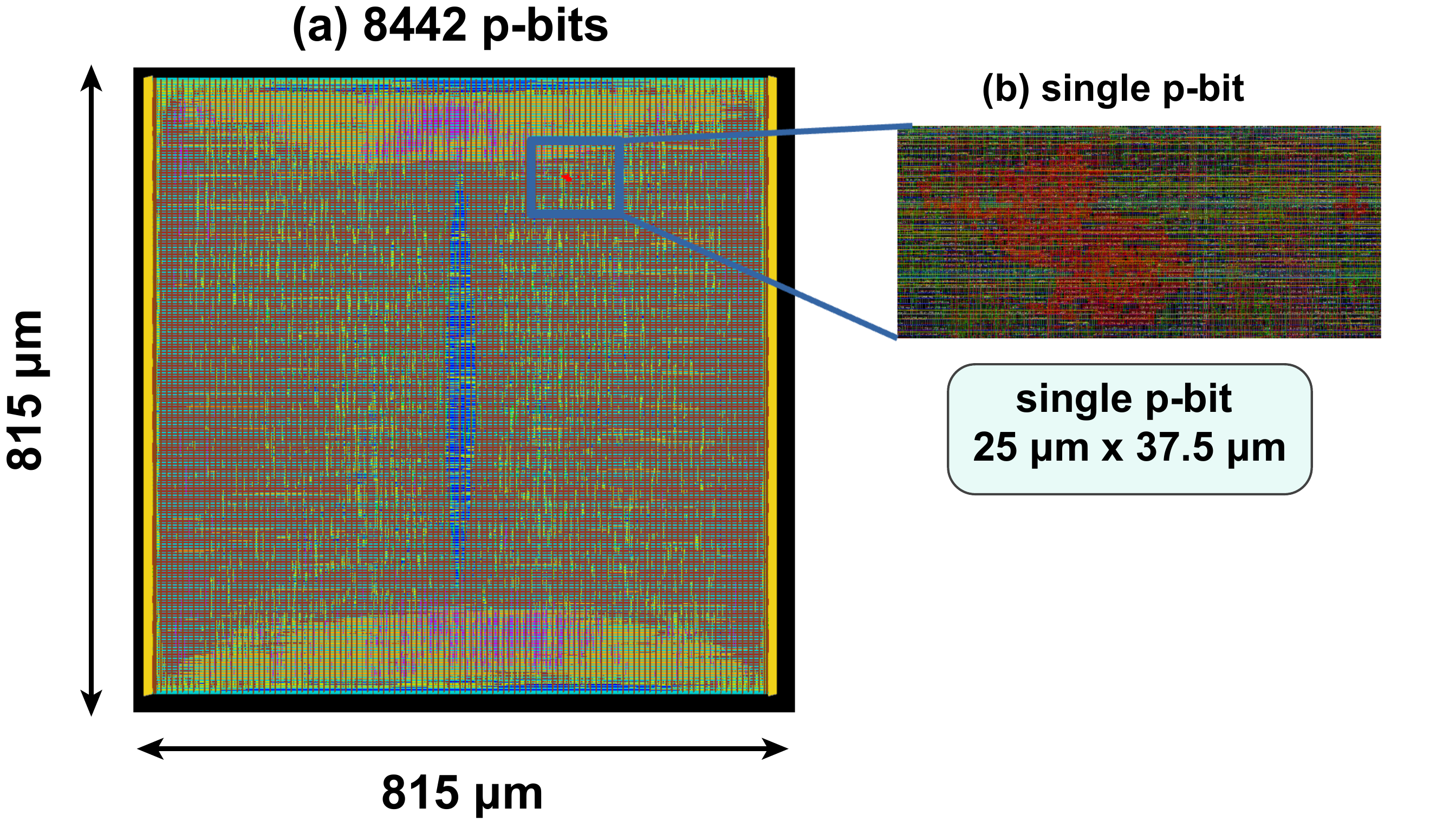}
  \caption{
  \textbf{7 nm physical design snapshot for one \(L^3=37^3\) partition.}
  \textbf{(a)} Place-and-route view of an 8442 p-bit partition (6 neighbors per p-bit) using s\(\{4\}\{1\}\) fixed-point, occupying a \(815\,\mu\)m \(\times\) \(815\,\mu\)m footprint. \textbf{(b)} The region labeled single p-bit, a $25\,\mu$m $\times$ $37.5\,\mu$m window with the logic of one p-bit highlighted in red. A p-bit's standard cells do not form a contiguous rectangular block, as the placement tools distribute them over the surrounding area during optimization. The design meets a 100~MHz target clock in ASAP7~\cite{clark2016asap7}.}
  \label{fig:supp_asic_layout}
\end{figure}

A link feasibility check is only meaningful if the compute core can match it.
To validate that the local update logic is not the limiting factor, we implemented one representative partition of the \(L^3=37^3\) system as a standard-cell design in the ASAP7 7 nm predictive process design kit (PDK)~\cite{clark2016asap7}.
The partition contains 8442 p-bits using the same fixed-point format s$\{4\}\{1\}$.
The resulting place-and-route closed timing at 100 MHz.
Representative metrics are summarized in Table~\ref{tab:supp_asic_summary}, and Fig.~\ref{fig:supp_asic_layout} shows the place-and-route layout. The shown physical dimensions are normalized (1/4 scale each) to account for how ASAP7 presents itself in the PDK.

\begin{table}[t]
  \centering
  \caption{\textbf{Representative metrics for one 8442 p-bit partition in 7 nm CMOS.}}
  \label{tab:supp_asic_summary}
  \begin{tabular}{l c}
    \toprule
    Metric & Value \\
    \midrule
    Partition size & 8442 p-bits \\
    Target clock & 100 MHz \\
    Timing status & Met constraints \\
    Total area & 0.66 mm\(^2\) \\
    Total power at 100 MHz & 248 mW \\
    Leakage power & 0.52 mW \\
    \bottomrule
  \end{tabular}
\end{table}

\FloatBarrier
\subsection{Max-Cut on the G81 Gset instance with 20,000 spins}
\label{sec:g81_app}

This section provides the full experimental details and a reproducible spin configuration for the G81 Max-Cut benchmark.
To demonstrate the distributed p-computer on a well-studied combinatorial optimization benchmark beyond spin glasses, we solve the G81 instance from the Gset library~\cite{Gset}.
G81 is a toroidal grid graph with $20{,}000$ spins and is among the largest and most intensively studied instances in the Max-Cut literature.
The algorithm used here is adaptive parallel tempering with isoenergetic cluster moves (APT+ICM)~\cite{chowdhury2025pushing}, a replica-based method in which multiple copies of the Ising graph run at different temperatures with periodic replica exchanges, augmented by non-local isoenergetic cluster moves that flip disagreeing spin clusters between replica pairs without changing their combined energy.
We employ 40 replicas organized as 10 inverse temperatures $\times$ 4 ICM replicas per temperature on DSIM-1 at $f_{\mathrm{p\text{-}bit}}=1$~MHz, the conservative limit that satisfies Eq.~\eqref{eq:extreme_fpbit_bound_main}, with $N_{\mathrm{color}}=2$ and fixed-point format s$\{4\}\{6\}$ (the wider format accommodates the distinct inverse temperatures required by adaptive parallel tempering), and run for $10^6$ Monte Carlo sweeps with one sweep per swap attempt.
Because the G81 graph with its 40 replicas does not fit on the 6-FPGA chain simultaneously, only one replica runs on-chip at a time. The 40 replicas are run sequentially via time-division multiplexing, and replica swap decisions are managed by a MATLAB host between successive on-chip runs.
The 10 inverse temperatures are $\beta = \{2.00, 2.05, 2.13, 2.22, 2.34, 2.52, 2.75, 3.13, 3.89, 5.61\}$, obtained from the APT preprocessing procedure~\cite{aadit2023accelerating}.

\begin{table}[t]
\centering
\caption{
\textbf{Comparison of Max-Cut results on the G81 Gset instance} ($20{,}000$ spins).
DSIM-1 matches the certified-optimal cut of $14{,}060$, first found by
the Cosm algorithm~\cite{zick2025performance,zick2026cosm} and proven
optimal with an exact solver~\cite{zick2026cosm}.
}
\label{tab:g81_comparison}
\vspace{0.5em}
\begin{tabular}{lc}
\toprule
\textbf{Solver} & \textbf{Highest G81 cut} \\
\midrule
This work (DSIM-1, APT+ICM) & 14{,}060 \\
Cosm~\cite{zick2025performance,zick2026cosm} & 14{,}060 \\
Algorithm portfolios~\cite{shylo2017algorithm} (2017) & 14{,}056 \\
GESPR~\cite{shylo2015teams} (2015) & 14{,}048 \\
Simulated bifurcation~\cite{goto2021high} (2021) & 13{,}992 \\
\bottomrule
\end{tabular}
\end{table}

\begin{figure}[t]
  \centering
  \includegraphics[width=0.7\textwidth]{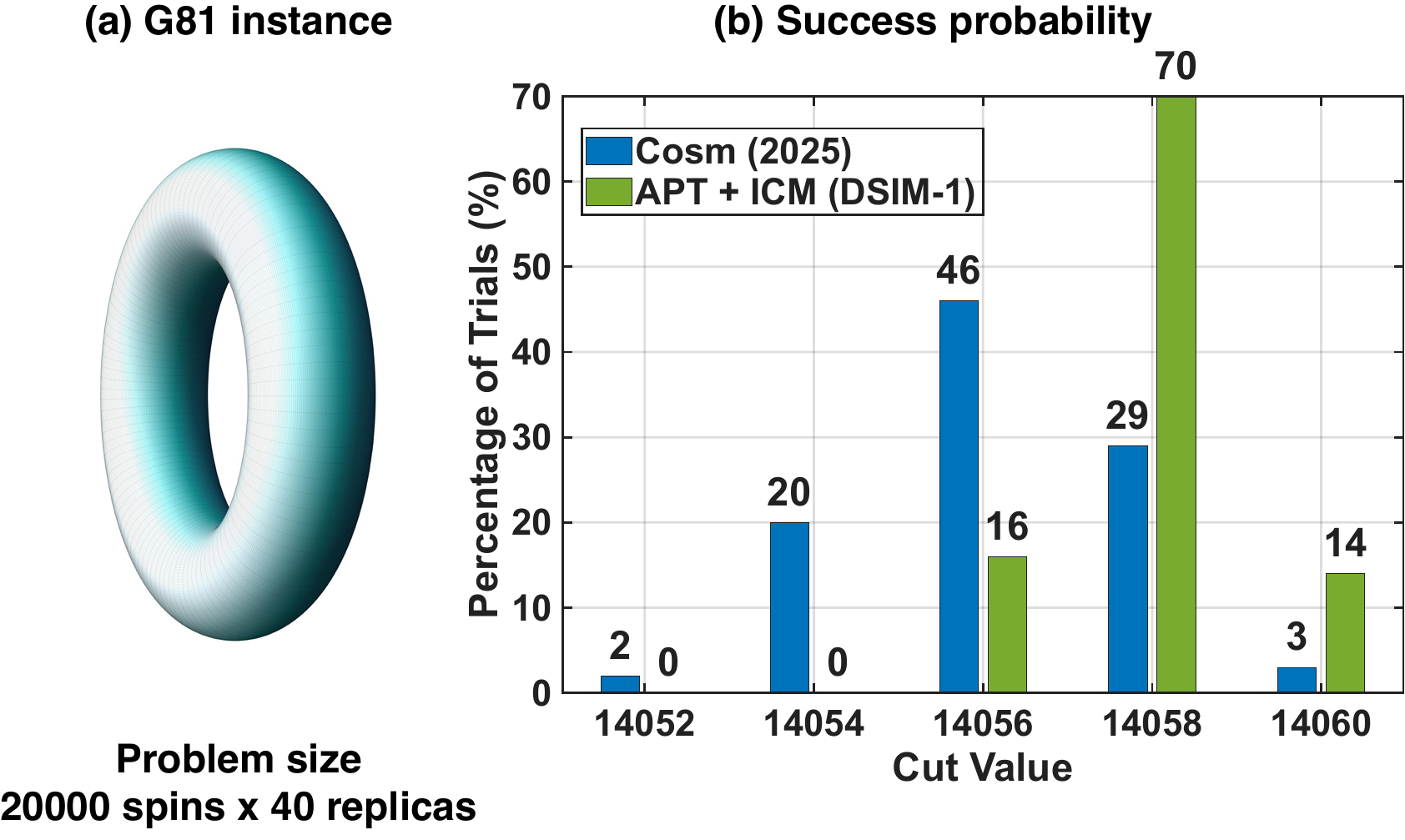}
  \caption{
  \textbf{Max-Cut on the G81 Gset instance with 20,000 nodes (DSIM-1).}
  \textbf{(a)} The G81 toroidal grid graph with 20,000 spins and 40 replicas.
  \textbf{(b)} Distribution of cut values found by APT+ICM on DSIM-1 across independent trials. The system achieves the certified optimal cut of $14{,}060$ in $14\%$ of trials and concentrates $70\%$ of trials at $14{,}058$. The Cosm algorithm reports $14{,}060$ in approximately $3\%$ of trials~\cite{zick2025performance} (different sweep definition).}
  \label{fig:supp_g81}
\end{figure}

Table~\ref{tab:g81_comparison} compares the DSIM-1 result against reported values from other solvers.
The APT+ICM run on DSIM-1 finds the maximum cut value of $14{,}060$ in $14\%$ of independent trials and the second-best value $14{,}058$ in $70\%$ of trials.
The ground state is degenerate: DSIM-1 identifies multiple distinct spin configurations that all achieve the $14{,}060$ cut value.
Fig.~\ref{fig:supp_g81} shows the G81 graph and the distribution of cut values across independent trials.
\begin{samepage}
For verification, we provide the hexadecimal representation of one spin configuration achieving the cut value of $14{,}060$.
Variables $1$ through $20{,}000$ are encoded in order. Expanding the hex string to binary and mapping $\{0,1\}\to\{-1,+1\}$ recovers the full spin assignment.

{\footnotesize
\begin{verbatim}
0F2D95FA4FCE926841977C9594C45766672BD108BAB90163DDA3AC738A0DA6A2CA24A4AED00A27357143D175551E3D
BBCCE4FAA4542E6AF69EA75E3A7FA996B68FCD499048C7FF44DA1F592B2CB9D6E71FC22468F20D0B911C380A442CC7
B554E09D1FC3C9B058A0AB60B6803FAEEF51254E00BD4F23DD2B82DF22F153B7E965CEEEB3ECC21E574A578C491352
530EF7ED762BB78B1EC651C5ED1F70D06494F2B487D481DBDE559348FBA86A39BC77494415CEDF4CC42FB2520FE0F1
6D847E9E3D0D38E700F6B5C86FDE9104F92ADBA0AC9EA7B0D24A21F897E6E5D4563E3B9BCED8E07BC329A814B04A83
EF538BDECE3CEE73C867BE7811121B86C138BEEC61868CBAFD908E36140EE9763E7D78C266DFF014B8CC6B63FD0BEA
566A5EE9084BB1176C352E1E30907A7BB78CD4B1876D55C401886E87025454DEEB760506FA780BBB2844EA955D2E83
75C76BD39972AA53344B0B12933B4E6289BDE4E31BA1EADDB9671C7F515452641B6D1B7A90381218FC46D0CF4F5641
261B0C7AAD8AFB1A7CE9DEA1EC2249DDC5C43710F69CDB5D9D74ED9772FAAF66B713F0D7593B283F686C8233CA8E70
B6929D66272EE40AEA488A62A6A669EFED41E6A1B6900E6A1D52B81FFD784B96AF8CDC9B9084A79046B4F8781C5D2E
0DB6B521DCEEF258823A372E681EE014DDEA211E8564B88625928BF1855B647610A32C562C9738648149312714FCB7
2CAB01A17F5E30C551AF22B70C3F8E4177C5B76F3794E07F441F1B3E46A140EB7142FD1D0B3C2CABB4F2D1E05F4D36
455AA60CC6305308592CD1C0D21FACCC03FD2602382FD7848217EB68A46F4FF497A38F812FF0967D0F7F877DB8CBCA
95C579FDCA351C2FA10F124B5B27B7DC9C6B7D888C203B31C88F9B3CD2289DE8AD9C5D8C0BE49585D91E2F5A6BD8F4
19DB38BBF429C70D805BB75312D96D5A739CB5E4CA174B26A17979D6822B1868876E84D5A4521747EB1DFB3F911A5C
308622823515350A29976C9D71BBAE69B3A1622DEE201FE2B7AFA22A2E99315ACB89DD685509BF9460F63F822CC1A8
BABFFBD522827926FB5CE7E5017771E07257C43A97D2F7626D0EDBC5876D2F5AFAD25644570D09D58466DE950D5FAA
1E57E9BE2C288B4128027202C15AA45C007ECB4FA84CDE3271142F1C005A92ADE964D3557624FC8722D2B4FD302466
4F84DF3600B272DD7DC36E1309501A3E9CBCA5781FB4B5AB5658A731B73758FCC7E15031C68CB98DB0C2609699FFDA
54954167965EF055D0395F8B0031FACDB2031EDB3A1347D073C905375C9A449E194CF310433C9A9D1660E75D25FBF5
5D84239DFCE251EB125644D82FF3AFE8A32E50095E0CF74B1ED9932DC7E1B19AF320F05CC7555AAFA83BC4D714BE86
4071FFDA56B834C9EE85ABF9A10006D18651F6C19786716BFCFE99FDC94FECC1E48FE68910C1B1C52983CC24E88668
3BE054BA16E99111D10DFDF9920E25120FE4A6003B8631E0B3575CC89CC9640794A780F8CA473F5927766CD9CEEC97
F2FED8E277BEE121EC24F94DEA744D0F74E5C4300EA91579D695B92E8087588FB2A3271E69B520B7D6D682273EEC4B
C3D420DE71912D17BA9CF5F208EBDB71961E6DF27915C4AAE20B6451D66E7605F3343C273B0DDB379E7FEEE81AF1BC
3D4AAEB7E70F24F5194DC6B19BD358265ED437793B66A46F91A745568D612513F277073CBAB64DC9248615ACBF668D
008E0E77B3A381A69D5394339D5B8762FA6099A2CFC08B21A152AFDD22CE68FCC2028E43542CE940FEB8ABEC5C5FD5
6135D783BD570B83ED7F0F0C6BF5EB5D2A2EF4D144442C97C2957DED615BF5BC2C13285AE8B250EBB5A7ECF475F1FD
0AAA087D24F20DDEA203C1EBA82DDD152BCC7B7917D71E5C675F301715012A60AD9B8C84C483E7BA7CCEF9DDE24D10
A445978D4FF5316DEA4CB5BB90D6D55264C1E6968BD3F51B7A08D13F403A67948EB9AE35115A75F198FC81DC24E411
383D03B23943B2148D7C1BA48A90475425A8C5F862ABDB5507DEBBBFA0002DD10D70EB6DFA8ACB065C00D153A777E0
A71C555C8438E4F649598E0808BE81910E2D32C3AEC5B86441E330156CB0DB2E9F7D7CB3A7937425BA27E10C222857
6CF81F496260EFCCCC53D5FDA2272EAF5C1921FE892C06BA27FF30796BCD14F217CAEB5A755B3D8830A215ACE4FD68
7B51DFF4CEA5F322D21723FF17D5422A0FB2B26B8B6B4CCE9E318E0C1682FC23CD77417A97D61366EFE1521E287595
E0F148E8D0E879A81D7EDBEA8216F7091F843B961F01BA93F1F5593FE0ED73DC66B6E5F113F6773F22183BB1AFA6E8
E798F572E15C38EB73CAE39A119B7EF9756FC9EFE111CF4DE9D7EEC95051468776A93032E7D0BD7493FEF7E4FA1817
16F7CE02D961B1E5A2593DD8E8DC170A319C6F650CE6CD23C393AF11B7B2EE274482FFDA13A1B14FADEF2443E93CD4
21B304D79197B9E3BA5512C3481400013A05AD8A1AD060905BBA1A8915F515B61EBEB98F52555B5F26C0A91F1FCBEE
7C7EBDF2006420228CBC6BB731C0929236C74497806876FDE4554364589E6AF15DE324AB711265F4F5621F3BF072E4
5CEB267F56ADFF390AE7CFFC52CB2FF37AC95974ABCF578238EBE1AF07715955867DD48CF20CFA099F80F1172B0B71
12F7280FF2048C3A88F9FC9AD8D5C2FA1FAC7185A9B694881D8DDE80B1F152A77FD81C029CEC0B3C869C107D01D04E
72DAB9E12B720CA7834894EE02CE67C4EA5AAD8C884BEE75C9DB24AB02005D64126A6CBFB36F9142F409B52B2729F1
B97A0785095F348138B852D73757B86D1C2D62DB10CA1C4004964252BDDBC732A954EB1082AA76EC5FCF23D204C760
1DF3BE8CEDD5C571DD314D67C60282B9595D557AE0A8A1D06CA5418D9639024721091969C06541AF82035F5C0E7672
40E572726AD2071E077CC6178D398FAF0CC0B0279B1ECF9755477D466DC1A9206C3ADD66AB2605EEACCF417DF609A3
59B651F681DC06866FE3DC12A87A72E30664C2B7383348824718631F14C60AC26E8B4BC9FEFC8BAE14AE9CC0577C79
ADB37F036B91714033C614ACA80098F0DCCBF950A08DF4DF20EF7B20BAFD5FA0A3071E0558817E4B40DE4F1BE14406
F1B8D852599E23D8425BC0E398BC31806E7E9C10A299CACDBC2BF44A6BAB2DDCE1BC9F76556DA76795C59C82F6C963
2DAF2F71BEC89EFFD9FFFE7DE2D28B4C33A9E15BEDB64ACAB0AFEE54447C4241B7D78F8CA7204896E6A471C4BFC78E
46CC6C4BA617CBB50C8D20C6B69B1F5D9375B384F0F5574CF087D48602A33EE804EB11C9483101F0B992AD51ED6A5C
2092D79C67CCE89688A9385CA2F50025E8DA76F02D93522BCBCB3857C2D23BE4F4A4A46DEF66C0F308801B7DE27ED6
E943CF526A7DCB73B3634437A168A4398FAA60D2620CC141518BEE671F6AAA33229898676E6169633F8DF4DEEDF6A0
713E449D862D28717CD30A9E5BC91F051A7C8D2DEE1E0FA0593578E07A8A0B89EFE2A73734AC445033091F36DFF460
32FAD75D7080B970BB
\end{verbatim}
}
\end{samepage}
\FloatBarrier
\subsection{SLR-driven partitioning at extreme scale}
\label{sec:slr_partitioning_app}

\begin{figure}[htbp]
  \centering
  \includegraphics[height=0.42\textheight]{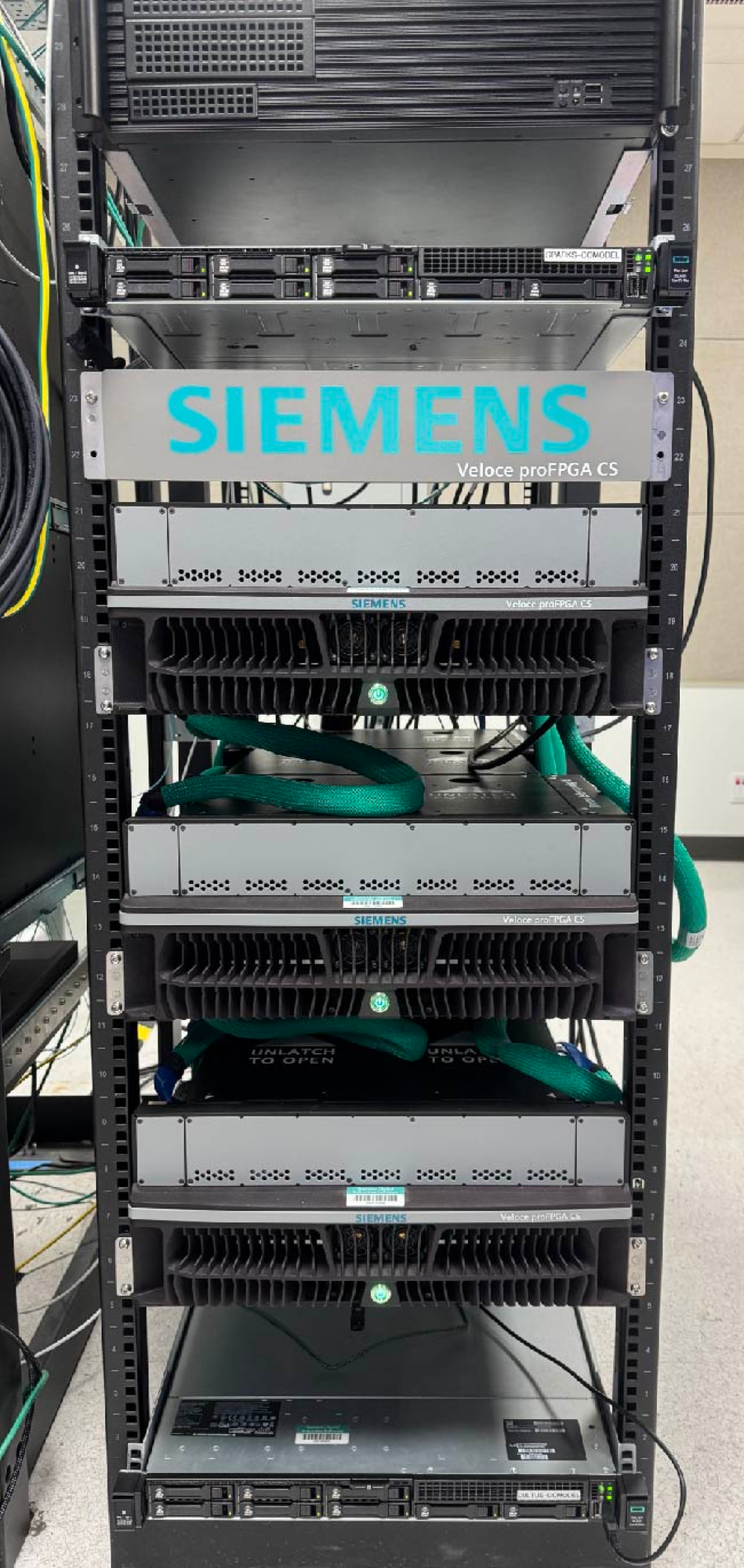}
  \caption{
  \textbf{DSIM-2 hardware.} The 18-FPGA Siemens Veloce proFPGA CS system used for the million-p-bit experiments.}
  \label{fig:supp_dsim2_photo}
\end{figure}

When scaling to one million p-bits on DSIM-2 (Fig.~\ref{fig:extreme_fig5} of the main text), partitioning across boards is not the only constraint.
On modern large FPGAs, physical resources are segmented into Super Logic Regions (SLRs), connected by a limited number of inter-SLR routing resources (super long lines).
Without an explicit min-cut at SLR boundaries, too many signals must cross between SLRs, exhausting the available inter-SLR routing and preventing the design from closing.
For the \(L^3=100^3\) EA benchmark, the quasi-2D spin glass on P103 and the large 3SAT benchmark, the implementation therefore introduces a second level of partitioning inside each FPGA.
Rather than treating each FPGA as a single partition, each SLR hosts a sub-partition with its own local weight memory and update logic.
With 18 FPGAs and 4 SLRs per FPGA, this yields 72 sub-partitions.
The same shadow-weight principle is applied at SLR boundaries: if an interaction crosses an SLR boundary, the associated weight is duplicated on both sides so that every SLR can compute its local fields from local memory, while only the required boundary p-bit states are exchanged across the internal boundary.
Fig.~\ref{fig:supp_dsim2_photo} shows the assembled 18-FPGA Siemens Veloce proFPGA CS system.

\subsection{Irregular planted Ising graphs on Pegasus and Zephyr topologies}
\label{sec:pegasus_zephyr_app}

The main text reports that the DSIM architecture handles non-cubic graph topologies without modification. This section presents the supporting data.

A planted Ising instance is one in which a ground-state spin configuration is chosen first and the coupling weights are then constructed so that this configuration is a ground state of known energy~\cite{navid2022nano,hen2019equation}. Because the ground-state energy is known by construction, planted instances provide a controlled benchmark for verifying that an optimizer reaches it.

\begin{figure}[t]
  \centering
  \includegraphics[width=0.95\textwidth]{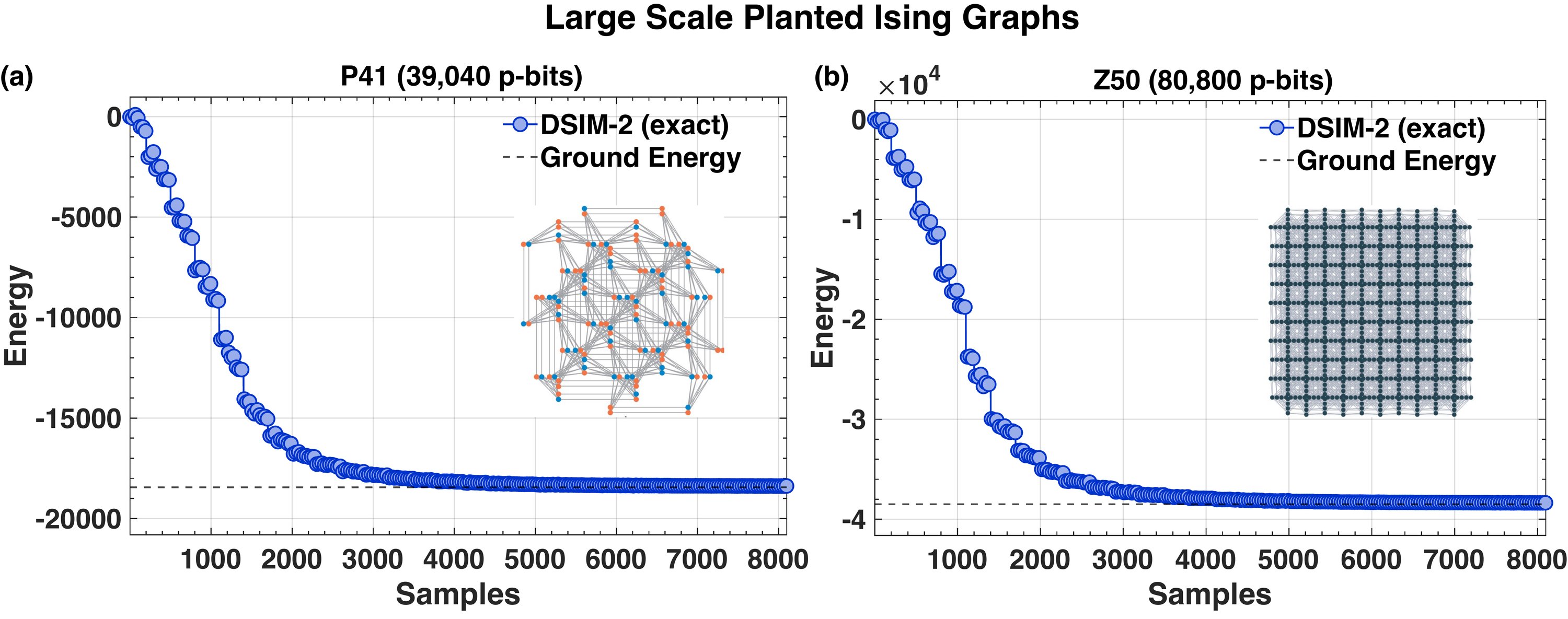}
  \caption{
  \textbf{Irregular planted Ising graphs on Pegasus and Zephyr topologies (DSIM-2).}
  \textbf{(a)} Energy trace for a Pegasus P41 instance with $39{,}040$ p-bits ($N_{\mathrm{color}}=4$, 2 FPGAs, $f_{\mathrm{p\text{-}bit}}=2.5$~MHz).
  \textbf{(b)} Energy trace for a Zephyr Z50 instance with $80{,}800$ p-bits ($N_{\mathrm{color}}=6$, 6 FPGAs, $f_{\mathrm{p\text{-}bit}}\approx 1.67$~MHz).
  Both are implemented on a subset of DSIM-2 with METIS partitioning, using fixed-point format s$\{4\}\{3\}$ and a simulated annealing schedule $\beta = 0.5, 0.625, \ldots, 10$.
  The horizontal axis is the readout sample index, with samples taken uniformly over a fixed $10^6$-sweep window. Dashed lines indicate planted ground-state energies.
  Single representative run for each topology.
  ``Exact'' denotes fresh boundary information at every update, as defined in main text Fig.~\ref{fig:extreme_fig4}.}
  \label{fig:extreme_fig6}
\end{figure}

Fig.~\ref{fig:extreme_fig6} shows representative energy traces for a Pegasus P41 instance ($39{,}040$ p-bits, $N_{\mathrm{color}}=4$) on 2 FPGAs at $f_{\mathrm{p\text{-}bit}}=2.5$~MHz, and a Zephyr Z50 instance ($80{,}800$ p-bits, $N_{\mathrm{color}}=6$) on 6 FPGAs at $f_{\mathrm{p\text{-}bit}}\approx 1.67$~MHz, both implemented on a subset of DSIM-2 with METIS partitioning.
Both use fixed-point format s$\{4\}\{3\}$ and a simulated annealing schedule $\beta = 0.5, 0.625, \ldots, 10$.
These traces are presented as single representative runs to demonstrate that the distributed architecture handles non-cubic topologies. A full statistical analysis with multiple instances and runs, as performed for the EA benchmarks, was not carried out for these planted instances because the primary goal here is a capability demonstration rather than a scaling measurement.
In both panels, the horizontal axis is the readout sample index, with samples taken uniformly over a fixed $10^6$-sweep window, and the dashed lines indicate the planted ground-state energies. The simulated annealing energy converges toward the planted ground state over the course of the run, confirming that the distributed p-computer successfully minimizes planted Ising instances on topologies native to quantum annealers.

\FloatBarrier
\subsection{Large invertible 3SAT instance with 250,011 p-bits}
\label{sec:3sat_app}

\begin{figure}[t]
  \centering
  \includegraphics[width=0.9\textwidth]{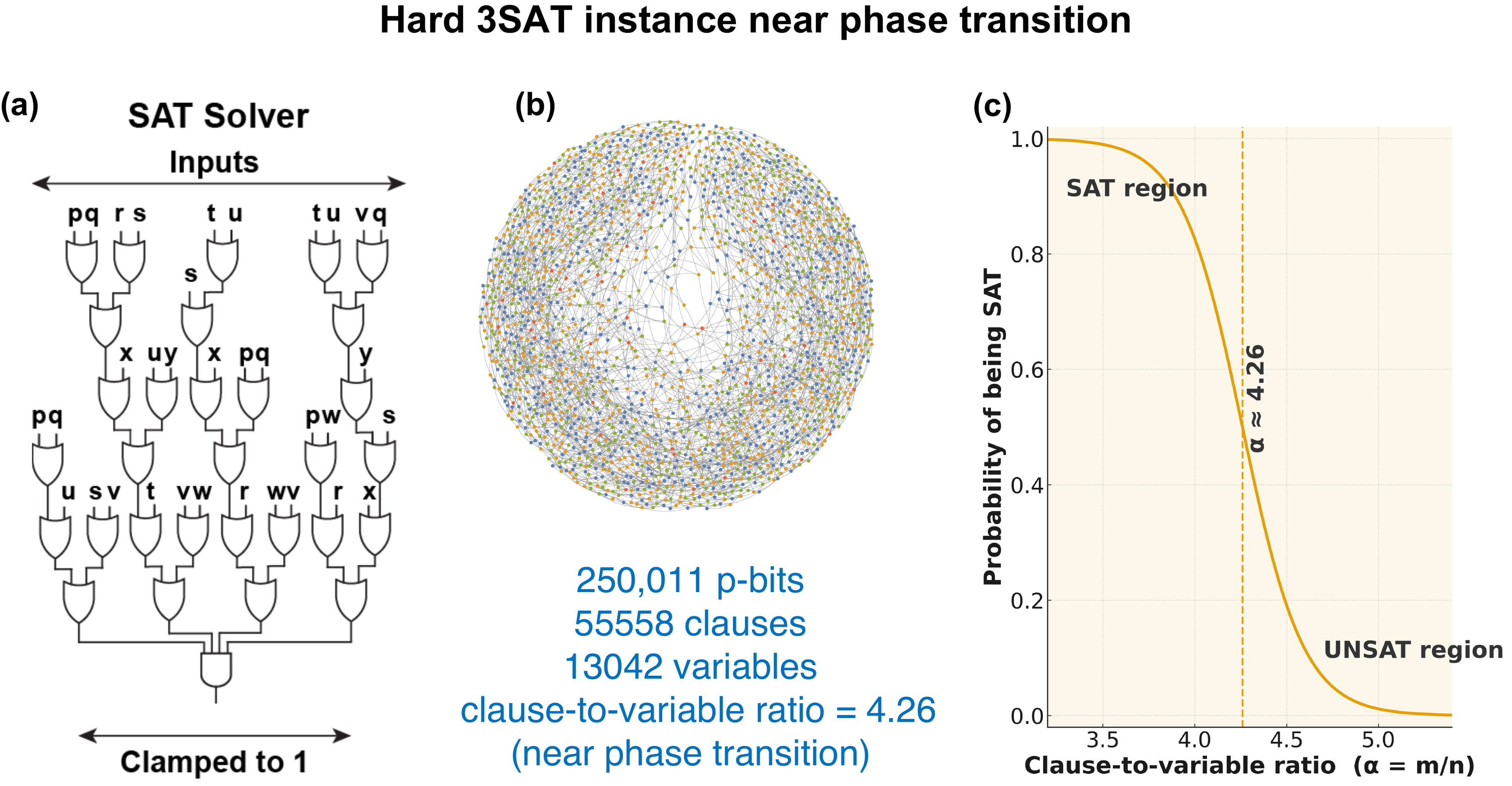}
  \caption{
  \textbf{Formulation of a large invertible 3SAT instance.}
  \textbf{(a)} Schematic of the invertible logic network used to encode 3SAT in an Ising circuit.
  \textbf{(b)} Visualization of the resulting sparse Ising graph after copy-gate sparsification, with 250,011 p-bits.
  \textbf{(c)} Schematic SAT probability curve versus \(\alpha=m/n\), with the chosen instance at \(\alpha\approx 4.26\) near the phase transition~\cite{mezard2002analytic}.
  Copy conflicts are resolved by majority vote over duplicated variables when decoding.}
  \label{fig:supp_3sat_formulation}
\end{figure}

We now move from structured Pegasus and Zephyr topologies to a highly irregular satisfiability benchmark.
The EA spin-glass benchmarks in the main text use a regular lattice geometry, but a distributed p-computer is not limited to that case.
Once weights are stored in local memories and only boundary states are exchanged, the same update logic can operate on highly irregular sparse graphs, provided the graph coloring schedule is chosen to respect the connectivity.
To stress this capability on a hard irregular instance, we consider a random 3-literal satisfiability (3SAT) problem near the satisfiability phase transition~\cite{mezard2002analytic}.
Prior pairwise Ising demonstrations of satisfiability (SAT) have been limited to much smaller scales~\cite{cilasun20243sat}; recent work has employed native higher-order interactions to encode clauses more compactly~\cite{kim20258k}.
The instance parameters are
\[
m = 55{,}558 \text{ clauses}, \qquad
n = 13{,}042 \text{ variables},
\]
so the clause-to-variable ratio is \(\alpha=m/n\approx 4.26\).
The instance was generated using the CNFgen tool~\cite{lauria2017cnfgen} as a uniform random 3SAT formula, where each clause is drawn independently and uniformly from all possible 3-literal clauses over \(n\) variables.
Because $\alpha\approx 4.26$ lies at the satisfiability phase transition for random 3SAT (theoretical threshold $\alpha_c\approx 4.267$~\cite{mezard2002analytic}), satisfiability of this particular instance is not guaranteed. The goal is to stress the DSIM on a large, hard, irregular graph.
After mapping the problem to an Ising model through an invertible logic circuit and applying copy-gate sparsification~\cite{camsari2017stochastic,aadit2022massively,grimaldi2022spintronics}, the resulting sparse graph contains \(250{,}011\) p-bits.
This scale is reached because the mapping introduces auxiliary copy p-bits~\cite{aadit2022massively} that preserve locality and sparsity while keeping the energy function consistent with the original SAT constraints.
Fig.~\ref{fig:supp_3sat_formulation} summarizes the formulation at a systems level.
The first panel sketches the invertible logic network used to represent clauses and variable literals, the second panel shows the resulting sparse graph, and the third panel places the chosen instance on the standard SAT probability curve versus \(\alpha\).
Copy conflicts introduced by sparsification are resolved by majority vote when decoding a logical assignment: when multiple p-bit copies represent the same logical variable, the logical value is taken as the majority across those copies.

\begin{figure}[t]
  \centering
  \includegraphics[width=0.7\textwidth]{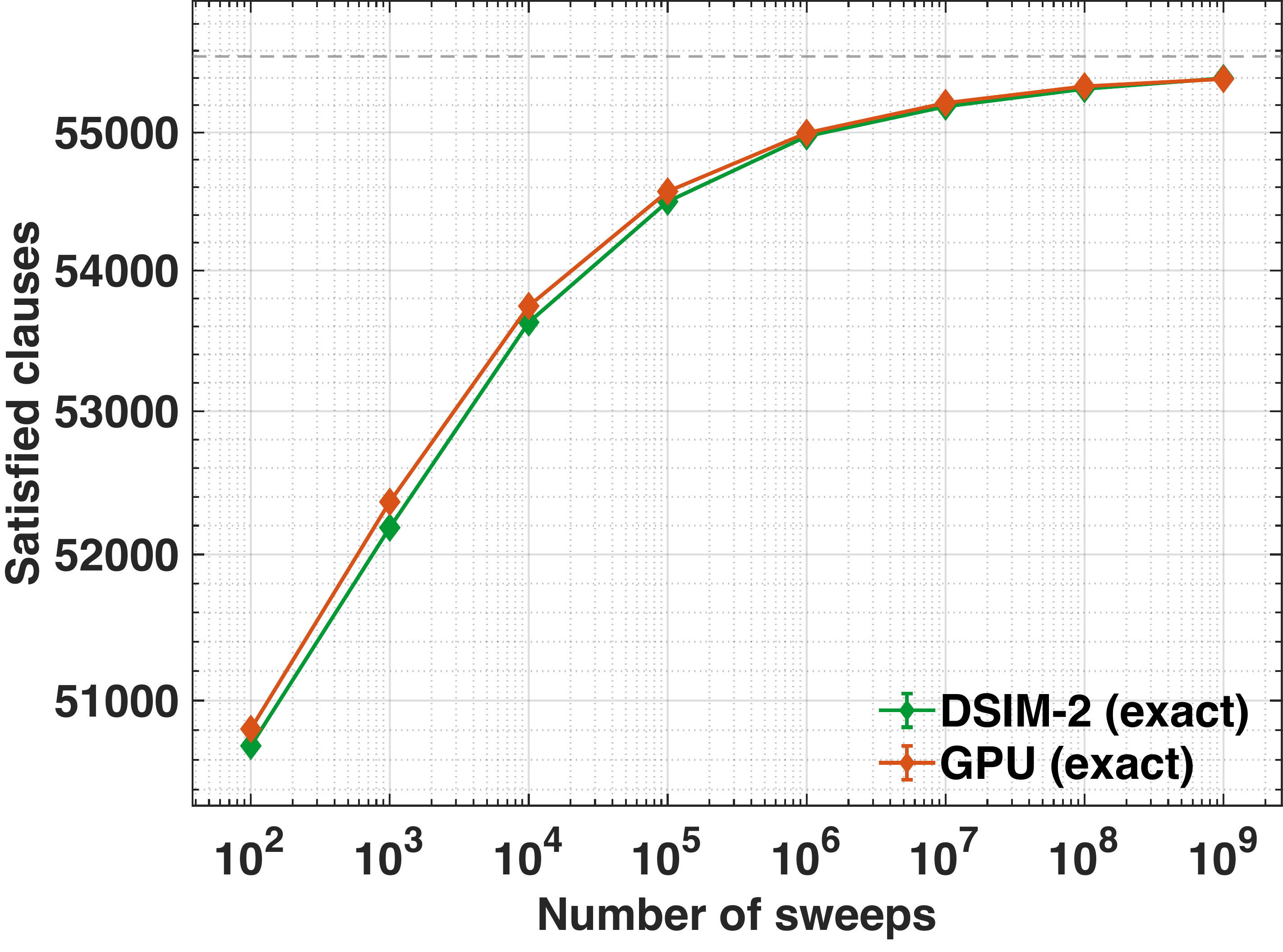}
  \caption{
  \textbf{Scaling on a large invertible 3SAT instance with 250,011 p-bits (DSIM-2).}
  Average best satisfied clauses per run versus sweeps for the instance of Fig.~\ref{fig:supp_3sat_formulation}, implemented on DSIM-2 at $f_{\mathrm{p\text{-}bit}}=0.5$~MHz and on an NVIDIA RTX 6000 Ada GPU under the same schedule.
Each point is the mean of the best (maximum) satisfied-clause count found within each of 10 independent runs. Error bars denote 95\% bootstrap confidence intervals across runs.
At $10^9$ sweeps the single best run (DSIM-2) satisfies $55{,}416$ of $55{,}558$ clauses ($99.74\%$), with $142$ clauses remaining unsatisfied. The dashed line marks the total number of clauses, $55{,}558$. ``Exact'' denotes fresh boundary information at every update, as defined in main text Fig.~\ref{fig:extreme_fig4}.}
  \label{fig:supp_3sat_scaling}
\end{figure}

To evaluate progress, we track the number of satisfied clauses versus sweeps under a fixed annealing schedule.
The instance is implemented on DSIM-2 at $f_{\mathrm{p\text{-}bit}}=0.5$~MHz with METIS partitioning, using a 4-color schedule (\(N_{\mathrm{color}}=4\)) and the fixed-point format s$\{4\}\{3\}$.
The simulated annealing schedule is $\beta = 0.5, 0.625, \ldots, 10$.
For comparison, we run the same schedule on an NVIDIA RTX 6000 Ada GPU.
Fig.~\ref{fig:supp_3sat_scaling} shows satisfied clauses versus sweeps on a semilog scale.
The curves track closely in their asymptotic improvement versus sweeps, indicating that DSIM-2 reproduces the scaling behavior of an optimized classical baseline even on a highly irregular graph with a larger coloring requirement than the EA lattice.

At $10^9$ sweeps the best single run on the FPGA satisfies $55{,}416$ of $55{,}558$ clauses ($99.74\%$) with $142$ clauses remaining unsatisfied. The GPU reaches a comparable level ($55{,}408$ best, $150$ unsatisfied).
The instance is not fully solved by either platform under the annealing schedule used, consistent with the extreme difficulty of random 3SAT near the phase transition at this scale.
Nevertheless, the close agreement between the two platforms over seven orders of magnitude in sweep count confirms that DSIM-2 preserves optimization scaling even on highly irregular graphs.

\subsection{Architectural comparisons}
\label{sec:arch_comparisons_app}

The main text argues that the DSIM tolerates boundary delay where synchronous multi-chip architectures do not. The difference lies in the update rule and its implementation, not in the randomness of the updates. A synchronous machine keeps the dependencies of every iteration, so every exchange between chips adds to the iteration time. The DSIM lets each device keep updating with the most recent boundary states it has received, and we measure how the dynamics change. This does not imply that deterministic algorithms cannot tolerate delayed information. Below we compare the DSIM with specific architectures.

GPU and tensor processing unit (TPU) clusters attack the memory wall with high-bandwidth memory and high-bandwidth interconnects~\cite{jouppi2017datacenter,nvidia2017nvidia}.
That strategy is well matched to dense deterministic workloads, but it depends on substantial communication and memory infrastructure to keep the computation synchronized across devices.
Wafer-scale systems such as Cerebras~\cite{lie2023cerebras} take a different route by minimizing package boundaries and keeping a very large amount of static random-access memory (SRAM) on one piece of silicon, reducing off-chip traffic at the cost of an unusually large monolithic implementation.

Other multi-chip accelerators keep the modularity of smaller dies while trying to soften the cost of communication.
Simba~\cite{shao2019simba} is an example of a tiled multi-chip accelerator, while Illusion and related architectures~\cite{radway2021illusion,srimani2024next} treat a network of chips as an approximation to a larger virtual system and identify inter-chip bandwidth as a central scaling constraint.
CHIMERA~\cite{prabhu2022chimera} further uses nonvolatile resistive random-access memory (RRAM) to reduce idle energy through temporal gating.
Across these systems, the common challenge is the same: once the model no longer fits on one device, communication and memory movement begin to shape performance.

The DSIM differs in both what is communicated and how much delay can be tolerated.
By storing all coupling weights on-chip through shadow weights, the DSIM limits inter-device traffic to 1-bit boundary states.
Delayed boundary information then changes the dynamics gradually. A single parameter, $\eta$, sets the tradeoff between accuracy and throughput.
As demonstrated in the main text, this allows the designer to trade a small exponent penalty for higher throughput, making multi-chip scaling a tunable design knob and opening a path to larger probabilistic computers built from networks of smaller devices.

\clearpage
\bibliographystyle{unsrtnat}
\bibliography{references}

\end{document}